\documentclass[prb,aps,epsfig,onecolumn,showpacs]{revtex4}
\usepackage{color}
\usepackage{epsfig,amssymb,amsmath,latexsym}
\usepackage{subfigure}
\usepackage{wrapfig}
\usepackage{float}
\usepackage{color}
\usepackage{bm,bbm,dsfont,ulem,nicefrac}
\usepackage[colorlinks=true,linktoc=page,linkcolor=magenta,citecolor=magenta]{hyperref}

\newcommand\bea{\begin{eqnarray}}
\newcommand\eea{\end{eqnarray}}
\newcommand\beq{\begin{equation}}  
\newcommand\eeq{\end{equation}}

\begin{document}
\title{\bf{Introduction to abelian and non-abelian anyons}}
\author{Sumathi Rao}
\affiliation{ Harish-Chandra Research Institute, Chhatnag Road, 
Jhusi, Allahabad 211 019, India. }

\begin{abstract}
In this set of lectures, we will  start with a brief pedagogical  introduction to abelian anyons and their properties. This will
essentially cover the background material with an introduction to  basic concepts in anyon physics, fractional statistics, braid groups and abelian
anyons. The next topic that we  will  study is  a specific exactly solvable model, called the toric code model, whose excitations have (mutual)
anyon statistics.  Then we will go on to discuss non-abelian anyons, where we will use the one dimensional
Kitaev model as a prototypical example to produce Majorana modes at the edge. We will then explicitly derive the non-abelian unitary matrices under exchange
of these Majorana modes.

 \end{abstract}
%\pacs{73.23.-b, 73.63.Nm, 71.10.Pm}
\maketitle

\section{Introduction}

The first question that one needs to answer  is  why we are interested in anyons\cite{leinaas}. Well, they are new kinds of excitations
which go beyond the usual  fermionic or bosonic modes of excitations,   so in that sense they are like new toys to play with! But it is not just that they
are theoretical constructs -  in fact, quasi-particle excitations have been seen in the fractional  quantum Hall (FQH) systems,  which 
seem to obey these new kind of statistics\cite{stern}. Also, in the last decade or so, it has been realised
that if particles obeying non-abelian statistics could be created, they would play an extremely important role in
quantum computation\cite{anyonsqc}. So in the current scenario, it is clear that understanding the basic notion of exchange statistics
is extremely important.

So if we want to start by explaining\cite{anyonprimer,anyonlerda,anyonkhare} what an `anyon' is, to someone who may be hearing the word for the first time,
we can tell them that just as fermions are particles obeying Fermi-Dirac statistics and bosons are particles obeying
Bose-Einstein statistics, anyons are particles  obeying `any' statistics. Clearly, they are not as ubiquitous as bosons and
fermion, else they would have been just as familiar to everyone as bosons and fermions.  But as we shall see  later in this
lecture, even theoretically, anyons can only occur in two dimensions, whereas the world is three dimensional.  So it is
only in planar systems, or in systems where the motion in the third dimension is essentially frozen, that excitations can 
be anyonic.

Hence, although the  theoretical possibility of anyons was studied as early as 1977\cite{leinaas}, it shot into prominence
only in the late eighties and early nineties, when not only the excitations in the FQH systems  were found to be anyonic, for
a while, there was also speculation that anyons could explain the unusual features of 
high temperature superconductivity\cite{anyonsuperconductivity}.

%--------------- Fig 1 ----------
\begin{figure}
 \includegraphics[width=0.45\textwidth]{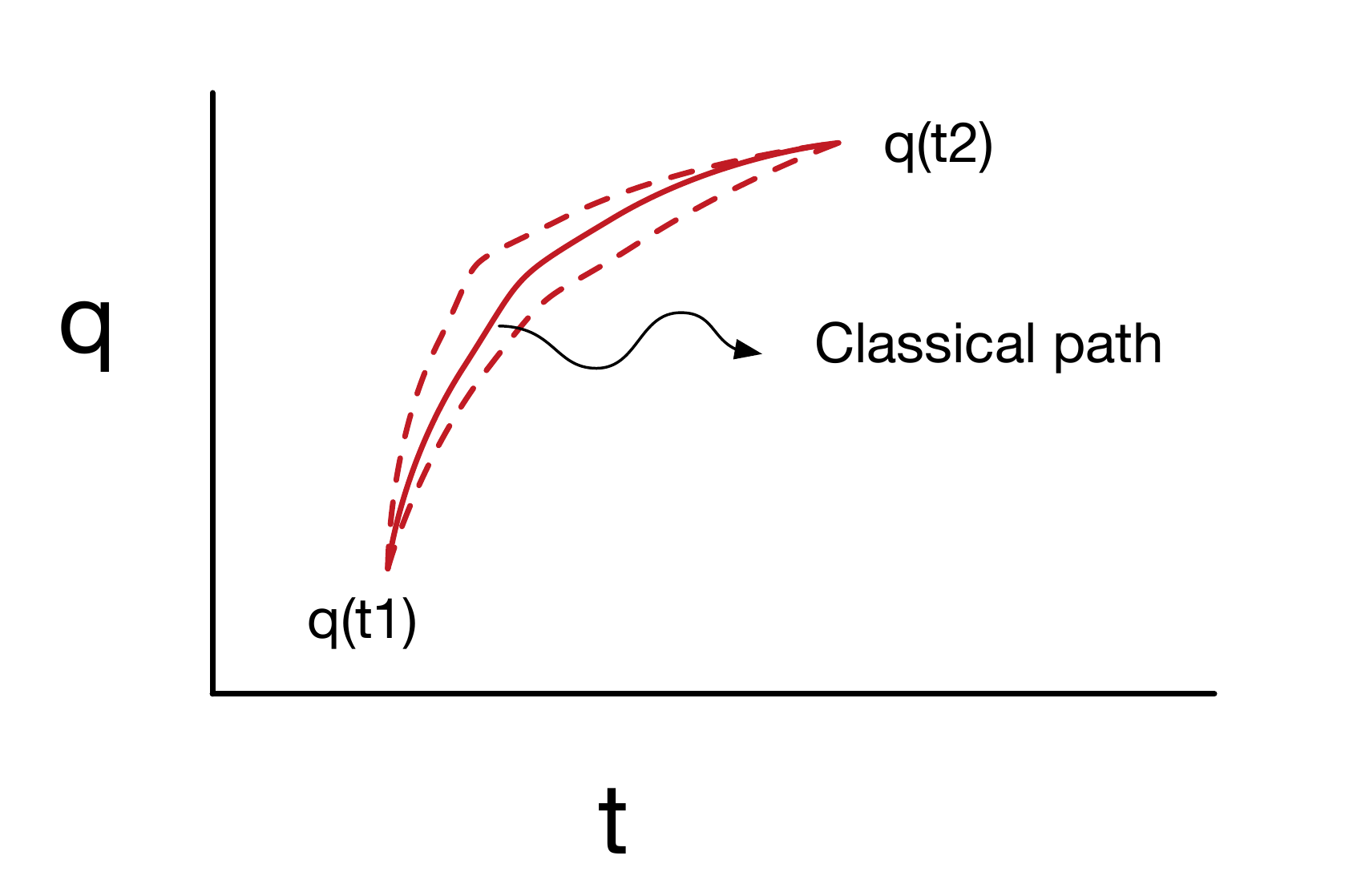}
 \caption{ Sum over paths from $q(t1)$ to $q(t2)$ }
 \label{lec1fig1}
	\end{figure}
%--------------------------------------- 

The easiest way to understand the notion of phases and statistics under exchange of particles in quantum mechanics
is to think about  how particles move around each other and from this point of view, the easiest way of understanding the
quantum motion of these particles is via path integrals. Here,  we will  assume that you have some familiarity with the idea
of path integrals, although not many details will be required.  To recollect it, we  just mention the following few things. 
In quantum mechanics, the probability amplitude to go from one space-time point to another is given by
\beq
A = \sum_{paths} e^{i\mathcal{S}}~,
\eeq
where $\mathcal{S}= \int \mathcal{L}dt$ is the action for the particular trajectory or path.
 In other words, quantum mechanically, we need to include all possible paths
 between the initial and final points of the trajectory sSee Fig.\ref{lec1fig1}).  But most of these paths
 will interfere destructively with each other and hence will not contribute
 to the probability amplitude. The only exception is the classical path and
 the paths close to it, which interfere constructively with the classical path - i.e.,
 the most important contribution will be from the classical path. That is all the
 information that we need here.

With this introduction, we come to the details of what we will study here. In Sec.(II), we
will explain the basic notions of anyon physics and why they can exist only in two spatial dimensions.
We will analyse  a simple physical model of an anyon and use it to understand the quantum mechanics
of two  anyons and see that even in the absence of any interactions, it needs to be studied as an
interacting theory, with the interactions arising due to the anyonic exchange statistics.
Then, in Sec.(III), we will  study the exactly solvable toric code model as an example of a system with
anyonic excitations. Finally, in Sec.(IV), we will discuss non-abelian statistics, where again, we will
explain many features of non-abelian anyons using the one-dimensional Kitaev model
as a typical example.

\section{Abelian anyons}

\subsection{Basic concepts  of anyon physics}

The term `exchange statistics'
 refers to the phase picked up by a wave-function when two identical particles are
exchanged. But this definition is slightly ambiguous. Does statistics refer to the phase picked up by the
wave-function  when all the quantum numbers of the particles are exchanged (i.e., under permutation
of the particles) or the actual phase that is obtained when two particles are adiabatically transported
giving rise to the exchange? In three dimensions, these two definitions are equivalent but not in two
dimensions. In quantum mechanics, we deal with interference of paths of particles and hence, it
is the second definition which is more relevant, and we will show how it can be different from the first definition.

%--------------- Fig 2  ----------
\begin{figure}
 \includegraphics[width=0.35\textwidth]{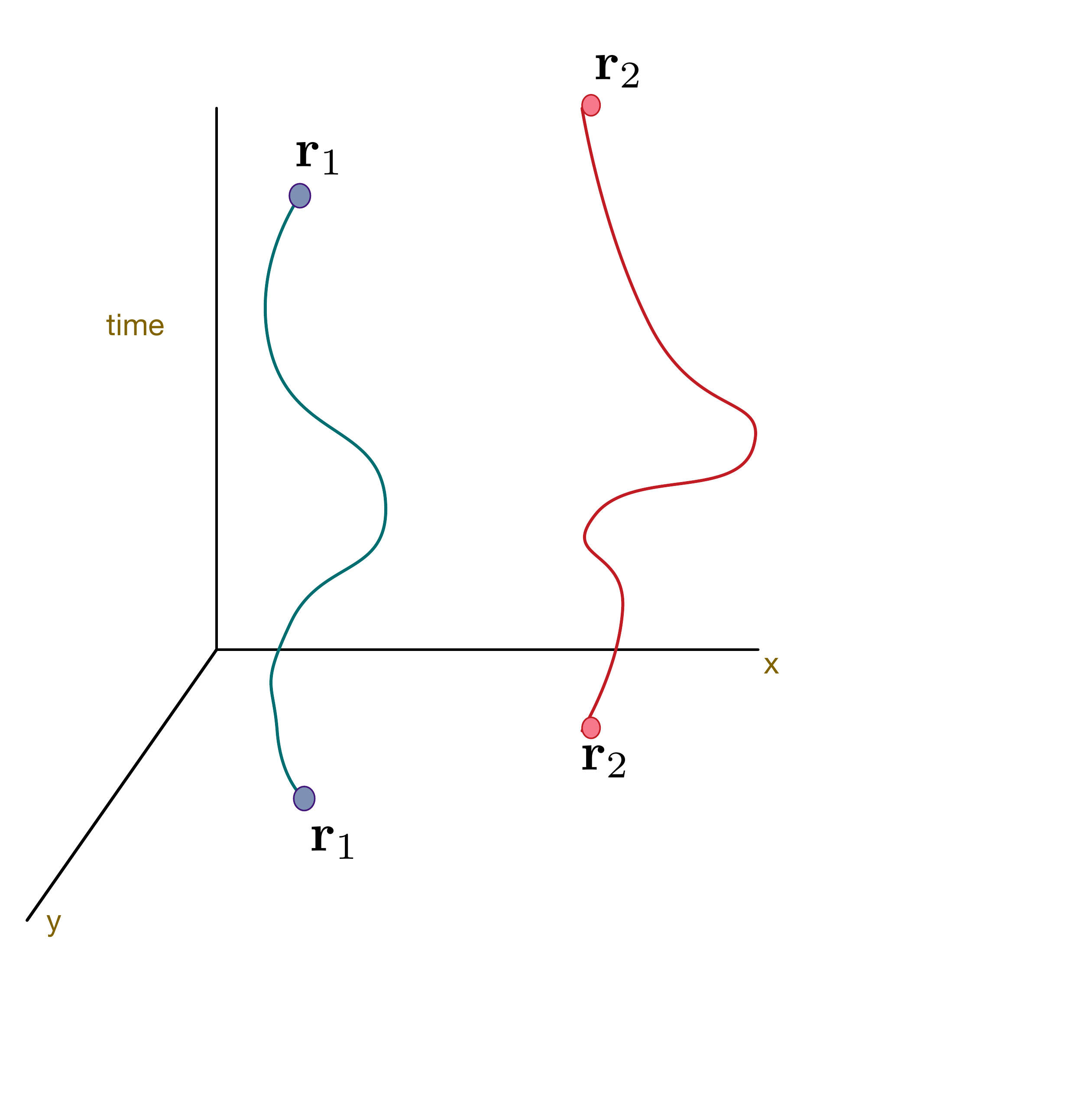}
 \includegraphics[width=0.35\textwidth]{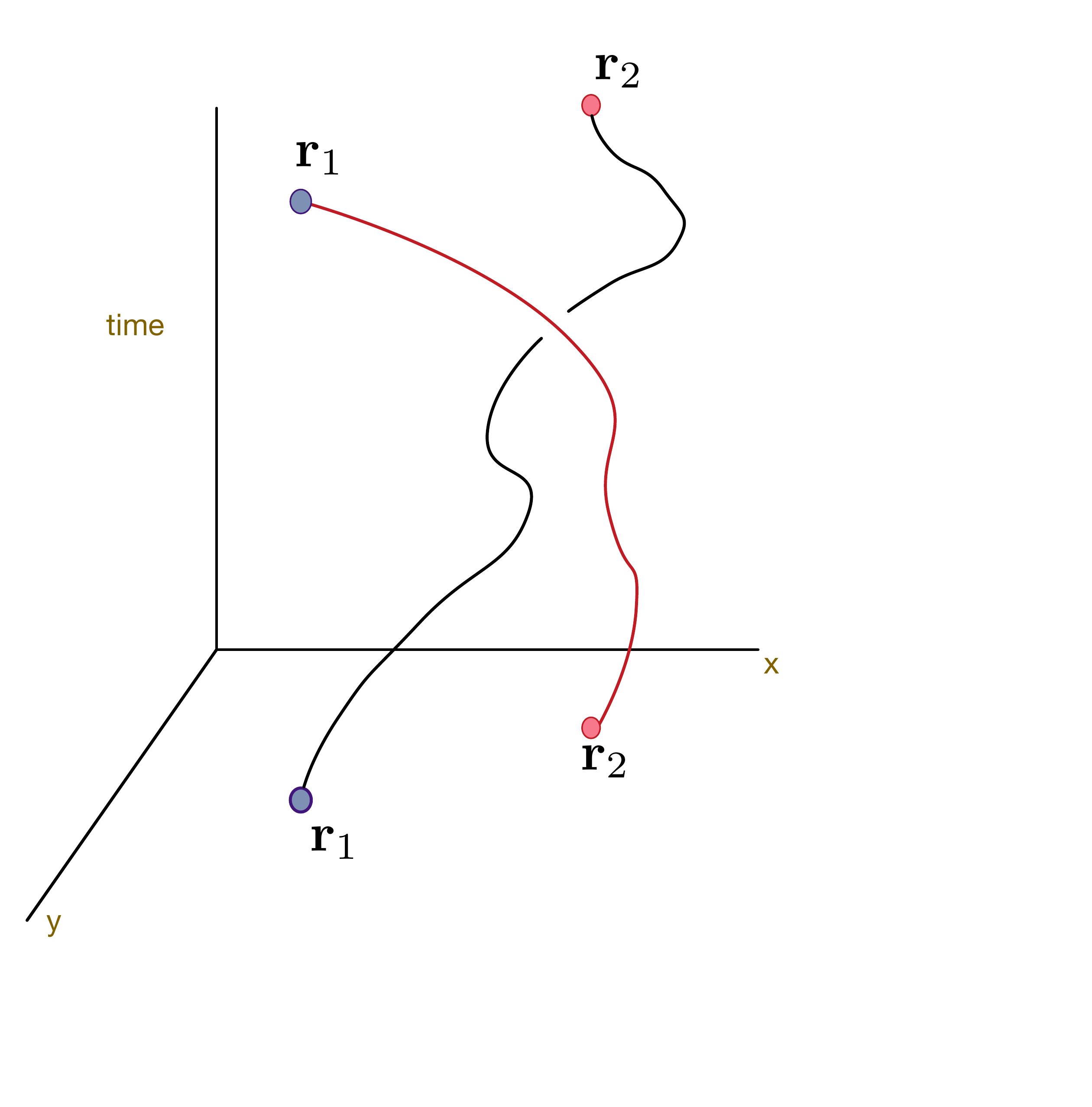}
 \caption{Direct and exchange paths}
 \label{lec1fig2}
	\end{figure}
%--------------------------------------- 

Let us first consider the statistics  under exchange of  two particles in three dimensions. By the path integral prescription,
the amplitude for a system of particles that moves from (${\bf r}_1(t_1),{\bf r}_2(t_1)$) to (${{\bf r}_1}^\prime(t_2),{{\bf r}_2}^\prime(t_2)$)
is given by 
\beq
A = \sum_{paths}e^{i\int_{t_1}^{t_2} dt \mathcal{L}[{\bf r}_1(t),{\bf r}_2(t)]} ~.
\eeq
If the two particles are identical, then there are two classes of paths (see Fig.\ref{lec1fig2}).
How do we see this?
If we use the convention that we always refer to the position of the first particle first and the second
particle second, then we see that the final configuration remains the same whether we have
 (${{\bf r}_1} (t_2),{{\bf r}_2}(t_2$)) or  (${{\bf r}_2}(t_2),{{\bf r}_1}(t_2)$)
 because the two particles are identical. Even though the particles are exchanged in one path and not
 in the other, the final configuration is the same.
In terms of the centre of mass (${\bf R} = ({\bf r}_1 + {\bf r}_2)/2$) and relative coordinates (${\bf r} = {\bf r}_1 - {\bf r}_2$),
we see that the centre of mass motion is the same for both the paths, but the relative coordinate changes for both the paths.
Also, since the CM motion moves both the particles together, it is independent of any possible phase under exchange.
For convenience in visualising the configuration space, let us keep $|{\bf r}|$ fixed and non-zero, i.e, the two particles
do not intersect. Then,  the vector ${\bf r}$ takes values on the surface of the sphere.

%--------------- Fig 3  ----------
\begin{figure}
 \includegraphics[width=0.75\textwidth]{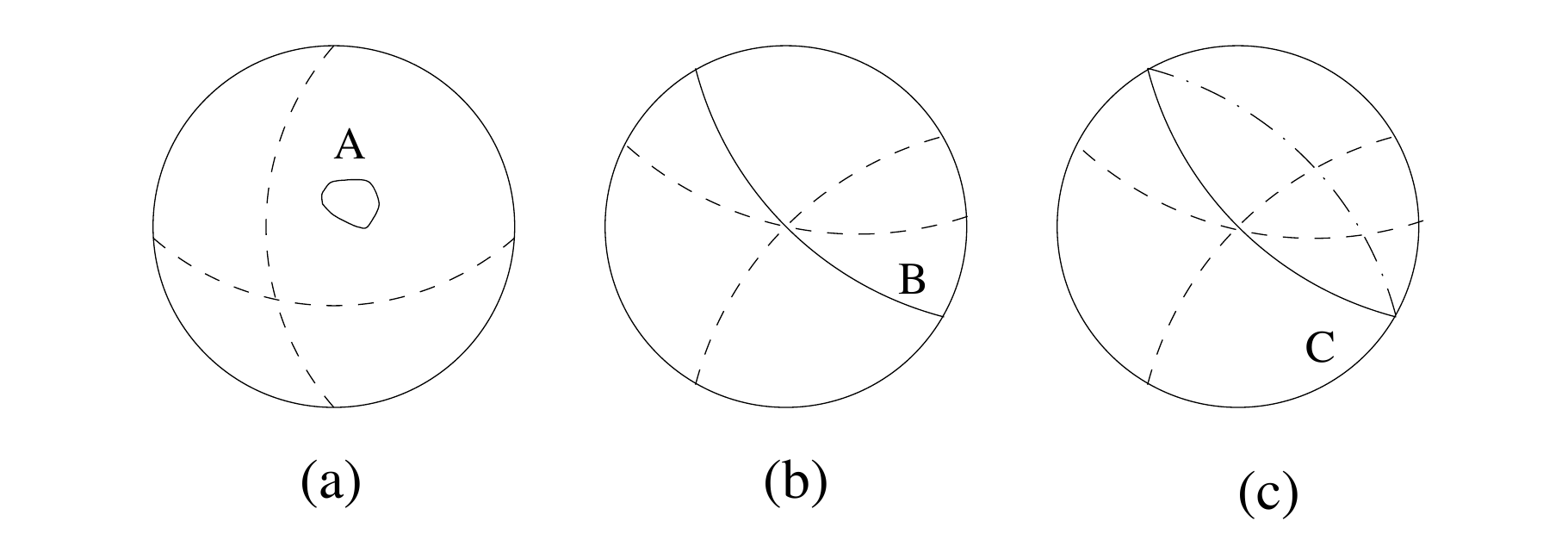}
 \caption{Paths in three dimensional configuration space with fixed radius}
 \label{lec1fig3}
	\end{figure}
%--------------------------------------- 

Now let us draw the paths in configuration space as shown in Fig.\ref{lec1fig3}.
 It is clear that the paths can only move along the surface
of the sphere as the two particles move, since  $|{\bf r}|$ is  fixed. But once they get back to their original positions
or get exchanged, since they are indistinguishable, the path is closed. In other words, closed paths on the surface
of the sphere are formed by the particles coming back to their original positions (no exchange)  or  going
to the antipodal point  (${\bf r} \rightarrow -{\bf r}$  or getting exchanged).  But if we exchange the particles another
time, then $\bf r$ comes back to itself, after having gone around the sphere once. In terms of diagrams, this is shown in
Fig.\ref{two}.
Since we have eliminated coincident points, the wave-function is non-singular and well-defined at all points in the configuration
space, and consequently on the surface of the sphere.  So the phase picked up by the wave-function is also well-defined and does
not change under continuous deformations of the path.
Let us consider the  possible phases of the wave-function when the motion of the particles is along each of the three paths, - 
A (no exchange), B (single exchange) and C (two exchanges) - depicted in Figs. \ref{lec1fig3}(a),(b) and (c). Path A is a closed path which
does not involve any exchange and can clearly be shrunk to a point. Hence, the wave-function cannot pick up any phase other than unity.
Path B  involves the exchange of the two particles and goes
from a point on the sphere to its diametrically opposite point. Since the two end points are fixed, this path cannot be shrunk to a single
point. So this exchange can have a non-trivial phase  in the wave-function .  However, path C which forms a closed loop on the surface
and involves two exchanges can be continuously shrunk to a point by imagining the path to be a physical string looped around a sphere.
So this again cannot pick up any phase.  Let $\eta$ be the phase picked up under a single exchange. Since two exchanges are
equivalent to no exchange, $\eta^2 = +1 \Longrightarrow \eta = \pm 1$. Hence, the only statistics possible in three dimensions are
Fermi statistics or Bose statistics.

With slightly more mathematical rigour, one can say that the configuration space of relative coordinates is given by $({\mathbb{R}}_3 - origin)/\mathbb{Z}_2$. Here ${\mathbb{R}}_3$ is just
the three dimensional Euclidean space spanned by the relative coordinate ${\bf r}$. We subtract out the origin because we have assumed
that paths do not cross (which is true for all particles other than bosons, because of hard core repulsion and for bosons, is not relevant
anyway, because the exchange phase is unity). The division by ${\mathbb{Z}}_2$ is because of the identification of ${\bf r}$ with $-{\bf r}$, which is 
because the particles are indistinguishable. To study the
phase picked up by the wave-function of a particle as it goes around another particle, we need to classify all paths in this configuration space.
The claim,  from the pictorial analysis above, is that there are just two classes of paths. Mathematically, this is expressed in terms of the
first homotopy group $\Pi_1$ of the space, which is the group of inequivalent paths (paths not deformable to each other), passing through
a given point in the space, with group multiplication being defined as traversing paths in succession and group inverse as traversing a path
in the opposite direction. Thus
\beq
\Pi_1(\mathbb{R}_3 - origin)/\mathbb{Z}_2)  =   \Pi_1 (\mathbb{RP}_2)  =  Z_2 
\eeq
where $\mathbb{RP}_2$ stands for real projective space and is the notation for the surface of the sphere with diametrically opposite
points identified and $\mathbb{Z}_2=(1,-1)$ is a group of just two elements. 

Now that we have determined that there are two classes of paths in three dimensions, in terms of path integrals, the amplitude can be written as
\beq
A[{\bf r}_1(t_1),{\bf r}_2(t_1)) \rightarrow  ({{\bf r}_1}^\prime(t_2),{{\bf r}_2}^\prime(t_2)]  % \nonumber \\
= \sum_{direct ~paths} e^{i\mathcal{S}} +\sum_{exchange ~paths} e^{i\mathcal{S}}~.
\eeq
The direct paths involve all closed paths which end at the same point and the exchange paths involve
all paths which end on antipodal points, (which are also closed paths).
In terms of the path integral, we can also introduce a phase between the two classes of paths and write
\beq
A[{\bf r}_1(t_1),{\bf r}_2(t_1)) \rightarrow  ({{\bf r}_1}^\prime(t_2),{{\bf r}_2}^\prime(t_2))]  %\nonumber \\
= \sum_{direct ~paths} e^{i\mathcal{S}} + e^{i\phi}\sum_{exchange ~paths} e^{i\mathcal{S}}~.
\eeq
Since have already seen that exchanging the particle twice leads again to the direct path, it is clear
that $e^{2i\phi} = 1$, which implies that $\phi$ can only be $0,\pi$ giving rise, as before, to  bosons and fermions.

%--------------- Fig 4  ----------
\begin{figure}
 \includegraphics[width=0.75\textwidth]{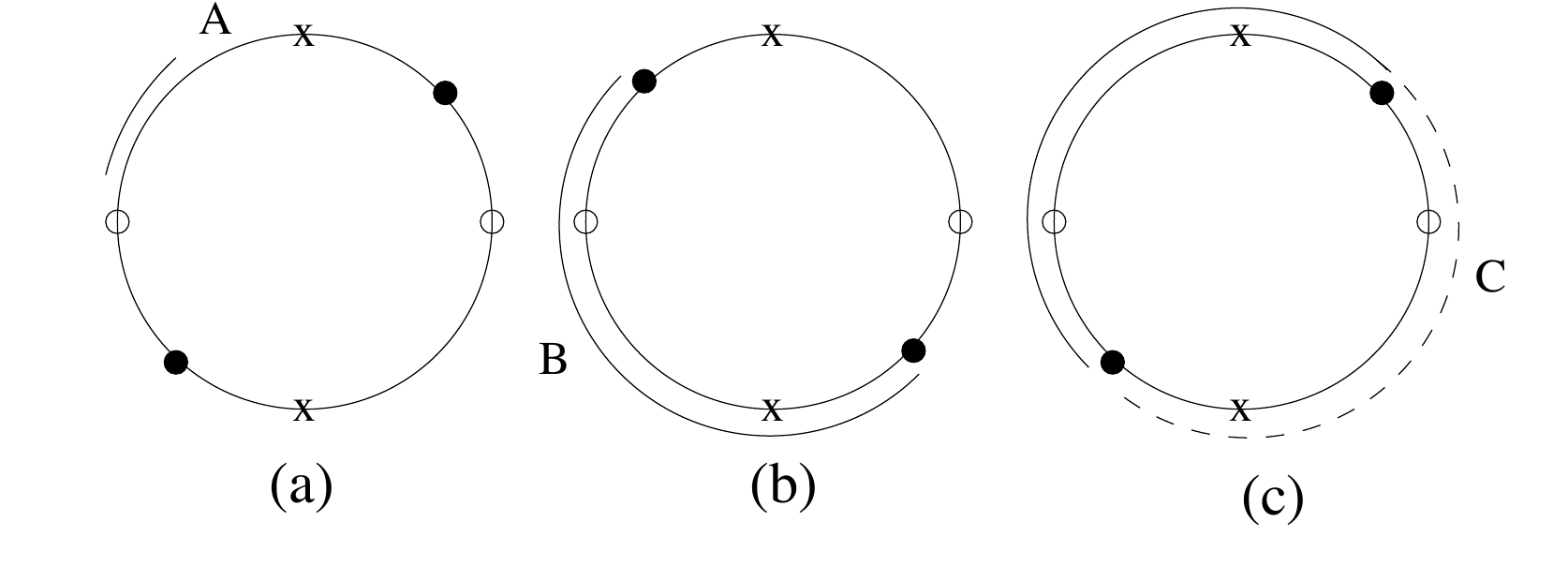}
 \caption{Paths in two dimensional configuration space with fixed radius}
 \label{lec1fig4}
	\end{figure}
%--------------------------------------- 

What changes in two dimensions? The point is that the topology of the configuration space is now different.
In two spatial dimensions, the configuration space of the relative coordinates is given by
$(\mathbb{R}_2-origin)/\mathbb{Z}_2$. Just as before, for ease of visualisation, we shall keep the magnitude of the relative
coordinate fixed, so that configuration space can be represented by a circle, and since the particles are
indistinguishable, diametrically opposite points are identified (see Figs.\ref{lec1fig4}(a),(b),(c)). Here, however,
several closed paths are possible. The path A that involves no exchanges can obviously be shrunk to a point,
since it only moves along the circle and back. But path B that exchanges the two particles is non-contractible
since the end-points are fixed. But even path C, where both the solid and dashed line are followed in the clock-wise
direction (or anti-clockwise direction) cannot be contracted to a point. This is easily understood by visualising the paths
as physical strings looping around a cylinder. Thus, if $\eta$ is the phase under single exchange, $\eta^2$ is the phase
under two exchanges, $\eta^3$ is the phase under three exchanges and so on. All we can say is that since the modulus
of the wave-function remains unchanged under exchange, $\eta$ has to be a phase - $\eta = e^{i\theta}$.
 This explains why we can get `any' statistics in two dimensions.
 
 %--------------- Fig 5  ----------
\begin{figure}
 \includegraphics[width=0.65\textwidth]{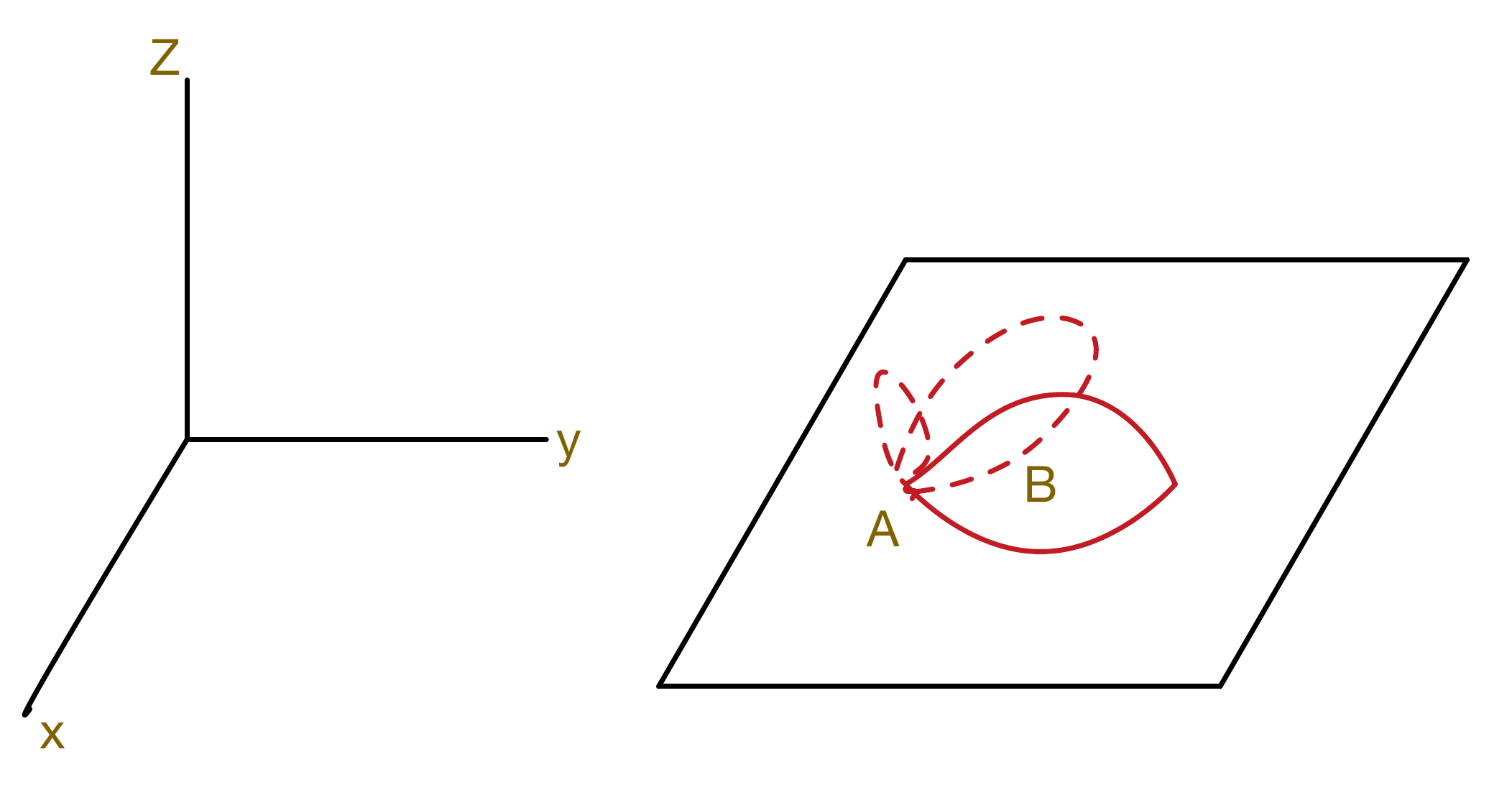}
 \caption{Path of A around B  being lifted off the surface and shrunk to a point}
 \label{lec1fig5}
	\end{figure}
%--------------------------------------- 

The distinction between the paths in two and three dimensions can also be seen as follows.  In three dimensions, the 
loop that is formed by taking a particle all around another particle (two exchanges) can be lifted off the plane and
shrunk to a point as shown in Fig.\ref{lec1fig5}.  This is not possible if the motion is restricted to a plane, as long as we disallow
configurations where two particles are at the same point (removal of the origin).

The mathematical crux of the distinction between configuration spaces in two and three dimensions, is that  the removal
of the origin in two dimensional space, makes the space multiply connected (unlike in three  dimensional space,
where removal of the origin keeps it singly connected). So it is possible to define paths that wind around the origin.
Mathematically, we can say that
\beq
\Pi_1((\mathbb{R}_2-origin)/\mathbb{Z}_2) = \Pi_1 (\mathbb{RP}_1) = \mathbb{Z}
\eeq
where $\mathbb{Z}$ is the group of integers under addition. $\mathbb{RP}_1$ is just the notation for the circumference of a circle
with diametrically opposite points identified. The different paths are labelled by integer winding numbers.

So in terms of path integrals, we now see that paths starting and ending at the same positions (upto exchanges due
to indistinguishability) can be divided into an infinite number of classes, all distinct. So we can write the total amplitude
as 
\beq
A = \sum_{direct~paths} e^{i\mathcal{S}} + e^{i\phi} \sum_{single ~exchange} e^{i\mathcal{S}} %\nonumber \\
+ e^{2i\phi}\sum_{two~exchanges} e^{i\mathcal{S}} + \dots
\eeq
where $\phi=0,\pi$ give the usual bosons and fermions, but since in general, $e^{in\phi} \ne 1$ for any $n$, $\phi$ can be anything
and as we said earlier, `any' statistics are possible in two dimensions.

Note that if we do want to understand exchange of particles in the Hamiltonian formulation without invoking
path integral ideas, we need to pin down the particles by using a confining potential - $i.e.$, by putting them 
in a box -
\beq
H = \sum_i\frac {{\bf p}_i^2}{2m} + \sum_i V_{\rm box} ({\bf x}_i - {\bf R}_i)
\eeq
The particles can now be moved around by changing ${\bf R}_i$ as a function of time. 
Since the particles are identical, exchanges are equivalent to closed paths, and do not depend on
the geometry of the paths ${\bf R}_i(t)$ involved. So the statistics of the particles under exchange
can be found by computing the  Berry phase when the particles are exchanged.
However, in this review, we shall basically use the path integral formalism.

\subsection{Anyons obey braid group statistics}

The distinction between the phase of the wave-function when the quantum numbers of the particles are exchanged
and the phase obtained under adiabatic transport of particles should now be clear. Under the former definition, the phase
$\eta^2$ after two exchanges is always unity, whereas the phase under the latter definition has many more possibilities
at least in two dimensions. Mathematically, the first definition classifies particle under the permutation group 
$\mathbb{P}_N$,
whereas the second one classifies particles under the braid group $\mathbb{B}_N$. The permutation  group 
$\mathbb{P}_N$ is the group
formed by all possible permutations of $N$ objects with group multiplication defined as successive permutations and
group inverse as undoing the permutation. It is clear that permuting two objects twice brings the system back to the
original configuration. Thus particles that transform as representations of the permutation group can only be bosons
or fermions.

%--------------- Fig 6  ----------
\begin{figure}
 \includegraphics[width=0.75\textwidth]{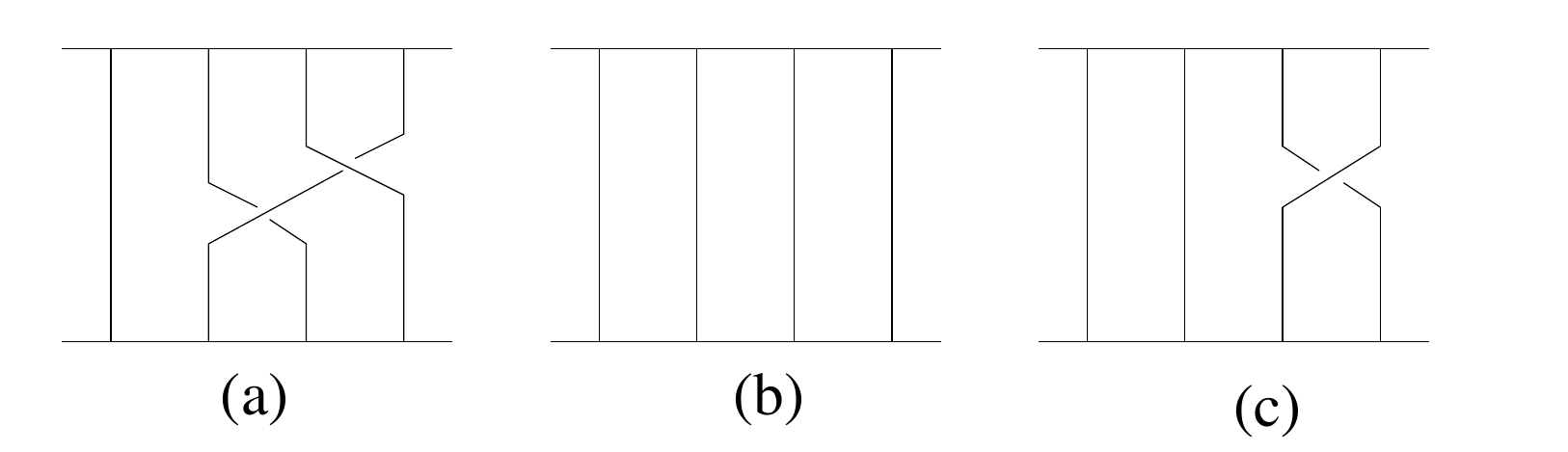}
 \caption{Elements of the braid group $\mathbb{B}_4$}
 \label{lec1fig6}
	\end{figure}
%--------------------------------------- 

\begin{figure}
 \includegraphics[width=0.45\textwidth]{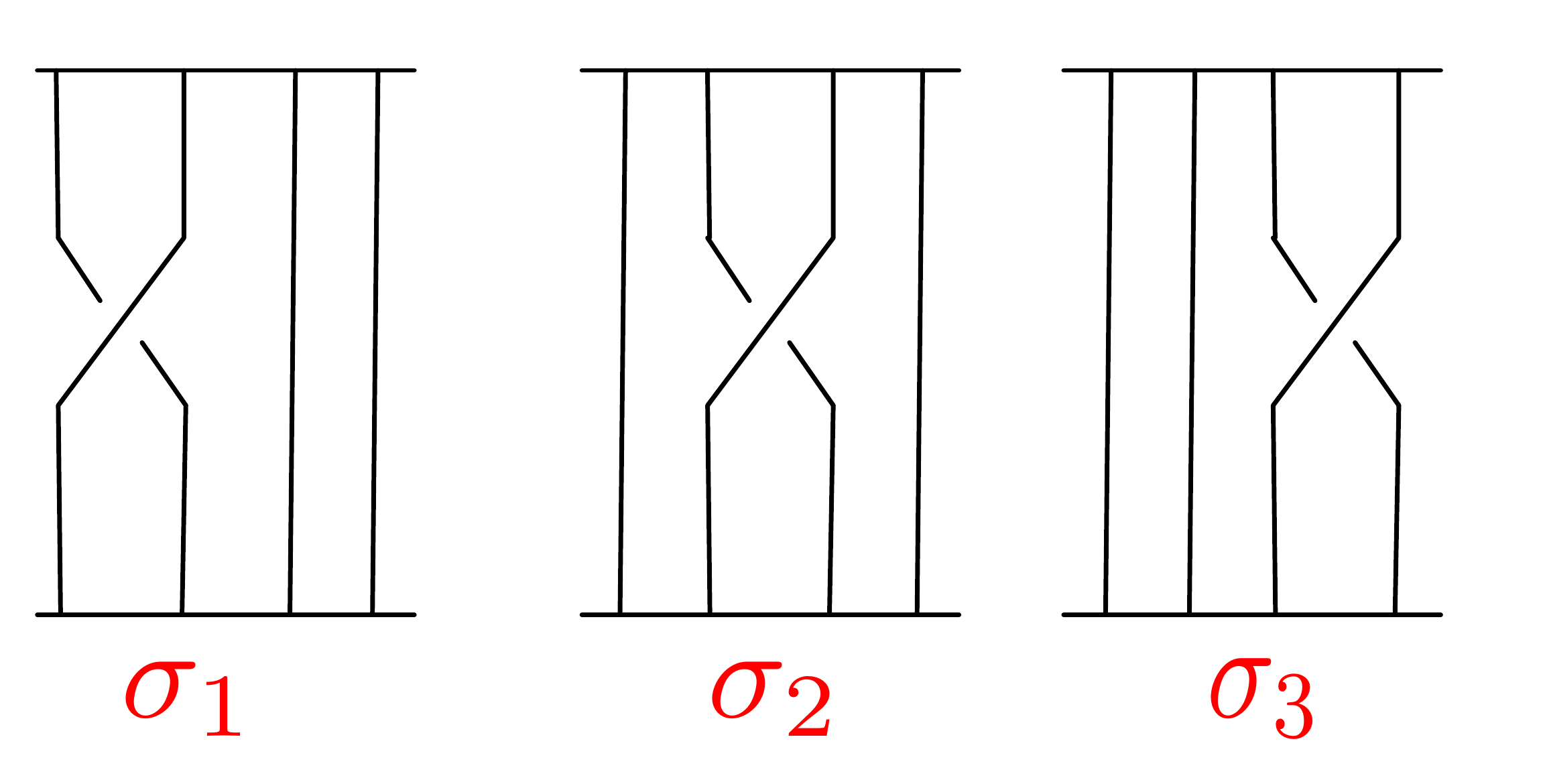}
 \caption{The three generators of the   braid group $\mathbb{B}_4$}
 \label{lec1fig7}
	\end{figure}

\begin{figure}
 \includegraphics[width=0.5\textwidth]{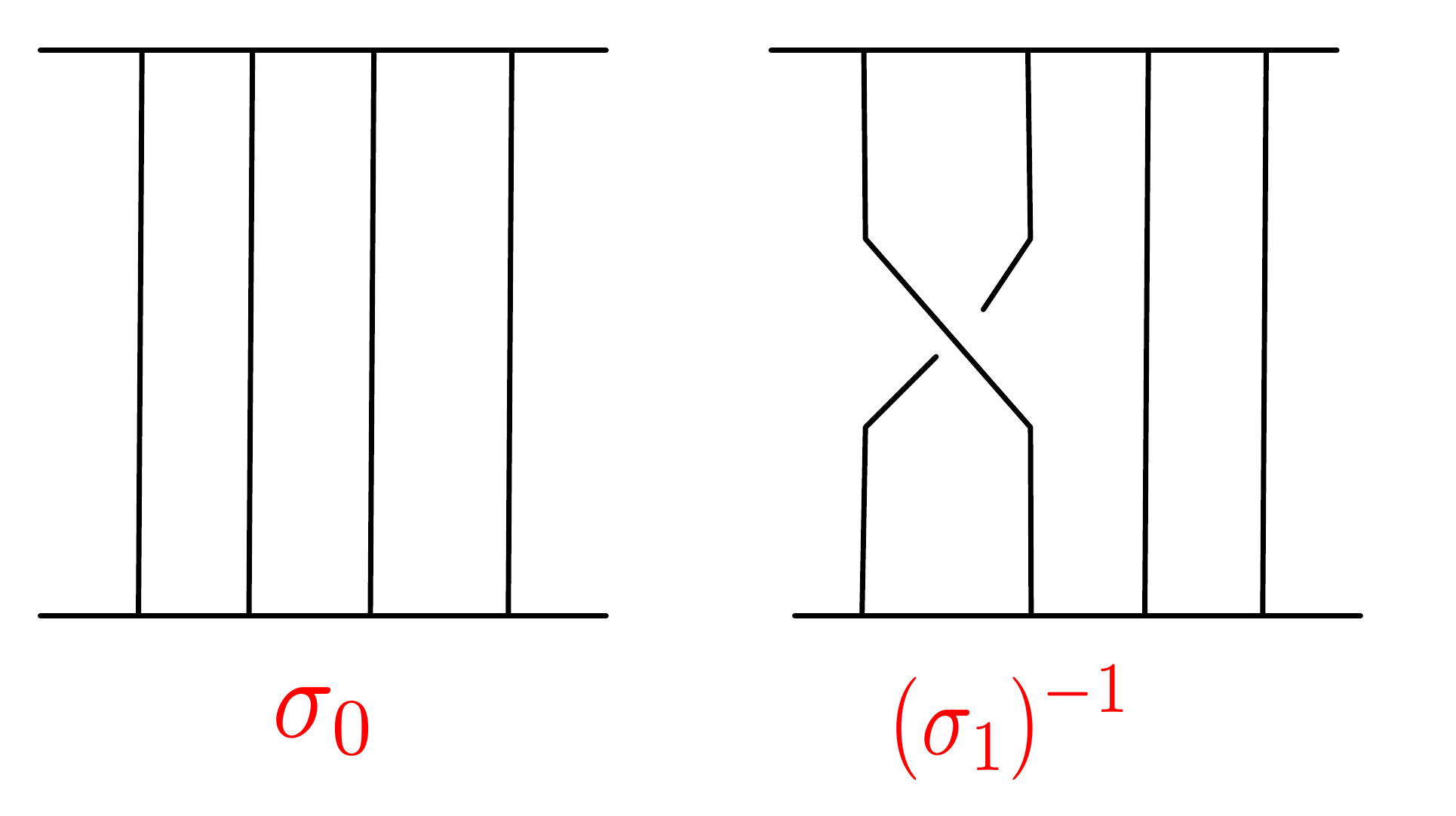}
 \caption{The identity and the inverse of the generator $\sigma_1$.}
 \label{lec1fig8}
	\end{figure}

\begin{figure}
 \includegraphics[width=0.5\textwidth]{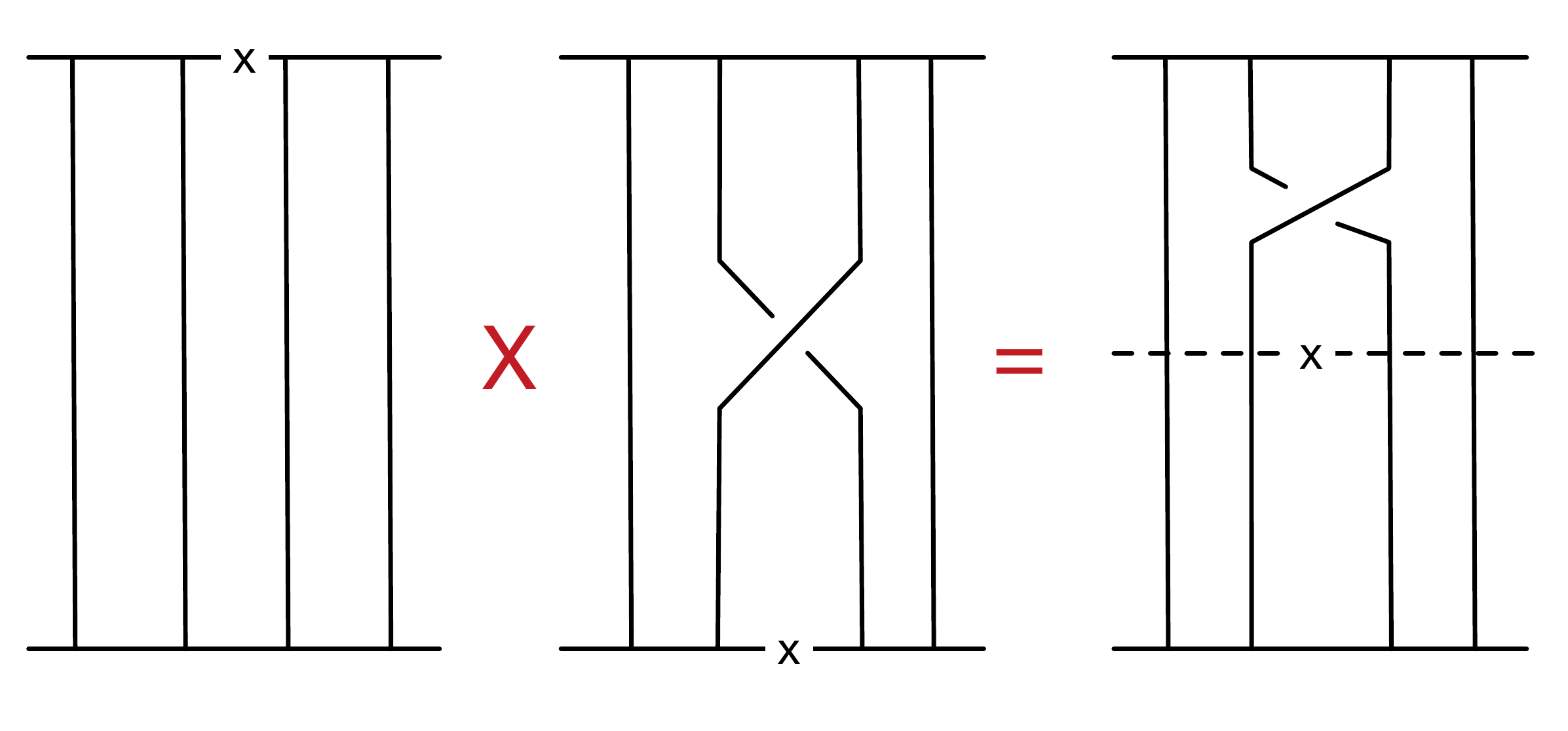}
 \caption{Group multiplication.}
 \label{lec1fig9}
	\end{figure}

\begin{figure}
 \includegraphics[width=0.7\textwidth]{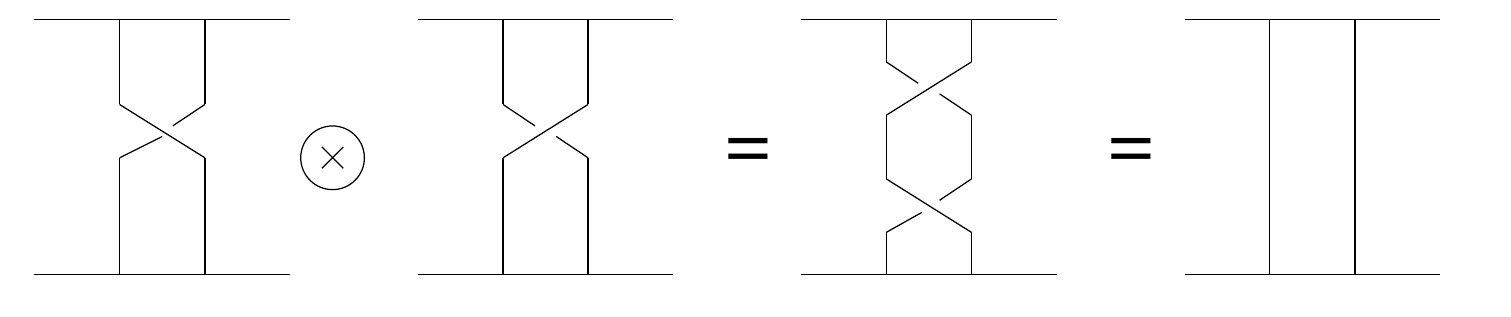}
 \caption{Product of $\sigma_1$ and $(\sigma_1)^{-1}$ giving rise to identity.}
 \label{lec1fig10}
	\end{figure}

\begin{figure}
 \includegraphics[width=0.35\textwidth]{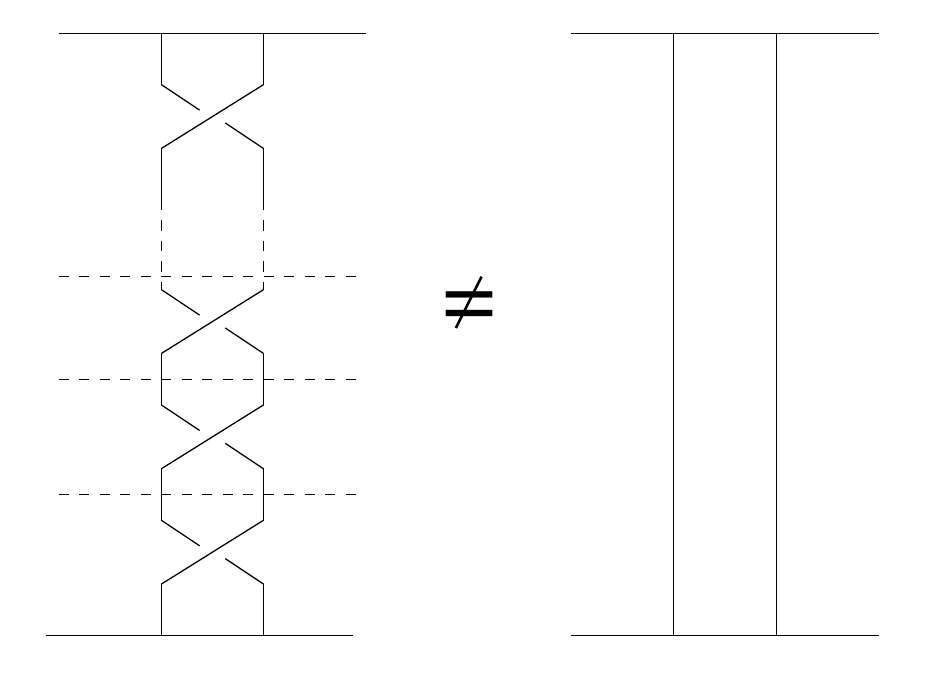}
 \caption{$(\sigma_1)^{n} \ne \sigma_0$ }
 \label{lec1fig11}
	\end{figure}

On the other hand, when we adiabatically exchange two particles, we can visualise the process  as paths in space-time
with time being the vertical axis and space being the horizontal axis as shown  in Fig.\ref{lec1fig2}. The  particles can circle around
each other and form closed paths by coming back to their original positions (upto permutations of the positions). 
The adiabatic exchange of particles classifies particles under the braid group. As we saw earlier, even under adiabatic exchange,
in three spatial dimensions, we only have fermions or bosons, whereas in two dimensions, there are many other possibilities.
Formally, the braid group $\mathbb{B}_N$ is
the group of inequivalent paths that occur when adiabatically transporting $N$ particles. Since they represent a
configuration of $N$ particles, at some particular time (say $t=0$),  evolving to a configuration of $N$ particles at some
later time $t=T$, the world lines cannot cross each other or form knots around each other or loop back. At each time,
we want to have only $N$ particles. Each history or set of trajectories of the $N$ particles becomes a braid. 
For example, in Fig.\ref{lec1fig6}, we show an example of some elements of the braid group $\mathbb{B}_4$, which is the braid
group of 4 particles. Exchanges of neighbouring  particles (by some counting rule, since the particles are in two dimensional space)
 form the generators of the group. For instance, the generators of the group $\mathbb{B}_4$ are given in Fig.\ref{lec1fig7} and are denoted as $\sigma_j$,
 $j=1,2,3$. 
 $\sigma_j$ describes exchange of $j^{th}$ particle with $(j+1)^{th}$ particle in a counter-clockwise direction (by definition),
 so that the clockwise exchange is denoted by $(\sigma_j)^{-1}$.
 The identity element is given by $\sigma_0$ where there is no exchange, and group inverse by the clockwise exchange  $(\sigma_j)^{-1}$ as shown in Fig.\ref{lec1fig8}.
Group multiplication is defined as following one trajectory by another in time as shown in Fig.\ref{lec1fig9}.
Note that we have put crosses on the time-lines which are identified (are at equal times) in the figure.
 It is now easy to check that $(\sigma_j)(\sigma_j)^{-1} =\sigma_0$ as shown in Fig.\ref{lec1fig10} (without the crosses).
It is also easy to see that $(\sigma_1)^n  \ne \sigma_0$ for any $n$, which is the reason that `any' statistics are allowed in two dimensions. (See Fig.\ref{lec1fig11}).

%--------------- Fig 7  ----------
%--------------------------------------- 

%--------------- Fig 8  ----------
%--------------------------------------- 

%--------------- Fig 9  ----------
%--------------------------------------- 

%--------------- Fig 10  ----------
%--------------------------------------- 

%--------------- Fig 11 ----------
%--------------------------------------- 

%--------------- Fig 12 ----------
\begin{figure}
 \includegraphics[width=0.47\textwidth]{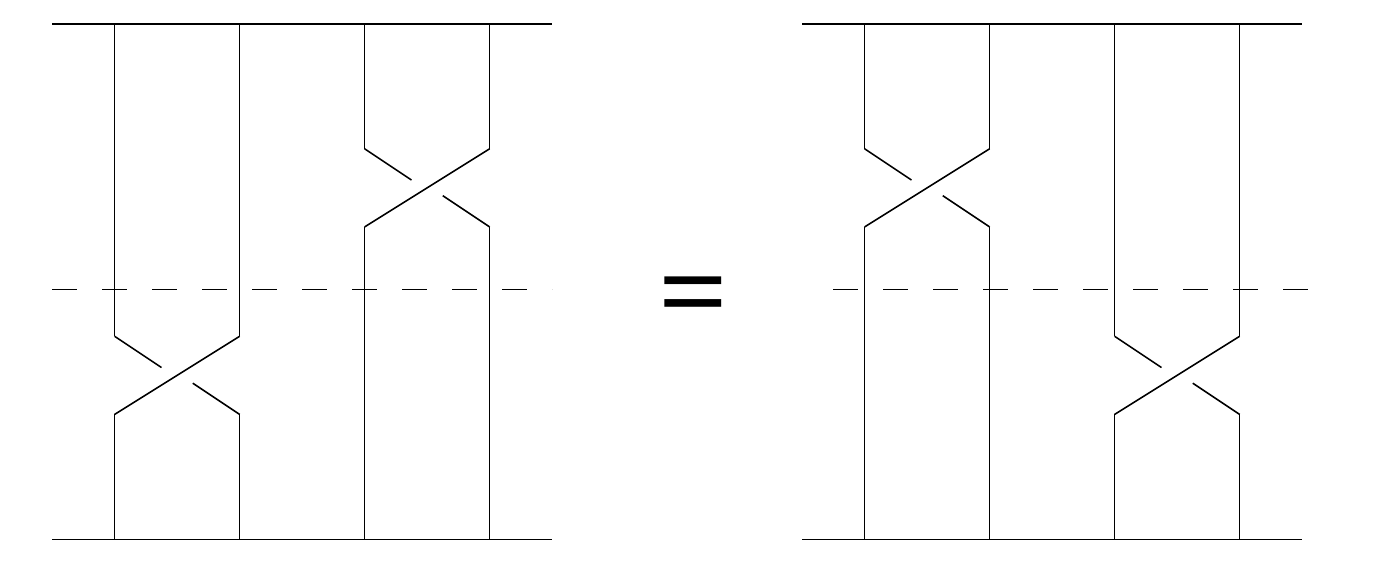} \quad
 \includegraphics[width=0.33\textwidth]{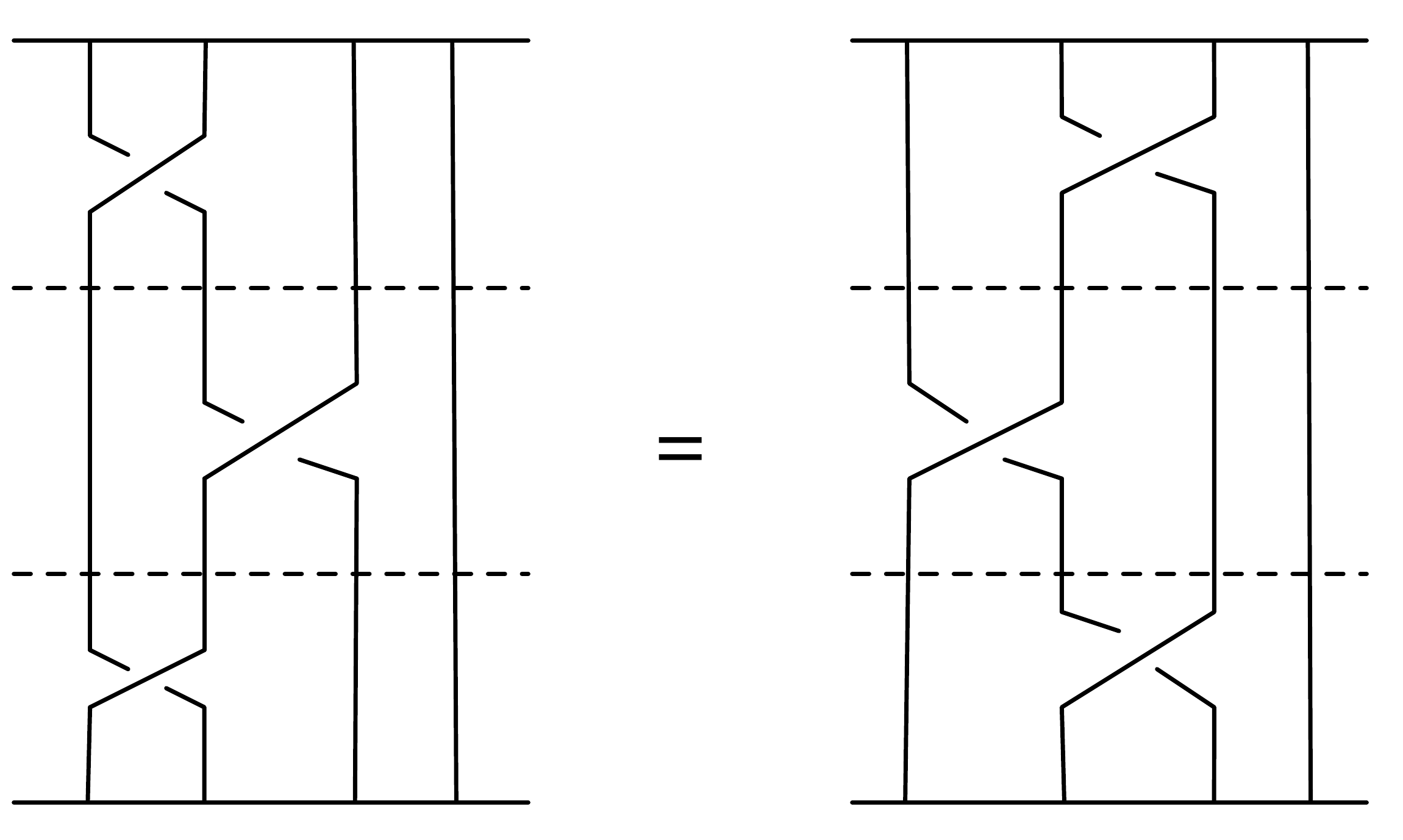}
 \caption{Yang-Baxter relations.}
 \label{lec1fig12}
	\end{figure}
%--------------------------------------- 

%--------------- Fig 13 ----------
\begin{figure}
 \includegraphics[width=0.5\textwidth]{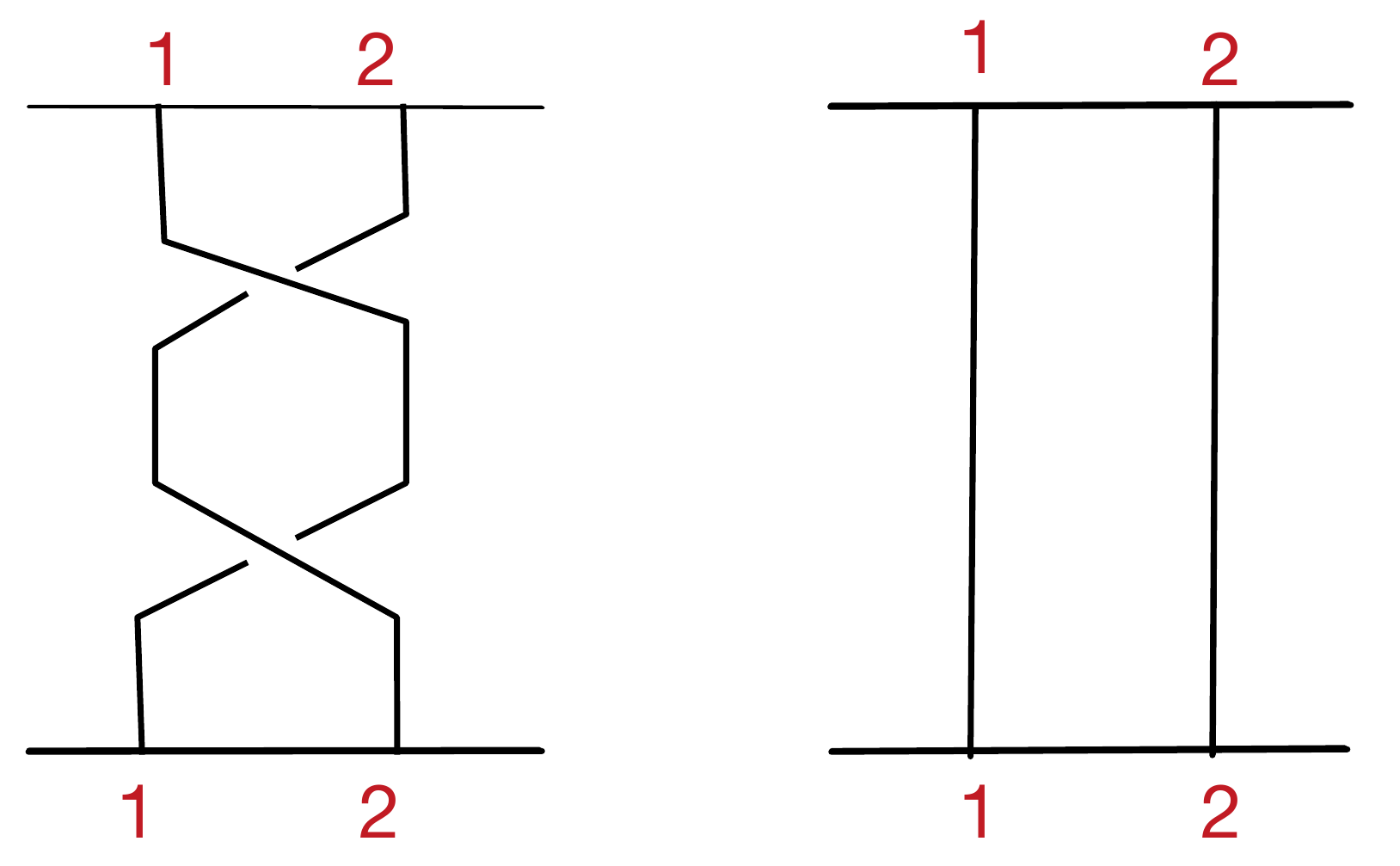} 
 \caption{Different elements of the braid group, but same element of the permutation group.}
 \label{lec1fig13}
	\end{figure}
%--------------------------------------- 

We will end this subsection by mentioning the two defining relations satisfied by the generators of the braid group.
\bea
\sigma_i \sigma_j = \sigma_j \sigma_i , \quad |i-j|\ge 2 \nonumber \\
\sigma_j\sigma_{j+1}\sigma_j = \sigma_{j+1} \sigma_j\sigma_{j+1}
\eea
The second one is called the Yang-Baxter relation.
Both these relations can be easily checked pictorially ( as we show in Fig.\ref{lec1fig12} 
for the generators of $\mathbb{B}_4$).

It should be clear by now that the braid group leads to a much finer classification than the permutation group.
For instance, the  two elements shown in Fig.\ref{lec1fig13}  are different elements of the braid group, but the same element of the permutation group.
So the quantum theory of anyons has the quantum states of the anyons transforming as unitary representations of 
the braid group. Abelian anyons form  one-dimensional representations of the braid group. There are an infinite number
of such representations, because under exchange, the phase that is picked up is $e^{i\theta}$ and $\theta$ can take any
value. $\theta =0$ and $\theta=\pi$ represent bosons and fermions respectively. We will discuss non-abelian representations in the last
section.

\subsection{Spin of an anyon}

Let us start with spin in the familiar three dimensional world. We know that spin is an intrinsic angular momentum quantum number
that labels different particles. The three spatial components of the spin obey the commutation relations given by 
\beq
[S_i,S_j]=\epsilon_{ijk}S_k
\eeq
We shall show that these commutation relations constrain the spin to be either integer or half-integer. Let $|s,m>$
be the state with $S^2|s,m> = s(s+1) |s,m>$ and $S_z|s,m> = m |s,m>$. By applying the raising operator, we may create the
state
\beq
S^+|s,m> = [s(s+1) - m(m+1)]^{1/2} |s,m+1> = |s,m'>
\eeq
Requiring this state to have positive norm for all $m$, leads to $m<s$. Thus, it is clear that for some integer $m' = m+$integer, $m'>s$ unless $s=m'$
or 
\beq
s-m = {\rm integer} \label{one}
\eeq
Similarly by insisting that $S^-|s,m>$ have a positive norm, we get $s(s+1) - m(m-1) \ge 0$, which implies that $m\ge -s$ for all $m$.
Once again, to avoid $m<-s$, we need to set 
\beq
m-(-s) ={\rm  integer} \label{two}
\eeq
Adding the equations in Eqs.\ref{one} and \ref{two}, we get
\beq
2s = {\rm integer} \Longrightarrow s = {\rm integer}/2~.
\eeq
Thus, just from the commutation relations, we can prove that the particles in three dimensions have either integer or half-integer spin.

However, in two dimensions, there exists only one axis of rotation, perpendicular to the plane of the two dimensions.
Hence, here spin only refers to $S_3$ which has no commutation relations to satisfy, and hence it can be anything!

\subsection{Physical model of an anyon}
\begin{figure}
 \includegraphics[width=0.35\textwidth]{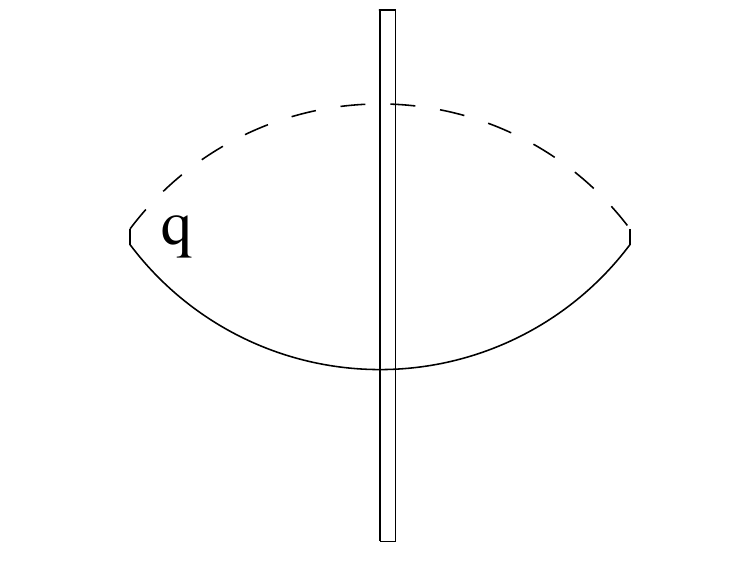} 
 \caption{Physical model of an anyon.}
 \label{lec1fig14}
	\end{figure}

Now let us construct a simple physical model of an anyon\cite{wilczek}. Imagine a spinless particle of charge $q$ orbiting around
a thin solenoid along the $z$-axis at a distance ${\bf r}$ as shown in Fig\ref{lec1fig14}. When there is no current through the solenoid,
the orbital angular momentum of the charged particle is quantitized as an integer - $l_z$ = integer. When a current is turned
on, the particle feels an electric field that can easily be computed using
\beq
\int (\nabla \times {\bf E}) d^2{\bf r} = \int B d^2 {\bf r} = - \frac{\partial\phi}{\partial t}
\eeq
where $\phi$ is the total flux through the solenoid. This is just the Aharanov-Bohm effect. Hence, 
\beq
\int {\bf E} \cdot d{\bf l} = 2\pi |{\bf r}| E_{\theta} = -{\dot\phi} \quad {\rm leading ~to} \quad {\bf E} = - \frac{{\dot\phi}}{2\pi|{\bf r}|} ({\hat z} \times {\bf r})~.
\eeq
Thus, the angular momentum of the charge particle changes with the rate of change given being proportional to the torque - i.e.,
\beq
{\dot l_z} = {\bf r}\times {\bf F} = {\bf r}\times q {\bf E} = -\frac{q{\dot\phi}}{2\pi} \quad {\rm leading~ to} \quad \Delta l_z =  -\frac{q{\phi}}{2\pi}.
\label{lz}
\eeq
Thus $\Delta l_z$ is the change in the angular momentum due to the fiux in the solenoid. In the limit where the solenoid becomes very narrow
and the distance between the charged particle and the solenoid is shrunk to zero, the system may be considered as a single composite object
- a charge-fluxtube composite. In fact in a planar system, there is no extension in the $z$ direction.  So this essentially point-like composite
object with fractional angular momentum can be considered as a model of an anyon. 
Note that we have denoted this angular momentum as the change in the angular momentum due
to the flux. So if we start with the original charge to be spinless, then the spin of the composite particle is given by $l_z=s_z = q\pi/2\pi$.
This is also sometimes referred to as a topological spin and is intrinsic to the anyon.
However, this is a little too naive. In an anyon, the charge and 
the fiux it carries are related - the charge gets turned on along with the flux. This implies that the $q$ in Eq.\ref{lz} is time-dependent, and $q(t) = c\phi(t)$ for
some constant $c$. Hence, we find 
\beq
\Delta l_z =\frac{ c\phi^2}{4\pi} = \frac{q\phi}{4\pi} 
\eeq
so that the angular momentum of a charge-flux composite with charge proportional to flux is less than what we
originally computed by a factor of 1/2. 
In the next subsection, we shall see that it has the right statistics, and complete the  identification of the charge-fluxtube composite as an anyon.

%--------------- Fig 14----------
%--------------------------------------- 

\subsection{Two anyon quantum mechanics}

We shall now study the quantum mechanics of two anyons in order to determine its statistics using the
simple physical picture of the anyon that we developed in the last subsection. The Hamiltonian for the system
is given by
\beq
H = \frac{({\bf p}_1 - q{\bf a}_1)^2}{2m} +  \frac{({\bf p}_2 - q{\bf a}_2)^2}{2m} \label{2anyqm}
\eeq
with 
\beq
{\bf a}_1 = \frac{\phi}{2\pi} \frac{{\hat z} \times ({\bf r}_1 -{\bf r}_2)} {|{\bf r}_1 -{\bf r}_2|^2}, \quad {\rm and} \quad {\bf a}_2 = \frac{\phi}{2\pi} \frac{{\hat z} \times ({\bf r}_2
 -{\bf r}_1)} {|{\bf r}_1 -{\bf r}_2|^2}
\eeq
where ${\bf a}_1$ and ${\bf a}_2$ are the vector potentials at the positions of the composites (anyons) 1 and 2 due to the fluxes in composites (anyons) 2 and 1 respectively. Let
us now work in the centre of mass (CM) and relative (rel) coordinates - i.e., we define respectively
\beq
{\bf R} = \frac{{\bf r}_1 +{\bf r}_2} {2} \Rightarrow {\bf P} = {\bf p}_1 +{\bf p}_2, \quad {\rm and} \quad  {\bf r} = {\bf r}_1 -{\bf r}_2  \Rightarrow {\bf p} = \frac{{\bf p}_1 -{\bf p}_2}
{2}~.
\eeq
In terms of these coordinates, the Hamiltonian can be recast as 
\beq
H=  \frac{{\bf P}^2}{4m} +  \frac{({\bf p} - q{\bf a})^2}{m} \quad {\rm with} \quad {\bf a}_{rel} = \frac{\phi}{2\pi}  \frac{{\hat z} \times {\bf r}} {|{\bf r}|^2}~.
\eeq
Thus the CM motion which translates both the particles rigidly and is independent of the statistics is free. The relative motion, 
on the other hand, which is sensitive to whether the particles are bosons, fermion or anyons, reduces to the problem of a single
particle of mass $m/2$ orbiting around a flux $\phi$ at a distance ${\bf r}$. Since the composites have been formed of a bosonic charge
orbiting around a bosonic flux, the wave -function of the two composite system has to be symmetric under exchange and the boundary
condition is given by
\beq
\psi({\bf r}_1,{\bf r}_2) = \psi({\bf r}_2,{\bf r}_1) \Longrightarrow \psi_{rel}( {\bf r})  =  \psi_{rel}( -{\bf r}) \Longrightarrow  \psi_{rel} (r,\theta+\pi) = \psi_{rel}(r,\theta) 
\eeq
where $\psi_{rel}$ is the wave-function of the relative piece of the Hamiltonian and ${\bf r} = (r,\theta)$ in cylindrical coordinates.

Now, let us perform  a (singular) gauge transformation so that 
\beq
{\bf a}_{rel} \longrightarrow  {\bf a}^\prime_{rel} = {\bf a}_{rel} - \nabla \Lambda (r,\theta) \quad 
{\rm where} \quad  \Lambda (r,\theta) = \frac{\phi}{2\pi} \theta ~.
\eeq
This gauge transformation is singular because $\theta$ is a  periodic angular coordinate with period $2\pi$ and is not single-valued.
In the primed gauge,
\beq
a'_{rel,\theta} = a_{rel,\theta} - \frac{1}{r} \frac{\partial\Lambda}{\partial\theta} =\frac{\phi}{2\pi r} - frac{\phi}{2\pi r} = 0, \quad {\rm and} \quad {a^{\prime}}_{rel,r} = a_{rel,r} -\frac{\partial}{\partial r} \Lambda = 0-0=0, 
\eeq
i.e., the gauge potential vanishes completely and the Hamiltonian just reduces to 
\beq
H = \frac{{\bf P}^2} {2m} +  \frac{{\bf p}^2} {2m}
\eeq
which is just the Hamiltonian of two free particles. However, in the primed gauge, the wave function
of the relative Hamiltonian has also changed. It is now given by
\beq
\psi_{rel}^\prime(r,\theta+\pi) = e^{-iq\phi/2}  \psi_{rel} (r,\theta)~
\eeq
-i.e., the particles obey anyonic statistics.
So the problem of two free anyons is equivalent to the problem of two interacting charge flux composites 
described by the Hamiltonian  in Eq.\ref{2anyqm}. The quantum mechanical problem can be solved - 
we need  to go back to the boson gauge ( since we do not know how to solve quantum mechanics
problems with non-trivial statistics) and write the Hamiltonian in terms of centre of mass (CM)  and relative coordinates and note
that the CM motion becomes free and the relative motion acquires an extra $q\phi/2m$ factor in the angular
momentum term, thus adding to the centrifugal barrier. The radial part of the relative motion can then
be identified as a Bessel equation. Thus the two anyon wave-function can be written as
\beq
\psi ({\bf R}, {\bf r}) = \psi_{\rm CM}({\bf R}) \psi_{\rm rel}({\bf r}) = e^{i{\bf P}\cdot {\bf R}} e^{i(l+q\phi/2m)\theta} 
J_{|l+q\phi/2m|} (kr)
\eeq
where ${\bf r} = (r,\theta)$. The two particle wave-function can be recast in terms of the original single
particle coordinates - $i.e.$, $\psi(({\bf R}, {\bf r}) = \psi({\bf r}_1,{\bf r}_2)$. However, unless $q\phi/2m$
is either integer or half-integer, the two particle wave-function cannot be factorised into a product of 
two suitable one-particle wave-functions. Also, the energy levels of the two anyon system cannot
be obtained as sums of one anyon energy levels. This is easier to see with discrete energy levels
and so we will  next solve the problem of two anyons in a harmonic oscillator potential.

The Hamiltonian for two anyons in a harmonic oscillator potential is given by
\beq
H = \frac {{\bf p}_1^2}{2m} + \frac {{\bf p}_2^2}{2m}  +\frac{1}{2}m\omega^2{\bf r}_1^2 + 
\frac{1}{2}m\omega^2{\bf r}_2^2~.
\eeq
The problem can be separated into CM and relative coordinates in terms of which the Hamiltonian
is given by
\beq
H = \frac {{\bf P}^2}{4m} + \frac {{\bf p}^2}{m}  +m\omega^2{\bf R}^2 + 
\frac{1}{4}m\omega^2{\bf r}^2~.
\eeq
The problem  can now be solved  in terms of the cylindrical $(R,\Theta)$
and $(r,\theta)$ coordinates.  The CM motion is independent of the statistics of the particles and one
can simply find the energy levels as
\beq
E_{ CM} = \omega(n+|L|+1)~.
\eeq
The Hamiltonian for the relative motion can be solved in the same way, except that because of
the phase under exchange, we need to go to the boson gauge, where there is a dependence on
the statistical gauge field - $i.e.$, we have
\beq
H_{rel} \psi_{rel}= [\frac{({\bf p} - q{\bf a}_{ rel})^2}{m}  + \frac{1}{4} m\omega^2{\bf r}^2]\psi_{ rel} 
= E_{rel}\psi_{ rel}~.
\eeq
In terms of the $(r,\theta)$ coordinates, we now find that the energy levels are given by
\beq
E_{rel} = \omega(n+|l+\alpha/\pi| +1).
\eeq
Note that $\alpha=0$ and $\alpha=\pi$ give the usual energy levels for bosons and fermions respectively.
Otherwise, they are given by 
\bea
E_j &=& (2j+1+\alpha/\pi), \quad {\rm degeneracy ~~factor} = j+1, \nonumber \\
E_j &=& (2j+1-\alpha/\pi), \quad {\rm degeneracy ~~factor} = j~.
\eea
Clearly, the levels are not equally spaced and the total energy of the two anyon system  given 
by 
\beq
E_{2anyons} = E_{ CM} +E_{rel} = (2j+p+2\pm\alpha/\pi)\omega,   \quad p,j = {\rm integers}
\eeq
is not a sum of the one particle levels $E=(n+1)\omega$, with $n=$ integer.  Similarly, the
two anyon wave function is also not a simple product of one anyon wave functions.
We find that
\bea
\psi({\bf R},{\bf r}) &=& e^{-m\omega(R^2+r^2/4)} r^{\alpha/\pi}e^{i\alpha\theta/\pi}
\implies \psi({\bf r}_1,{\bf r}_2) \propto e^{-m\omega(r_1^2+r_2^2)/2} ({\bf r}_1 - {\bf r}_2)^{\alpha/\pi}
\eea
which does not factorise into a product of two single particle wave-functions except when
$\alpha =0,\pi$.  This is why even a system of free anyons needs to be tackled as an interacting
problem. For more details, see Ref.\onlinecite{anyonprimer}.

\subsection{Many anyon systems}

Finally, we briefly mention what happens when we have many anyons. As we have seen above, even the two  free anyon 
system is an interacting system because of the statistical interactions. Hence, any many anyon system needs
to be treated as an interacting system, where each particle has long-range statistical interactions with each
of the other particles.

Here, we will just mention one important concept of many anyon systems, which is that of fusion rules.
A system which has anyons must have many types of anyons.  If we have an anyon with statistics parameter
$\theta$, then we can combine two such anyons  or make  a bound state  of two such particles. 
What would be the statistics of the bound state?  You may naively think that it should be $2\theta$.
But that is not correct.  One can see this by thinking of the anyon as a charge-flux composite in two ways.
(1) In terms of angular momentum, we can add up the individual spins of the two anyons  - $\theta/2\pi + \theta/2\pi = \theta/\pi$. But we also need to include the orbital angular momentum of the anyon pair. Normally, the orbital angular momentum is an integer because the wave-function of the two particle system is single-valued, if you take one particle
around the other. But here, the counter-clockwise transport of one anyon around another in a full circle
leads to the phase $e^{-2i\theta} = e^{i2\pi L}$ where $L$ is the orbital angular momentum. Now adding the
spin and orbital angular momentum, we find that the total angular momentum is $ \theta/\pi+\theta/\pi=2\theta/\pi = 4\theta/2\pi$
which gives the statistics parameter as $4\theta$. (2) Here, we note that a single anyon exchange leads to a 
phase of $e^{i\theta}$. So when a two anyon molecule exchanges with another 2 anyon molecule, there
are $2^2=4$ exchanges and hence the phase acquired will be $e^{4i\theta}$, which agrees with the total
spin of the bound state as well.

Now, we can generalise this to say that more particles can be bound together to form new
bound states or new types of particles.  This is called fusion. The statistics parameter
when $n$ such particles are bound together is given by $e^{in^2\theta}$. One can think
of this new particle as an $n$-charge-$n$-flux composite. The formation of a different type
of anyon by bringing together two anyons is called fusion. If we bring together an anyon
and anti-anyon with opposite statistics parameter ( $\theta$ and $-\theta$), the result
has statistics zero, which is equivalent to having no particles.
The system with no particles (called the vacuum)  is often denoted by the identity $I$. It
is also called a trivial particle.

For abelian anyons, it is clear that if we bring together 2 anyons with statistics parameter $\theta$,
they give rise to an anyon with statistics parameter $4\theta$ and in general $n$ such particles
give rise to anyons with statistics parameter $n^2\theta$. But for non-abelian anyons, this is no
longer true. There is no unique way of combining anyons to form new anyons and one can have different
outcomes by bringing them together (just like two spin1/2 particles can be brought together to form
spin 0 or spin1 particles). These are called fusion channels. The probability of the different outcomes
is specified by a set of numbers which give rise to the  fusion rules. 

In the next section, instead of studying anyons abstractly as we have done in this section, we
shall study an explicit lattice model, whose excitations turn out to be abelian anyons. We will come back to this later when we study non-abelian anyons.

\section{Toric code model as an example of abelian anyons}

Let us begin this section, by first answering two questions - what is the toric code and why do we want to study it.
The toric code is actually a spin model defined on a two dimensional lattice. It was engineered by Kitaev\cite{kitaevtc} to be exactly solvable and 
to have low energy
excitations that are anyonic. The reason that this model became so important was because it was shown by Kitaev that these anyons could
be used, in principle,  to perform fault tolerant quantum computation - fault tolerant because information could be stored in the fusion properties of anyons, 
which  could not be destroyed by local perturbations.  This model is thus, the prototypical concrete lattice model for many of the more
abstract ideas of  quantum computation. 

%--------------- Fig 1  ----------
\begin{figure}
 \includegraphics[width=0.45\textwidth]{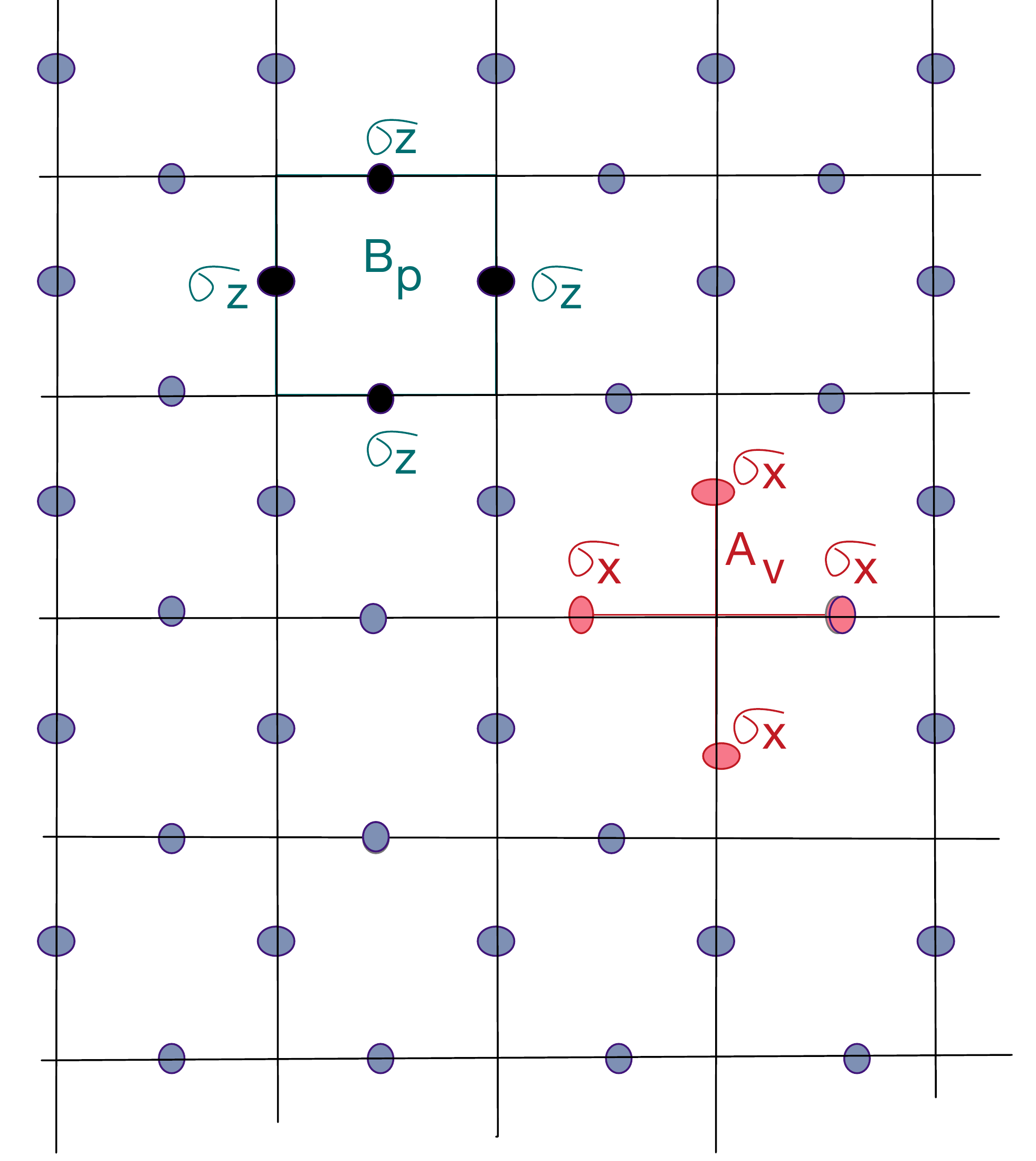}
\caption{The Kitaev toric code model on a square lattice. Spins are placed on the links. 
The operator $A_v$ is the product of the $x$-components of the spin of the four links
that cross at the vertex and the operator $B_p$ is the product of the $z$-component of the spins
around the perimeter of the square.   }\label{kitfig1}
	\end{figure}
%---------------------------------------

\subsection{Toric code on a square lattice}

In general, the toric code model can be defined on any lattice, but here we will work on 
the simplest model which  is an exactly solvable  spin 1/2  model on a two dimensional square lattice of $N\times N$ points\cite{kitaev2,burello}.
%We will take periodic 
%boundary conditions, so essentially the lattice is defined on a torus.
 The spins are placed on the edges or links  of an open  lattice as shown in Fig.\ref{kitfig1} and it is easy to check that we have twice as many spins as the number
 of lattice points - $2N^2$. The components of the spins
 on different links commute with one another. On a given link, the spins satisfy the usual anti-commutation relations $\{\sigma_\alpha,\sigma_\beta\} = 2\delta_{\alpha\beta}$
 where $\alpha,\beta = x,y,z$.
The Hamiltonian for the model is given by\cite{kitaev2}
\beq
H = -J_e \sum_v A_v -J_m\sum_p B_p
\eeq
where 
\beq
A_v  = \Pi_{j\in v} \sigma_j^x  ~~{\rm and } ~~B_p = \Pi_{j\in p} \sigma_j^z
\eeq
 are the vertex operator that acts on the four spins surrounding the vertex and 
the plaquette operator involving the four spins around the plaquette, respectively,
as shown in Fig.\ref{kitfig1}.  Note that this is  not a very common or physical looking Hamiltonian since it has four spin interactions, and has explicitly been engineered for a purpose! As can be easily seen from the diagram,
every spin is a part of two vertex and two plaquette operators.  The eigenvalues of both $A_v$ and $B_p$ are $\pm 1$. However, 
the product of the eigenvalues over all the plaquettes or all the vertices is always unity - i.e.,
\beq
\Pi_{j\in V} A_V = \Pi_{j\in P} B_p= +1 \label{totalconstraint}
\eeq
where $V =$ total number of vertices in the model is $N^2$ and $P=$ total number of plaquettes in the model is also $N^2$.

\begin{figure}
 \includegraphics[width=0.45\textwidth]{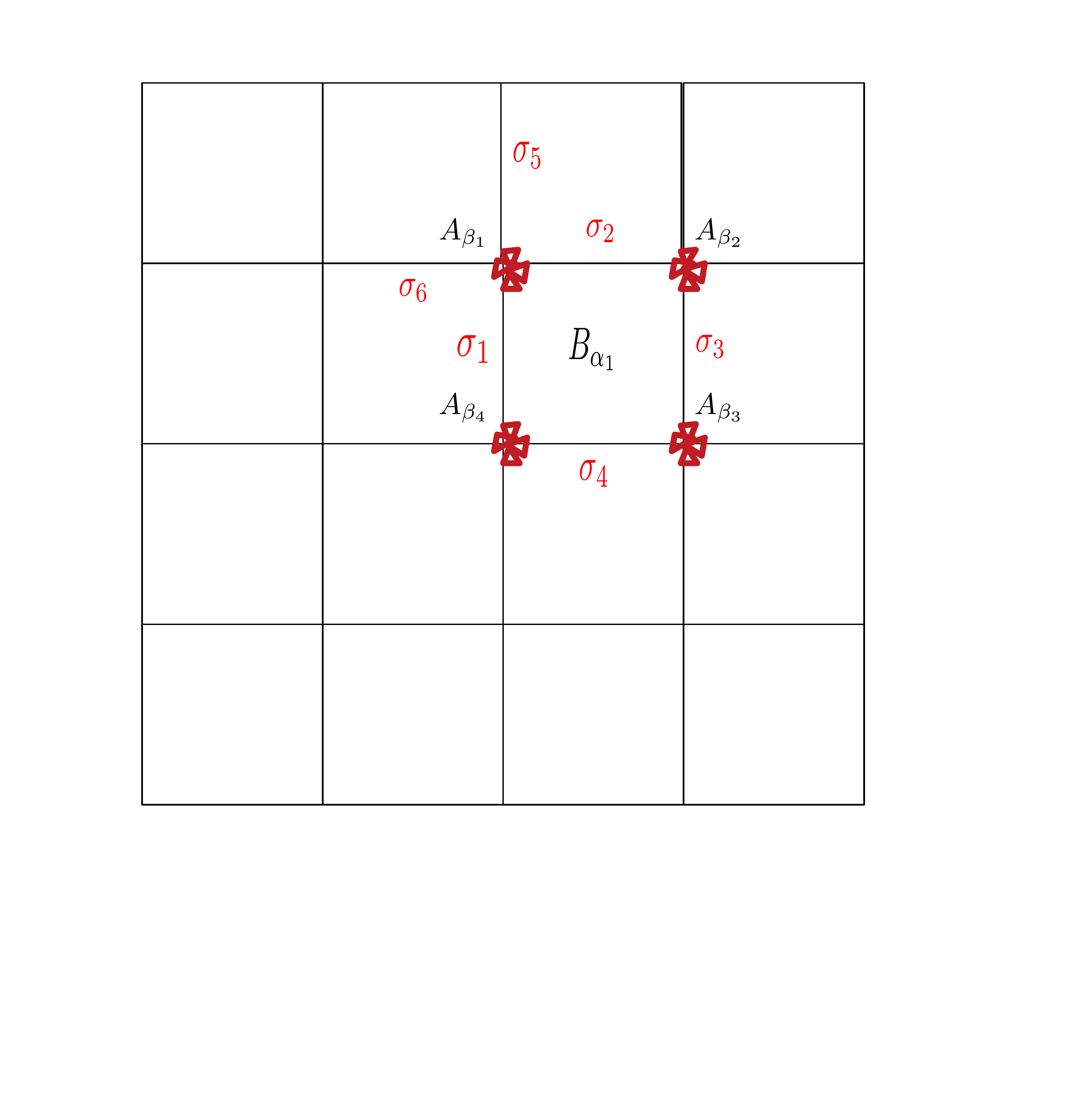}
\caption{Commutation relations of the operators on different links and same links. See text for details.}\label{kitfig2}
	\end{figure}

Note that the vertex operators contain the $x$-component of the spins and
the plaquette operators contain the $z$-component of the spins. Hence, it is easy to check that the vertex operators and the plaquette operators
commute among themselves
\beq
[A_v,A_{v'}]= [B_p,B_{p'}]=0
\eeq
But it is also true that $[A_v,B_p]=0$.  This is trivially true if they do not have any spins in common, since spins on
different sites commute. But in case, they do have a spin in common, they will always have two spins in common.
For example, from Fig.\ref{kitfig2}, it is clear that the plaquette $B_{\alpha_1}$ shares spins with the four vertex
operators $A_{\beta_1}, ...A_{\beta_4}$. But with each of them, it shares 2 spins. For instance, we have
\beq 
[B_{\alpha_1},A_{\beta_1}]= \sigma_1^z  \sigma_2^z  \sigma_3^z  \sigma_4^z  \sigma_1^x  \sigma_2^x   \sigma_5^x  \sigma_6^x  - 
 \sigma_1^x  \sigma_2^x   \sigma_5^x  \sigma_6^x \sigma_1^z  \sigma_2^z  \sigma_3^z  \sigma_4^z  
\eeq
which is zero because $\sigma_1^z\sigma_1^x= -\sigma_1^x\sigma_1^z$ and $\sigma_2^z\sigma_2^x= -\sigma_2^x\sigma_2^z$.
So the two negative signs cancel each other. The same thing goes through for the other three vertex operators as well. So the
bottomline is that all the terms in the Hamiltonian commute with each other, and commute with the Hamiltonian. So all the $A_v$ and $B_p$ operators
 can be simultaneously diagonalised  and their values can be used to label the states.
 
 %--------------- Fig 2 ----------
%---------------------------------------

\begin{figure}
 \includegraphics[width=0.45\textwidth]{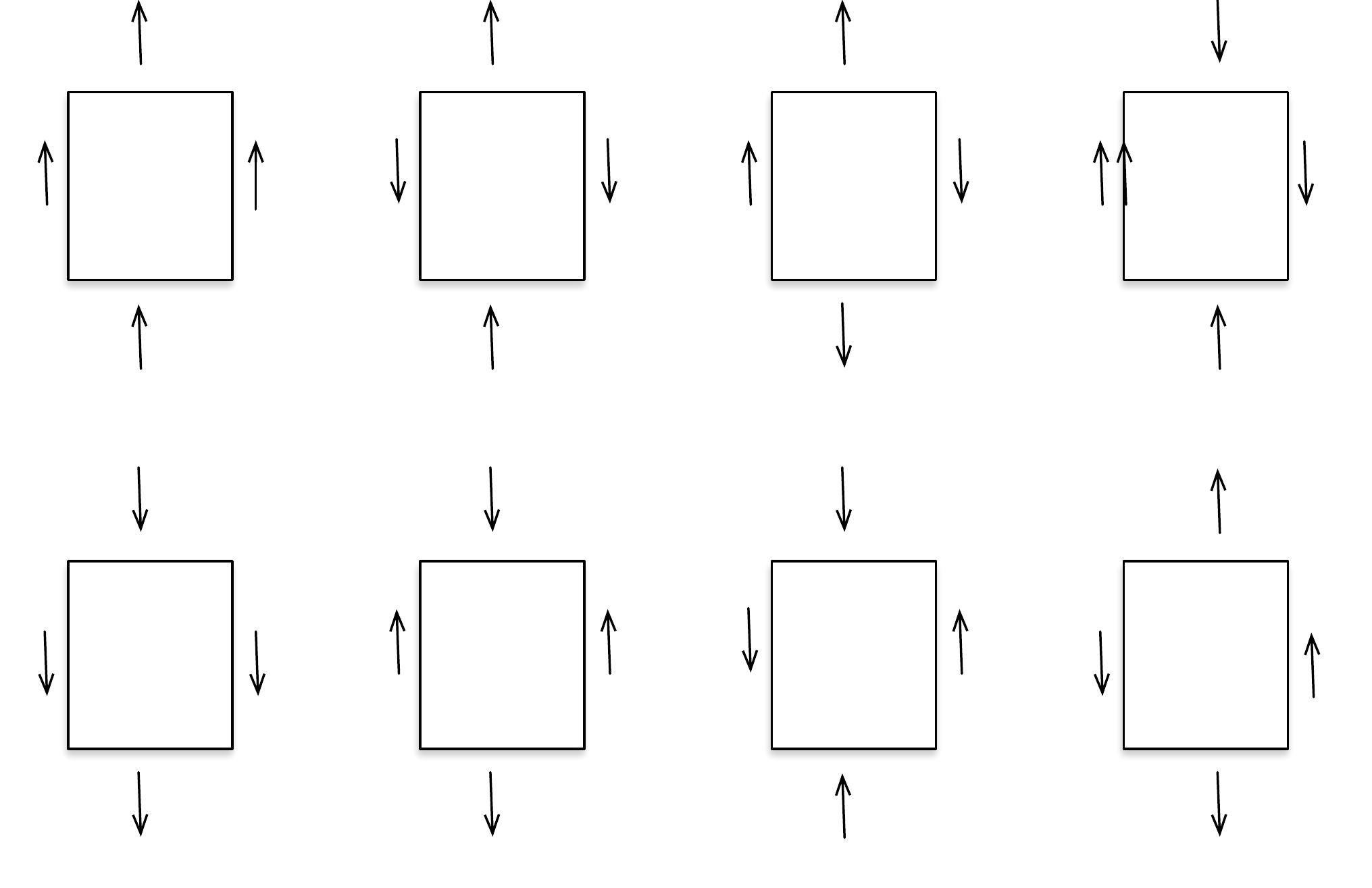}\hspace{0.5in}\includegraphics[width=0.45\textwidth]{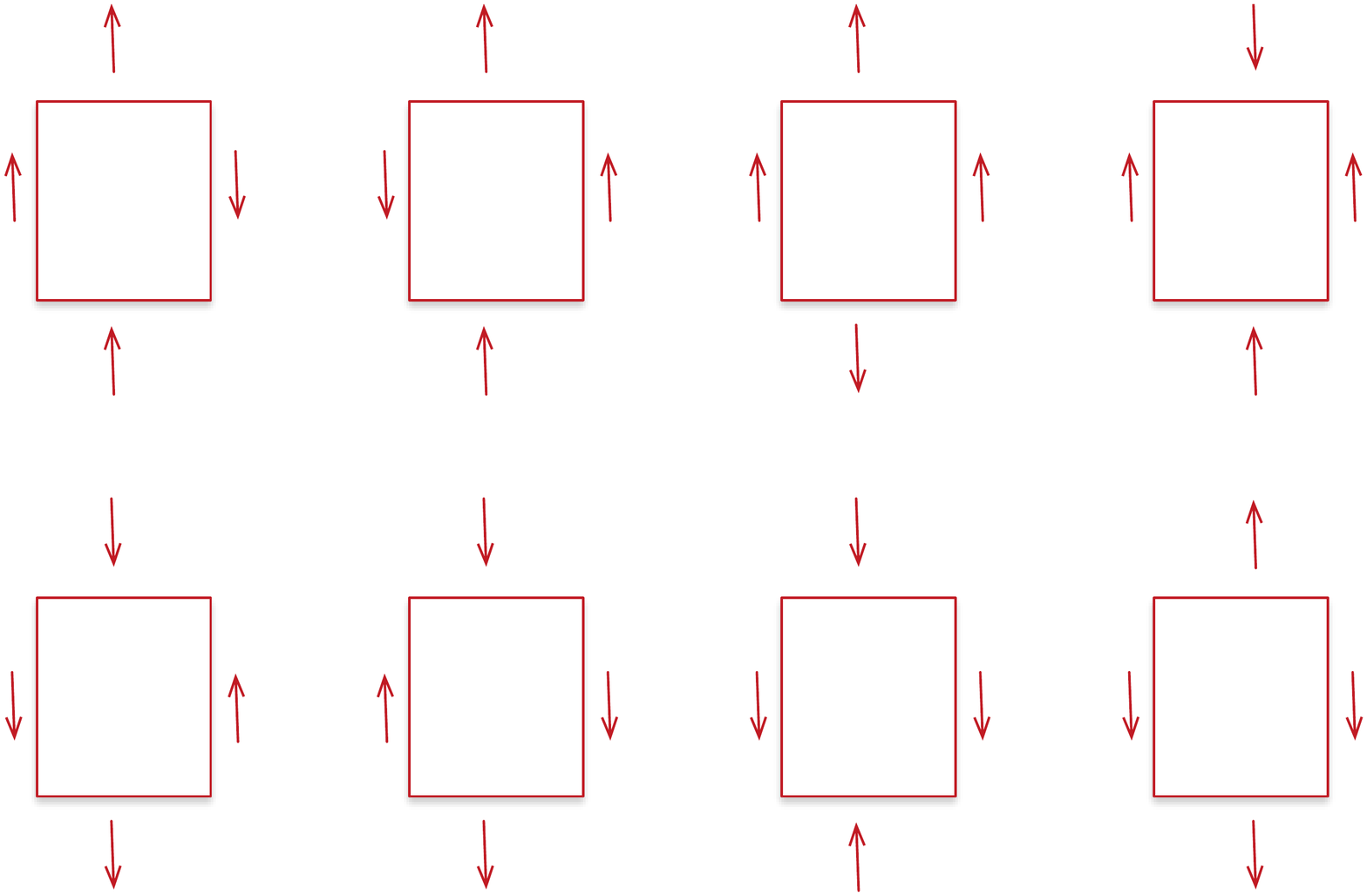}
\caption{ The configurations of spins on each plaquette on the left have $\omega_p (s)=1$ and
the configurations on the right have $\omega_p (s)=1$}\label{kitfig3}
	\end{figure}

The next step is to find  the ground state of the Hamiltonian. This means that the energy of all the terms in the  Hamiltonian
have be minimised,  which, in turn means that each of the $A_v$ and $B_p$ terms have to be maximised. Let us work in
the $\sigma_z$ diagonal basis. The eigenvalues of $\sigma_j^z$ are $s_j=\pm 1$. Let us also define $\omega_p(s)$ , the product of
the eigenvalues around the plaquette $p$ (where $s$ stands for  the configuration of $\{s_j\}$). This can also be just +1 or $-1$.
We will call this the flux through the plaquette. A configuration where $\omega_p=-1$ is called a vortex (or flux) configuration.
(See Fig.\ref{kitfig3} for examples.)
Now suppose that we have only $B_p$ terms in the Hamiltonian - i.e. $H=-J_m\sum_p B_p$.  Here, again the only possible eigenvalues
for the operator $B_p$ are +1 and $-1$ and the configurations are all given in Fig.\ref{kitfig3}. We note that there are eight configurations with `no
flux' ($\omega_p=+1$)  and eight configurations with non-zero flux (vortex configurations with $\omega_p=-1$)
as shown in Fig.\ref{kitfig3}.  As  to why these are called flux configurations, it turns out that this toric code model turns out
to be the same as the $Z_2$ gauge theory on a lattice, and the flux is that of the $Z_2$ gauge field. But the study of this connection
is beyond the scope of these lectures and for us here,  the word flux can be thought of as just nomenclature. The ground state is now
clearly given by any linear combination of configurations which have no vortices -i.e, the ground state is given by
\beq
|\psi> = \sum_{\{s,\omega_p(s) = +1 \forall p\}} C_s |s>
\eeq
where $C_s$ is arbitrary. All we know about the ground state is that it has no vortices and 
\beq
B_p|\psi>=|\psi> ~~ \forall ~ B_p
\eeq
because the eigenvalue of  $B_p$ acting on $|s>$ is always $+|s>$.   For $2N^2$ spins, the ground state degeneracy would be $2^{N^2}$ (because
pairs have to be either $\uparrow$ or $\downarrow$). In other words, of the total number of configurations $2^{2N^2}$, $2^{N^2}$ would have
$\omega_p=+1$ ( ground state configuration) and $2^{N^2}$ would have $\omega_p=-1$ ( vortex configuration).

%--------------- Fig 3 ----------
%--------------------------------------- 

Now let us add back the vertex terms to the Hamiltonian.
The $A_v$ acts on $|s>$ by flipping spins, since $\sigma_x\sigma_z\sigma_x^{-1} = -\sigma_z$, but in any plaquette, it
will always flip two spins. So it will keep configurations with $\omega_p=1$ in configurations with $\omega_p=1$.
Hence, operation of $A_v$ on $|s>$ will only take it to some other $|s'>$, which will also belong to the same set of vortex free configurations
with $\omega_p(s) =1$. But since $A_v$ can act on any of the lattice points,  $A_v$ can be an eigen-operator for $|\psi>$ only if $C_s=+1$ for all $s$.
So now, we define 
\beq
|\Psi_0>= \sum_{\{s,\omega_p(s) = +1 \forall p\}}  |s>
\eeq
as the ground state with all $A_v$ and $B_p$ acting on it with eigenvalue 1.  (We could have worked in a $\sigma_x$ diagonal basis,
and defined a  state $|s'>$ with `no ($x$) vortices'  in the dual lattice  configuration so that  $A_v$ acting on the state would give +1 for all $v$ and
repeated the same argument. So it is clear that 
this ground state is a sum over all spin configurations that
have neither $x$ nor $z$ `vorticity'.) It has an energy $E=- 2N^2J_e -2N^2J_m $. Any excitation over this ground state
 must have non-zero vorticity which would imply that at least one of the spins would have to flip in either the $z$ or $x$ directions. 
 In that case,  2 of the $B_p$'s or  2 of the $A_v$'s would
have eigenvalues $-1$ and hence, the energy of the excitation would be either  $4J_m$ or $4J_e$, but we will come 
back to these excitations later.

So we have found the  ground state of an interacting Heisenberg spin model in two dimensions exactly, 
essentially because the Hamiltonian has been constructed to be exactly solvable.
The ground state can also be written as 
\beq
|\Psi_0> = N \prod_v (1+A_v) |\xi>
\eeq
where $|\xi>$ is some reference state. For example, we can take $|\xi>$ to be the state with all spins pointing $\uparrow$.
The easiest way to check that this is the ground state is to check that for all $A_v$ and $B_p$, we get  the eigenvalue +1, when they
 act on this state. Let us check this. 
 \beq
 A_{v'} |\Psi_0> = A_{v'}N \prod_v (1+A_v) |\xi>
 \eeq
First consider the terms where $v\ne v'$. Then $A_{v'}A_v = A_vA_{v'}$.  But for $v=v'$, $A_{v'}(1+A_{v'}) = A_{v'} +1$, since $A_{v'}^2=1$.
Hence 
\beq
 A_{v'} |\Psi_0>  = +1|\Psi_0>
 \eeq
 Furthermore, we already know that $A_v$ acting on any state does not change its $z$-vorticity since it always flips two spins.
 Since the reference state has  vorticity  = +1 and $B_p=+1$ on all states with vorticity =+1, we also have
 \beq
 B_p|\Psi_0> = +1|\Psi_0> ~.
 \eeq
 This is essentially a unique ground state on a plane or with open boundary conditions.  
 We shall see later what happens when we have periodic
 boundary conditions which is equivalent to putting the model on a torus.
 But before that, let us see how to create excitations over the ground state. 
 
 \subsection{Excitations over the ground state and fusion rules}
 
 %--------------- Fig 4 ----------
\begin{figure}
 \includegraphics[width=0.45\textwidth]{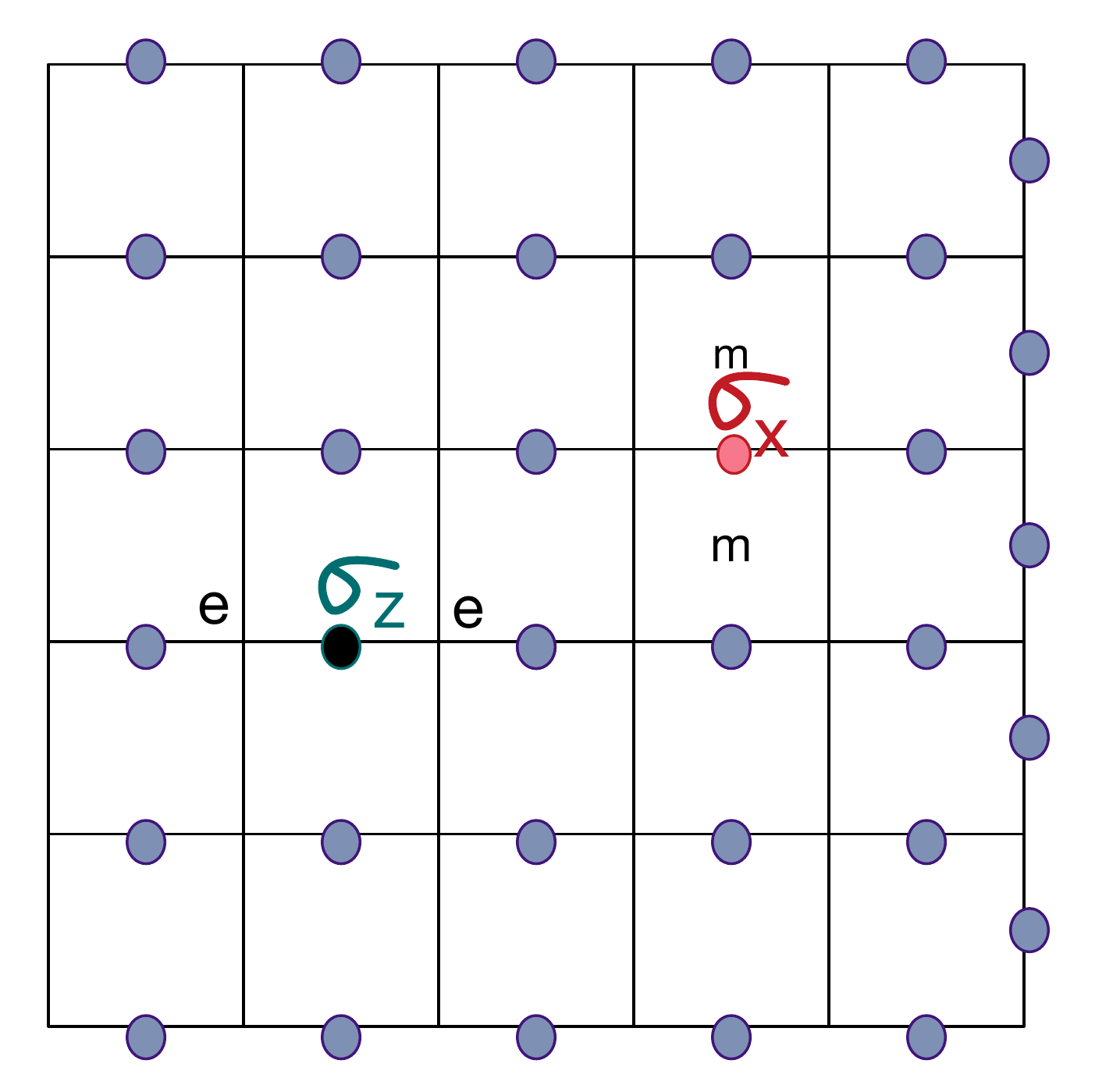}
\caption{ Flipping the $z$-component of the spin by $\sigma_x$ creates 2 monopoles, in adjacent plaquettes
whereas flipping the $x$-component of the spin by $\sigma_z$ creates 2 charges at adjacent
vertices.}\label{kitfig4}
	\end{figure}
%--------------------------------------- 

%--------------- Fig 5 ----------
\begin{figure}
 \includegraphics[width=0.3\textwidth]{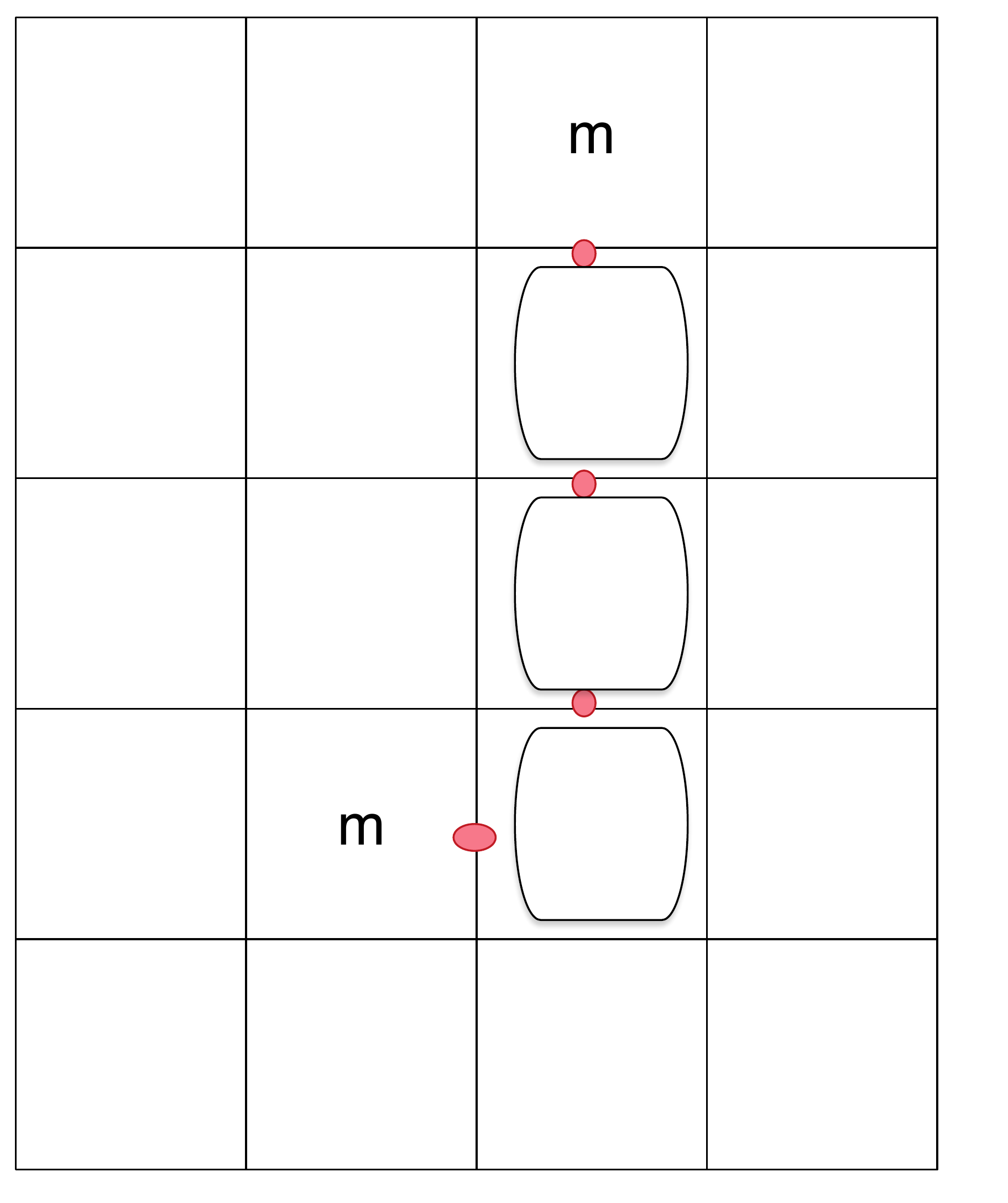}\includegraphics[width=0.3\textwidth]{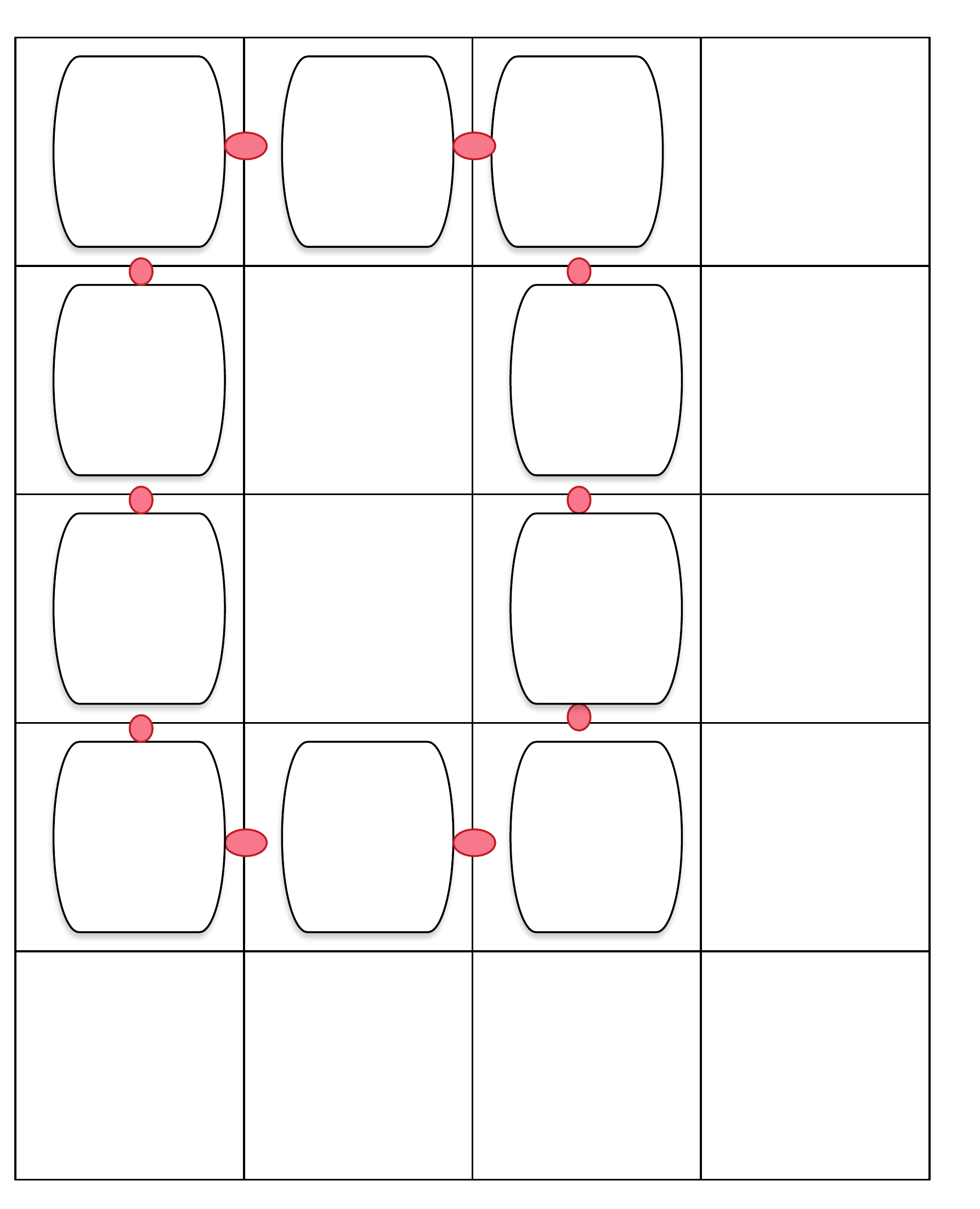}
\caption{(a) A pair of monopoles created by flipping the $z$-component of the spin by $\sigma_x$
along a string on the dual lattice. 
(b) Annihilating the monopoles  by closing the string. This configuration again has zero vorticity and commutes
with the Hamiltonian. (Note that the spins that do  not change are not shown in this diagram and in most  further diagrams, where they are obvious, 
in the interest of not cluttering the diagrams.)}\label{kitfig5}
	\end{figure}
%--------------------------------------- 

%--------------- Fig 6 ----------
\begin{figure}
 \includegraphics[width=0.3\textwidth]{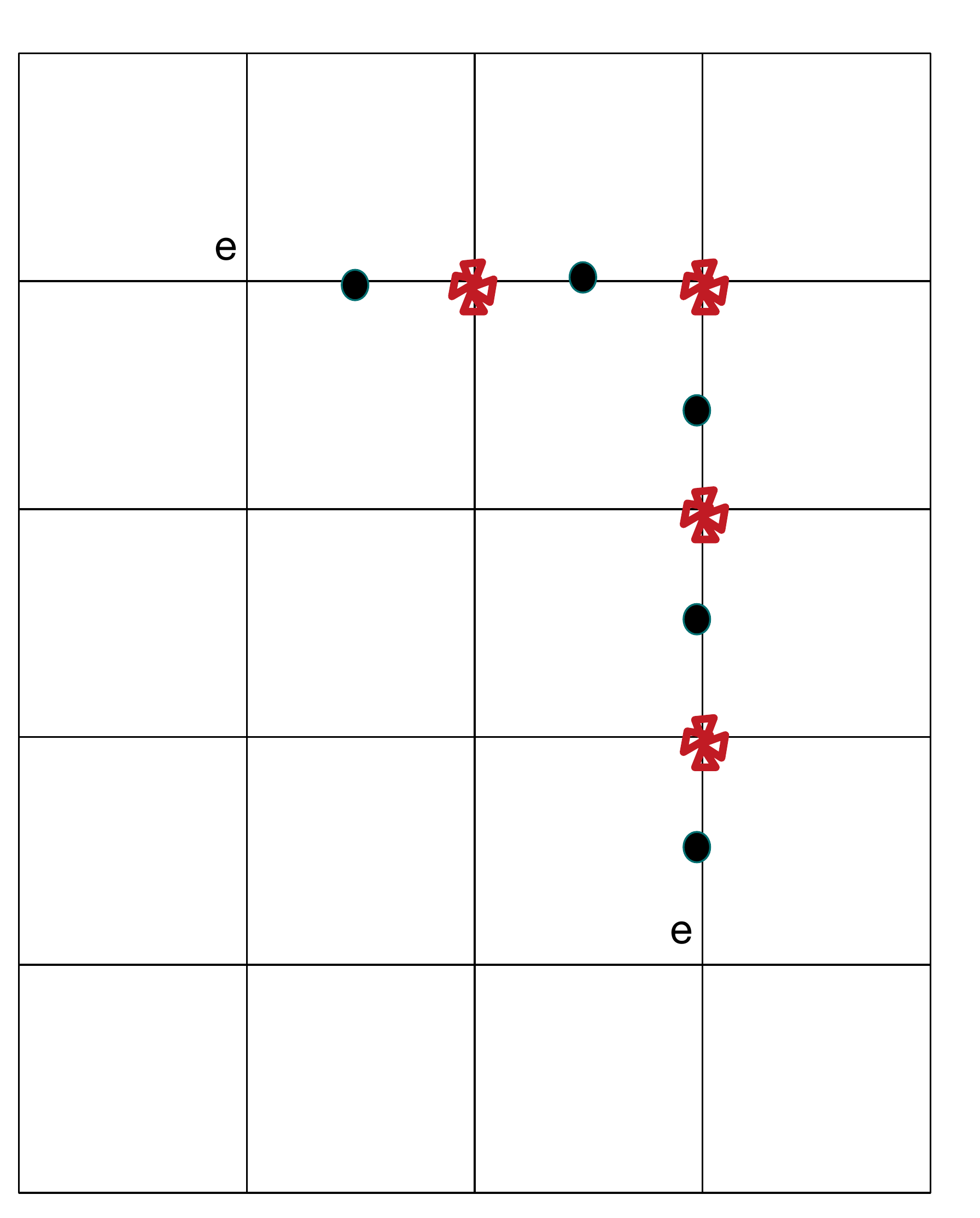}\includegraphics[width=0.3\textwidth]{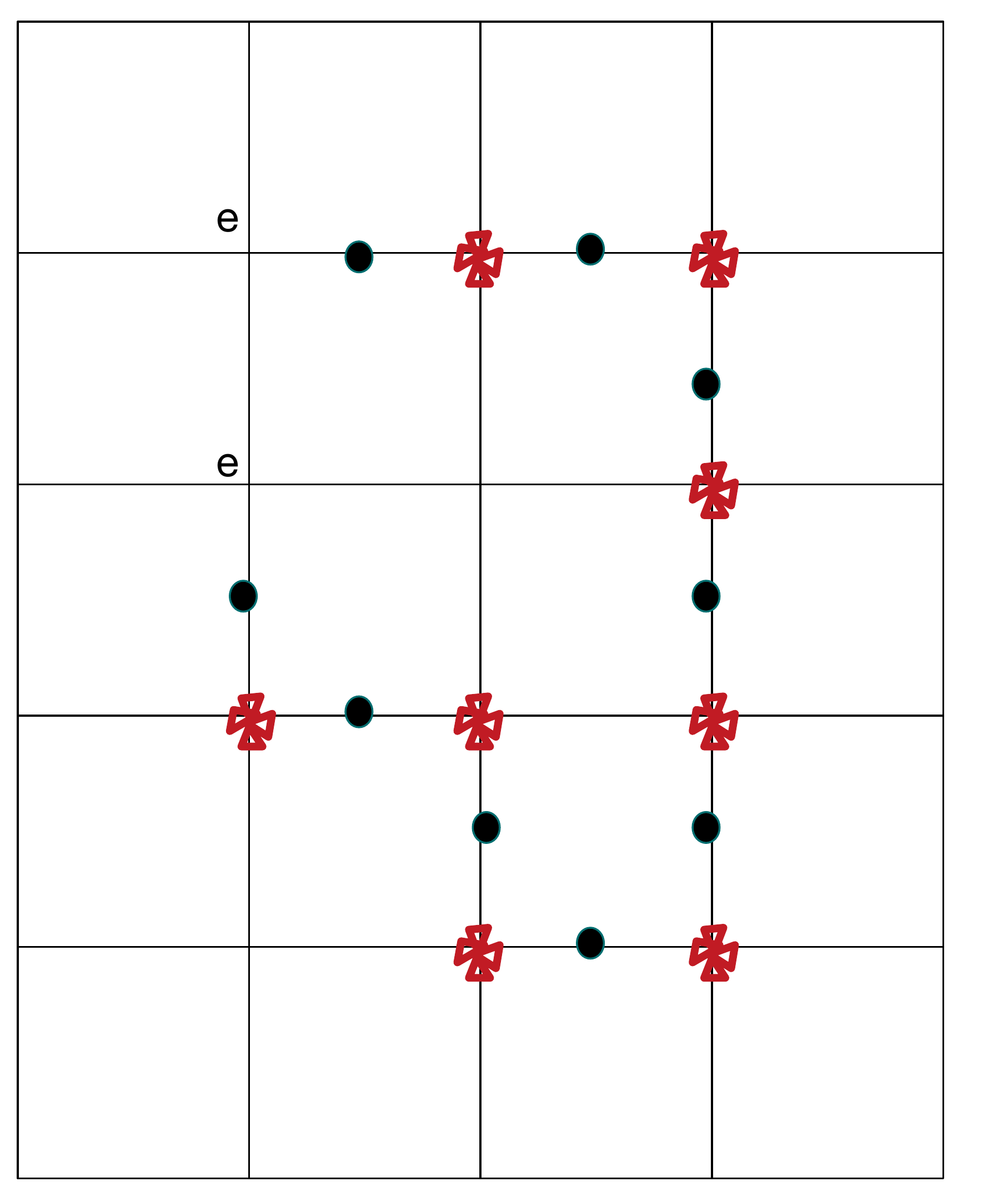}
 \includegraphics[width=0.3\textwidth]{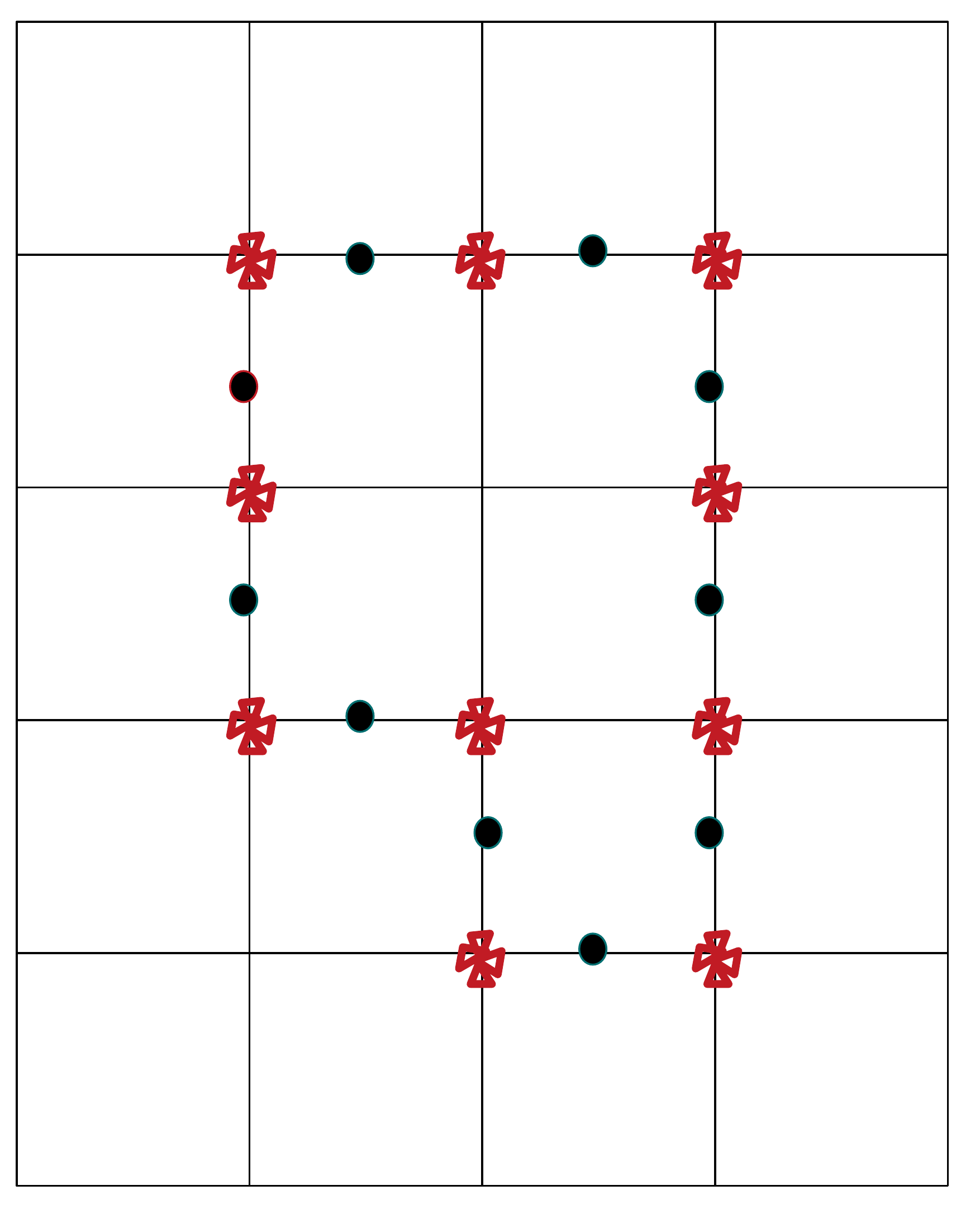}
\caption{(a) A pair of electric charges created by flipping the $x$-component of the spin by $\sigma_z$
along a string.  Note that the vertices in between the spins all have spins flipped on two links connected
to them. (b)  Moving  one of the charges by flipping more spins and increasing the length of the string
(c) Annihilating the charges by closing the string. This configuration has zero vorticity and commutes
with the Hamiltonian.   }\label{kitfig6}
	\end{figure}
%--------------------------------------- 

 There are two types of excitations that we
 can create on the ground state - one by applying $\sigma_x$ and the other by applying $\sigma_z$. Another way
 of thinking about excitations is to note that both $A_v$ and $B_p$ acting on the ground state have eigenvalues
 +1. One can make excitations if any of these values for some vertex or some plaquette becomes $-1$. 
 First, let us try to make the vorticity in a given plaquette become $-1$. 
 To do this, we need to flip the $z$-component of the spin
 on one link, which can be done by applying  $\sigma_x^i$ on the $i^{th}$ link.  So $\sigma_x^i|\Psi>$ flips
 $\uparrow$ to $\downarrow$ on the $i^{th}$ spin. However, this gives the value for $B_p$ to be $-1$ on two
 plaquettes, since the link is common to two plaquettes. The energy of this excitation is clearly $2J_m+2J_m=4J_m$
 since the sign change of a single plaquette costs $2J_m$.  These excitations are shown in
 Fig.\ref{kitfig4} where 2 e-excitations have been created on neighbouring vertices and 2 m-excitations 
 in neighbouring plaquettes.

   Note that the the pairs of excitations can be moved away from one another at no cost in
 energy. This is most easily seen diagrammatically.   The string $\Pi_{j\in t'} \sigma_x^j$  between the two magnetic excitations 
 (or monopoles)  changes the $z$-component on all the links between the end points as shown in Fig.\ref{kitfig5}(a), but essentially
 all the intermediate plaquettes have $B_p=+1$. If we consider drawing a line between the two
 monopoles,  note that this line  (or contour   or string) is defined on the dual lattice.

 Similarly, $A_v=-1$ if the $x$-component of the spin
 on one of the links gets flipped. This can be done by applying $\sigma_z$ to the ground state. But here again,
 the change of the $x$-component of the spin on a link affects two vertices, and hence creates a pair of electric
 excitations with energy $2J_e+2J_e=4J_e$. 
  Once again,  the two members of the pair of excitations can be moved away from one another at no cost in
 energy as is shown in Figs.\ref{kitfig6}(a,b)  by applying a string of operators, 
  $\Pi_{j\in t} \sigma_z^j$ between the two electric excitations.
 Here if we draw a line between the e-excitations,   the line or string  is defined on the lattice. The string
  changes the eigenvalues of the $x$ components of all the spins belonging to the string. However, except
 at the end points of the string, every vertex  will have two spins flipped and will continue to have $A_v=+1$. 
 It is only at the two ends of the string, that the vertices will have $A_v= -1$. For instance in Fig.\ref{kitfig6}(a), 
 we have flipped
 the $x$-component of 5 spins. However, we have created only two electric excitations.

We can even make closed string loops  that create, move and annihilate the electric or magnetic charges as shown in Figs.\ref{kitfig5}(b) and \ref{kitfig6}(c).
 These loops have $z$ or $x$ vorticity =+1  -  i.e., they commute with all $A_v$ and $B_p$.
 In other words, they commute with the Hamiltonian and can be thought of as symmetries, and the existence of these symmetries is the
 reason for the exact solvability of the model. The ground state can also be thought of 
 as a superpositions of all these loops.
 
 %--------------- Fig 7 ----------
\begin{figure}
 \includegraphics[width=0.5\textwidth]{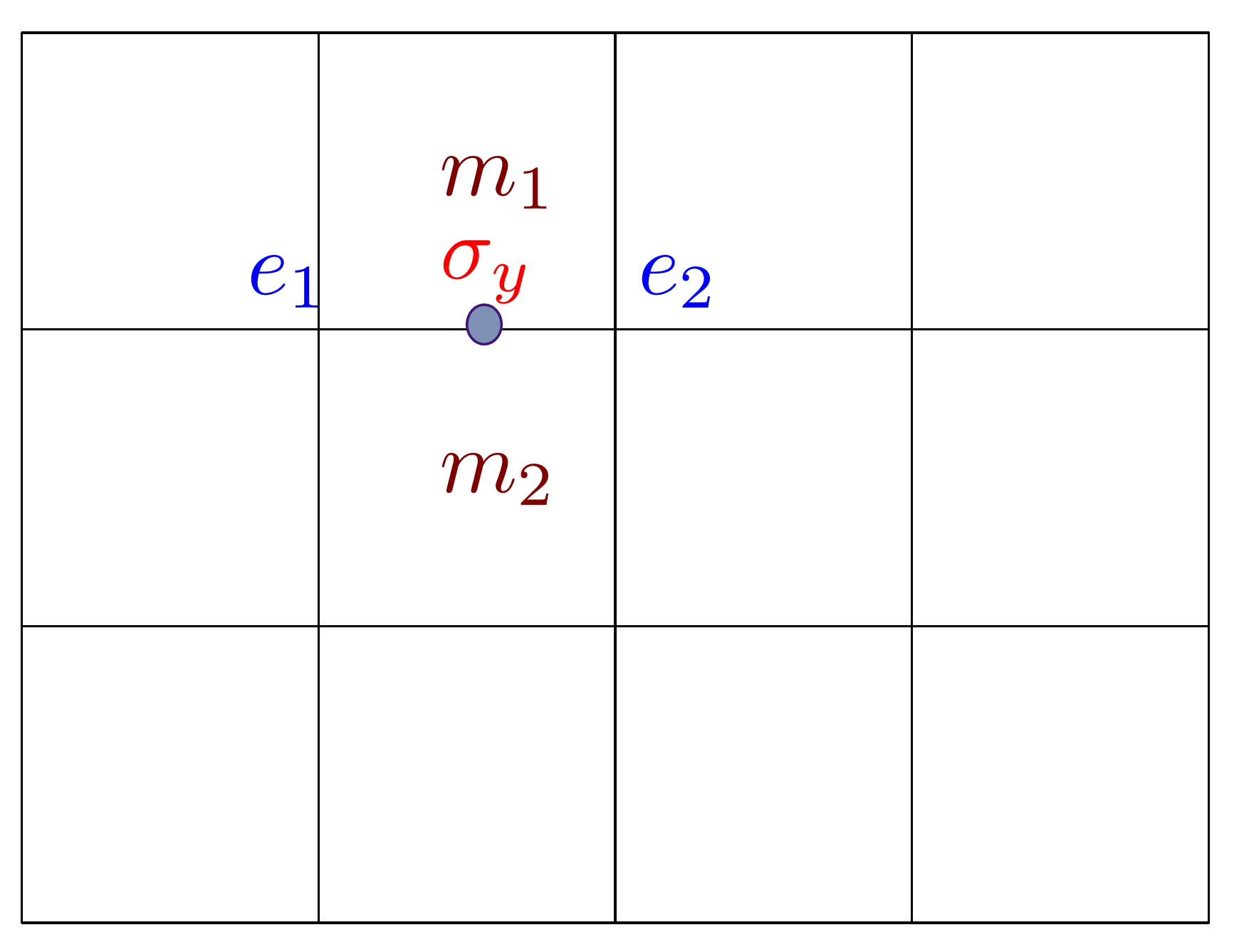}
\caption{A pair of monopoles  and a pair of electric charges created by flipping both the $x$-component 
and $z$-component of the spin by $\sigma_y$. This is equivalent to a pair of $\epsilon$ excitations.}\label{kitfig7}
	\end{figure}
%--------------------------------------- 

 Are there any other kind of excitations in the toric code? One might think that by applying $\sigma_y$ on the
 ground state, we may get new excitations. In fact, it turns out that there is a new excitation, which is a composite
 of the electric and magnetic excitations, which is called the $\epsilon$ excitation -
 \beq
 |\epsilon> = \sigma_z^i\sigma_x^i|\Psi_0> = i\sigma_y^i|\psi_0>
 \eeq
 Essentially, $\sigma_y^i$ acting on a spin flips both its $\sigma_x$ and $\sigma_z$ components. 
 So by applying it on a link as shown in Fig.\ref{kitfig7}, it creates a pair of electric and magnetic excitations, or
 a pair of $\epsilon = e\times m$ excitations. There are no further excitations that can be created. So the
 particle content of the model is given by $I$ (no particles), $m$-particles, $e$-particles and $\epsilon = e\times m$-particles.

 Now let us see what happens if we apply two $\sigma$ operators on the same plaquette. We know
 that $\sigma_x^j$ applied on a state creates a pair of excitations ( $m$- particles)
  on two adjoining plaquettes. But if we apply $\sigma_x^i$ twice, we do not get two pairs
  of excitations, because $(\sigma_x^j)^2 = I$ and $I$ operating on a state does not give rise
  to any excitation. So one cannot have more than one $m$-particles in 
  each plaquette. This leads us to what are called fusion rules.  We find that
  \bea
  e\times e &=& I \\ 
    m\times m &=& I \\ 
     \epsilon \times \epsilon &=& I ~.
     \eea
Further we had already seen that the first line of the following set of equations are true and it is not hard
to check that the others are true as well - 
  \bea
  e\times m &=& \epsilon \\ 
    e\times \epsilon&=& m \\ 
     m \times \epsilon &=& e~.
     \eea

   We had earlier seen that we can define string operators to move particles away from one
another and even annihilate them by forming closed loops. We said that all these closed loops
formed by creating, moving and annihilating particles, cost  no energy and commute with the 
Hamiltonian. They are products of $A_v$'s or $B_p$'s and 
can be thought of as trivial symmetries of the Hamiltonian because they map the
Hamiltonian onto itself. The ground state is unique and a linear combination of all these vortex-free states.

\subsection{Toric code on a torus and topological degeneracy}

  %--------------- Fig 8 ----------
\begin{figure}
 \includegraphics[width=0.4\textwidth]{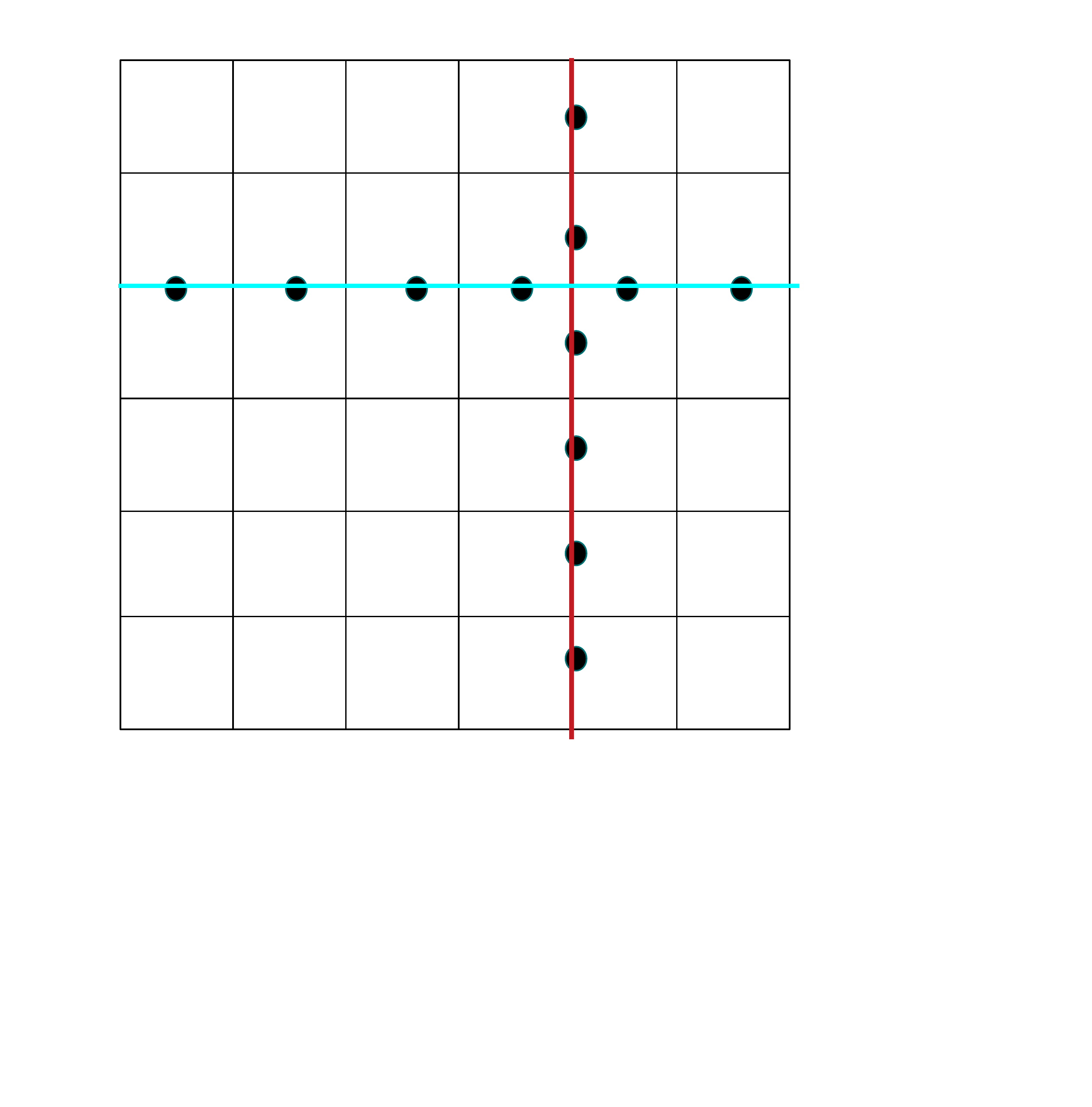}
\caption{Operators $W_{1\gamma_1}$ and $W_{2\gamma_2}$ which are defined by taking the product of the $z$ component of the spins,
$\sigma_z^j$, 
along the paths $\gamma_1$ (red) going
from top to bottom and $\gamma_2$ (cyan) going from left to right }\label{kitfig8}
	\end{figure}
%--------------------------------------- 

\begin{figure}
 \includegraphics[width=0.4\textwidth]{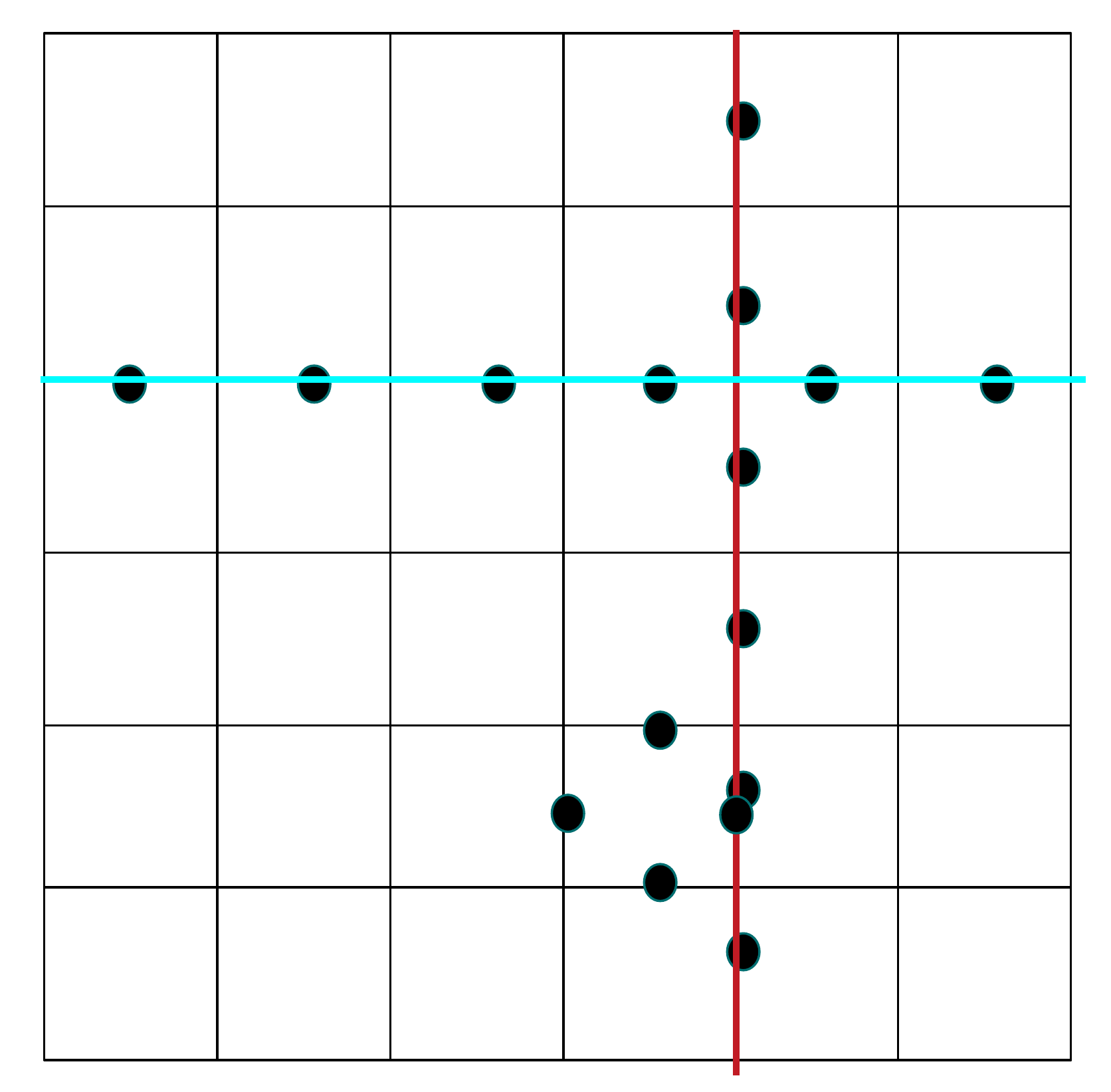}
  \includegraphics[width=0.4\textwidth]{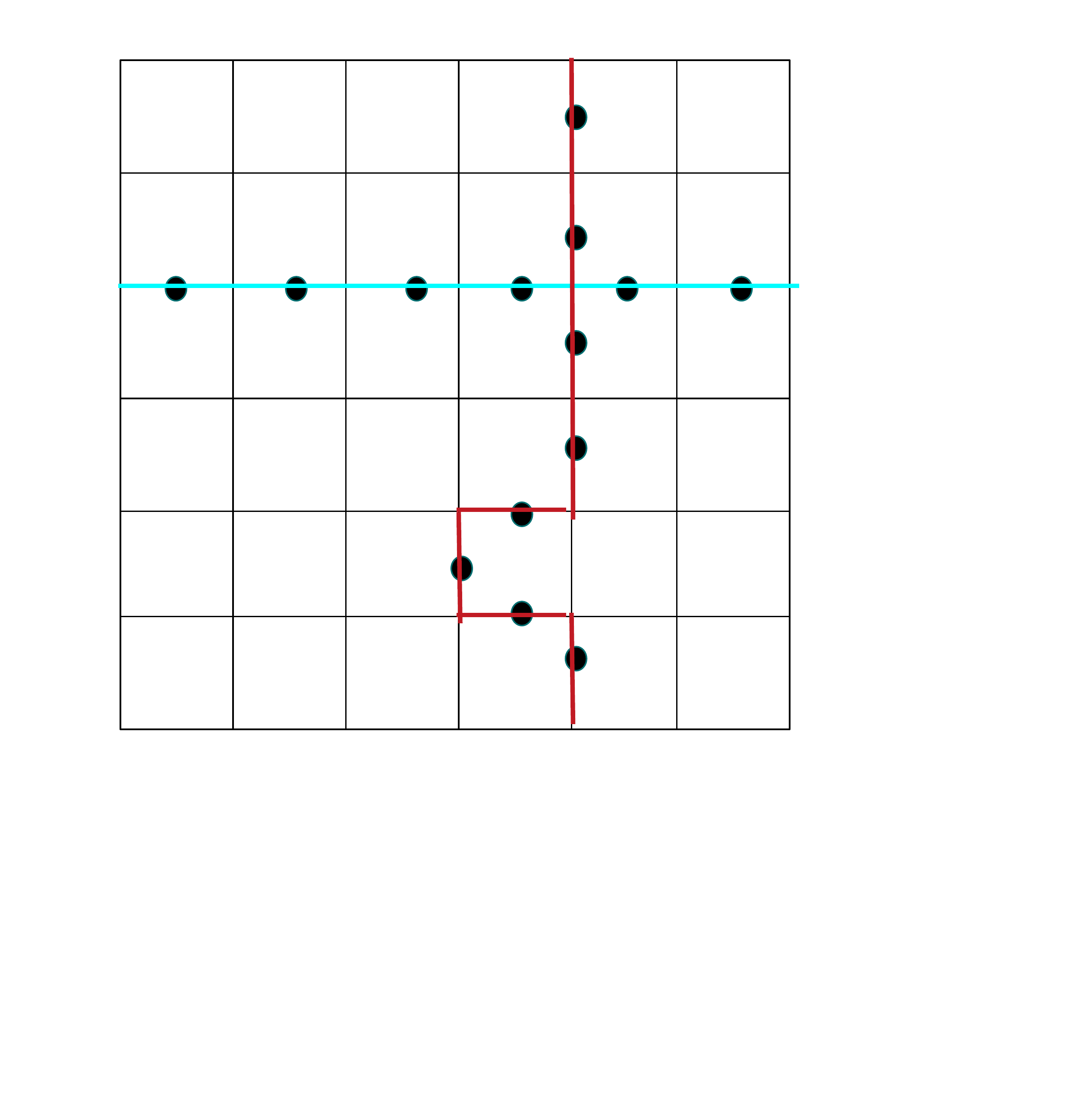}
\caption{The operator $W_{1\gamma_1}$ can changed  by multiplying the operator
by $B_p$. This essentially changes the path   $\gamma_1$, which is now no longer
the straight vertical path. 
 }\label{kitfig9}
	\end{figure}

But a new element is introduced if we have periodic boundary conditions or equivalently, consider
the toric code model on a torus. In this case,  there are two other  independent operators,  that
we can define, which are
not the products of the $A_v$ and $B_p$ operators of the Hamiltonian, and which can take
the values $+1$ and $-1$.
We can write them as 
\bea
W_{1\gamma_1} &=& \prod_{j\in \gamma_1} \sigma_z^j \\
W_{2\gamma_2} &=& \prod_{j\in \gamma_2} \sigma_z^j 
 \eea
 where $\gamma_1$ and $\gamma_2$ are paths which go from one edge of the torus 
 to the other in the two orthogonal directions  - for definiteness,
 let us assume that the loop $\gamma_1$ is in the vertical direction and the the loop
 $\gamma_2$ is in the horizontal direction - as shown in Fig.\ref{kitfig8}. They form non-contractible loops.
 The paths can be moved around by multiplying the $W_i$'s  with  $B_p$'s, (because
 they do not change anything since they just give +1 on the ground state as shown in Fig.\ref{kitfig9}). 
 But there is precisely one non-contractible loop in each direction,
 which we can take to be the shortest path. It is easy to see that 
 $W_1^2 = W_2^2 = 1$ which says that $W_1,W_2$ have eigenvalues $\pm 1$.
 They are symmetries, because they commute with the Hamiltonian.
 They also commute with one another and hence, they
 give rise to a four-fold degeneracy of the ground state  - 
 \beq
 |W_1,W_2>_{gs} = |1,1>,~~|1,-1>,~~|-1,1>,~~ |-1,-1>
 \eeq
 because each of the $W_i$ can take values $\pm1$.
 
   %--------------- Fig 9 ----------
%--------------------------------------- 

Now let us show that this degeneracy is topological and there is no local operator that can cause
transitions between these four ground states. 

We shall first see  if there are any operators that connect the degenerate states
 in the ground state manifold. We have already defined $W_1$ and $W_2$. 
  Let us now define also 
 \bea
 W_{3\gamma_1} &=& \prod_{j\in \gamma_1} \sigma_x^j \\
W_{4\gamma_2} &=& \prod_{j\in \gamma_2} \sigma_x^j 
\eea
where $W_3^2=W_4^2=1$. These operators also commute with the Hamiltonian, but they do
not increase the degeneracy, because as we shall see below, they do  not  commute with the earlier
operators. There are only 4 mutually commuting operators that commute with the Hamiltonian.
Clearly, $[W_1,W_2]=0$ and $[W_3,W_4]=0$. It is also easy to see that if both the loops are
in the vertical or horizontal direction, they will commute, since they can always be displaced,
i.e., we 
see that 
 $[W_1,W_3]=0$ and $[W_2,W_4]=0$. However, this is not true when we consider $W_1$ and $W_4$.
 With even the simplest choice of path, they must have at least one spin in common, since one of
 the paths is vertical and the other horizontal. (More complicated paths will also always give
 odd number of common spins). So  if this common spin is at location $0$, then it is the anticommutator
 of the two operators which vanishes - $\{W_1,W_4\} = \{\sigma_{(0)}^z,\sigma_{(0)}^x\} =0$ 
 (and similarly $\{W_2,W_3\} =0$).  So it is as if we can define $W_1 =\sigma_1^z, W_2 = \sigma_2^z,
 W_3 = \sigma_2^x, W_4 = \sigma_1^x$, so that the $W_i'$s can be represented as Pauli matrices.
 Thus, $W_3$ and $W_4$ can change the eigenvalues of $W_1$ and $W_2$ by acting on the
 spins one at a time.
 
 Now how do we confirm that the ground state degeneracy 
  is a topological degeneracy and is topologically protected?
 The idea is that no local operator allows transitions between the different ground states.
 Suppose $\Omega$ is a local operator -i.e., it is of the form
 \beq
 \Omega \equiv \sigma_i^\alpha \sigma_j^\beta \sigma_k^\delta \dots
 \eeq
 where the links $i,j,k$ are nearby in the sense that the maximum distance is small compared
 to the thermodynamic limit $N$.  Then, we can always ensure that 
 \beq
 [\Omega,W_i] =0
 \eeq
 This means that $\Omega$ commutes with both $\sigma_i^x$ and $\sigma_i^z$ - i.e., with both
 Pauli matrices at a given site. This means that it has to be proportional to the identity. So
 it cannot directly cause any transition between different ground states. It can only lead to 
 transitions if it can cause indirect transitions which will take it out of the ground state manifold.
 This would cost an energy $E_0$ which is the energetic distance to the next state. Moreover,
 since the operator is local, it would have to be applied $N$ times since the $W_{3,4}$ operators
 change the state of 1 link at a time. So if the relevant matrix element of $\Omega$ is $\omega$,
 we see that the total transition amplitude is of $O((\omega/E_0)^N)$ which goes to zero
 in the thermodynamic limit as long as $\omega/E_0<1$. In other words, no local operators can
 cause transitions between the degenerate states. One needs a non-local measurement to 
 distinguish between the four degenerate ground states.   This is why the ground state is said
 to have topological order, and this  
 is what makes it relevant for quantum computation.
 
    %--------------- Fig 10 ----------
\begin{figure}
 \includegraphics[width=0.2\textwidth]{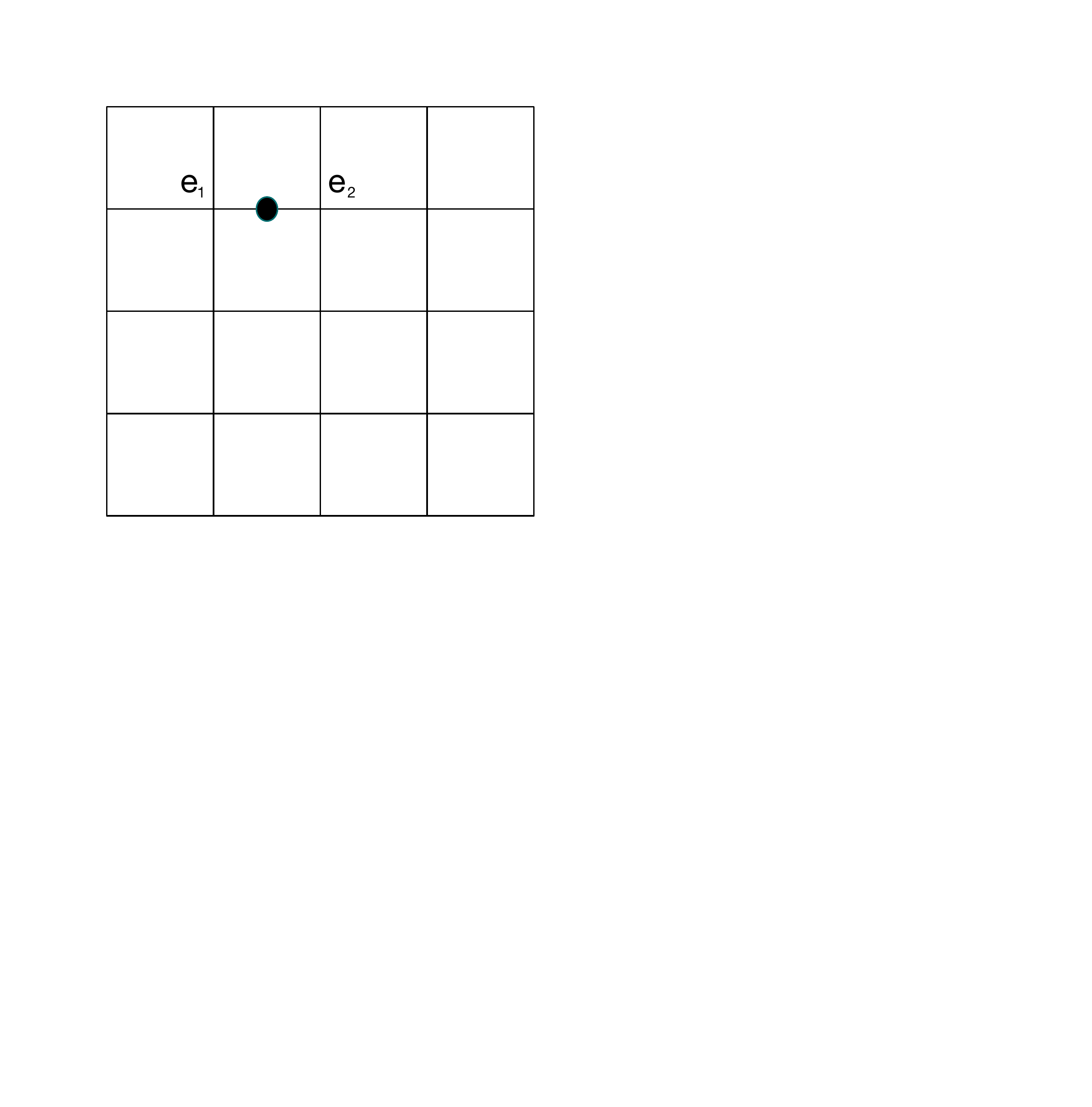}
  \includegraphics[width=0.2\textwidth]{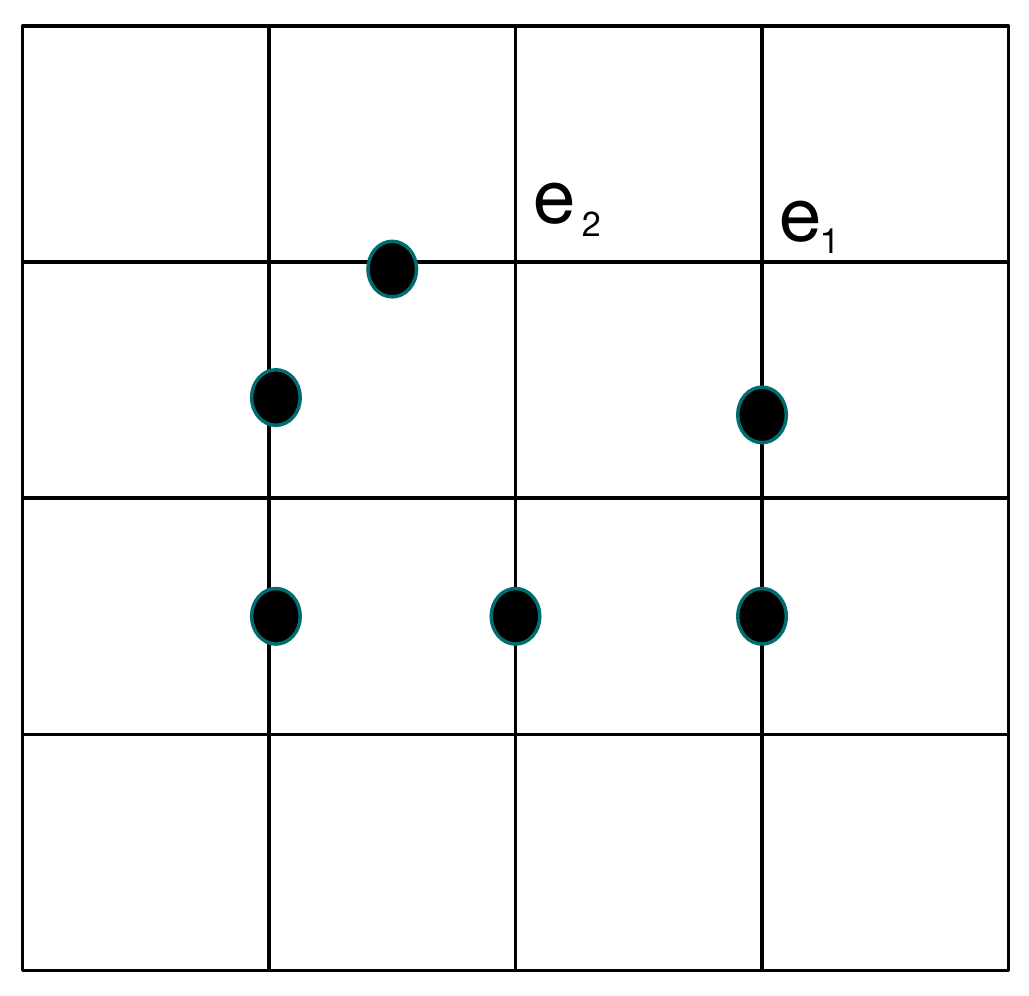}
  \includegraphics[width=0.2\textwidth]{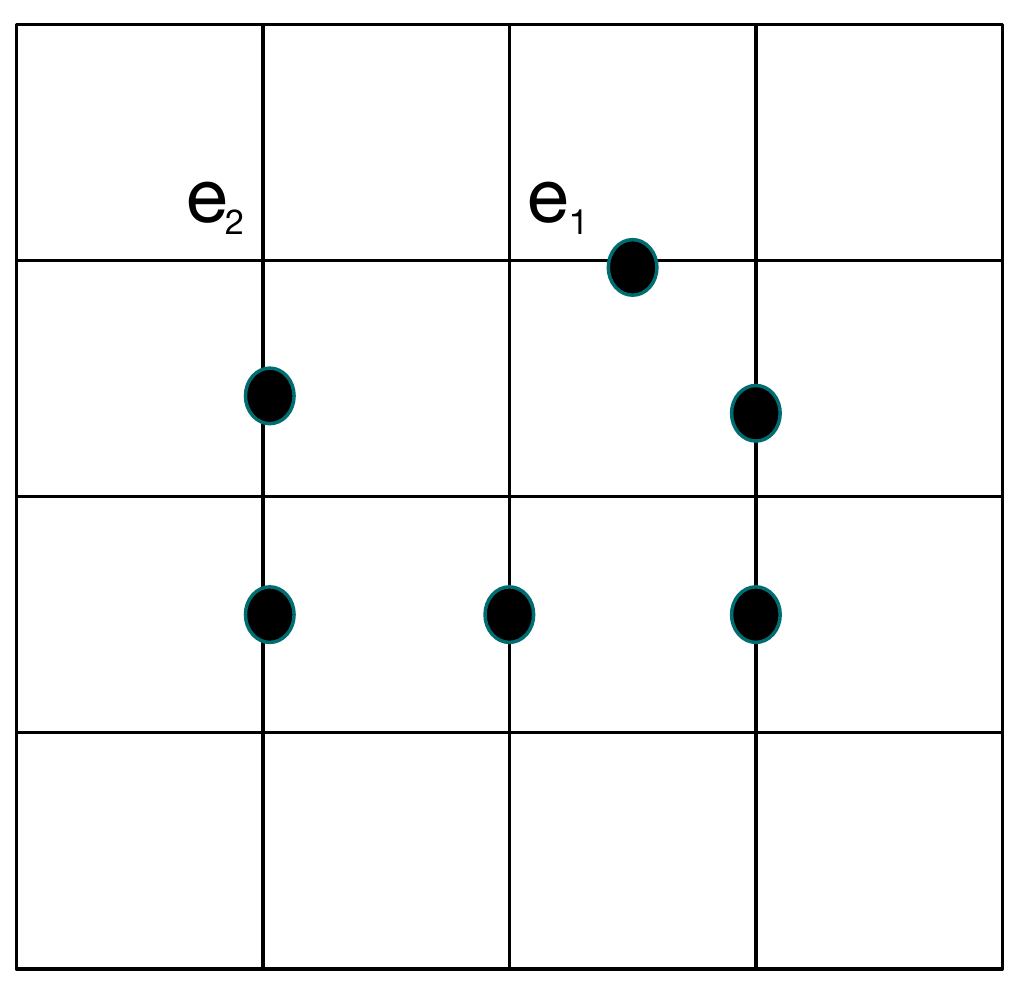}
  \includegraphics[width=0.2\textwidth]{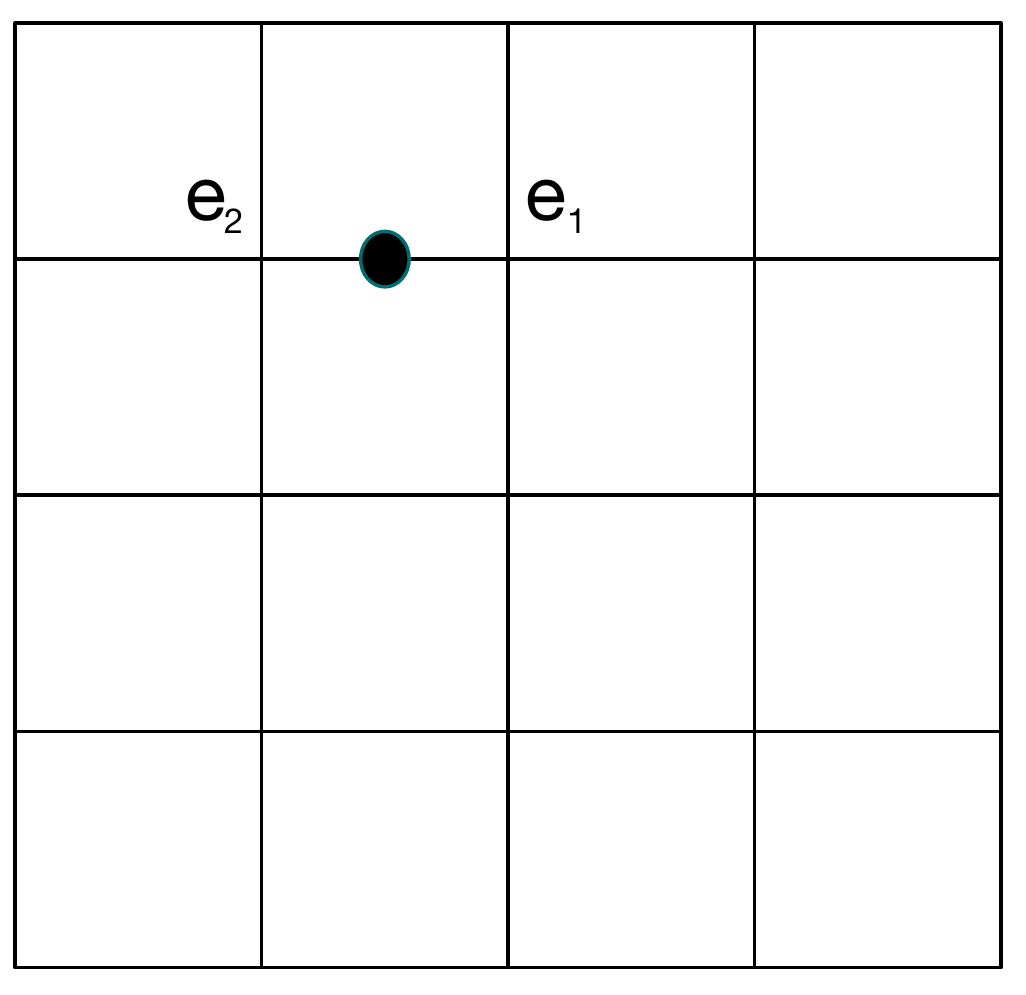}
\caption{The exchange of the charges $e_1$ and $e_2$ by operating with the string
operator $\Pi_j\sigma_j^z$. In the last step, we have simply removed a closed string loop of vorticity zero.
 }\label{kitfig10}
	\end{figure}
%--------------------------------------- 

\subsection{Statistics and braiding properties of the excitations}

 Finally, we want to understand the statistics and braiding properties of the excitations.
 Let us first look at the statistics of $e$-particles. $\sigma_i^z|\zeta>$, where $|\zeta>$ is the
  ground state,   creates two $e$-particles in the adjacent vertices to the edge $i$. A string $\Pi_j\sigma_j^z$
  can separate the two excitations. As shown in Fig.\ref{kitfig10}, the two particles can even be
  exchanged by applying the string operators. But since all $\sigma^z$'s commute with one another,
  whichever way we move them, we do not get any phase. Thus, the $e$-excitations are bosons.
  Similarly, it is easy to argue that all $m$-excitations are also bosons.

\begin{figure}
 \includegraphics[width=0.18\textwidth]{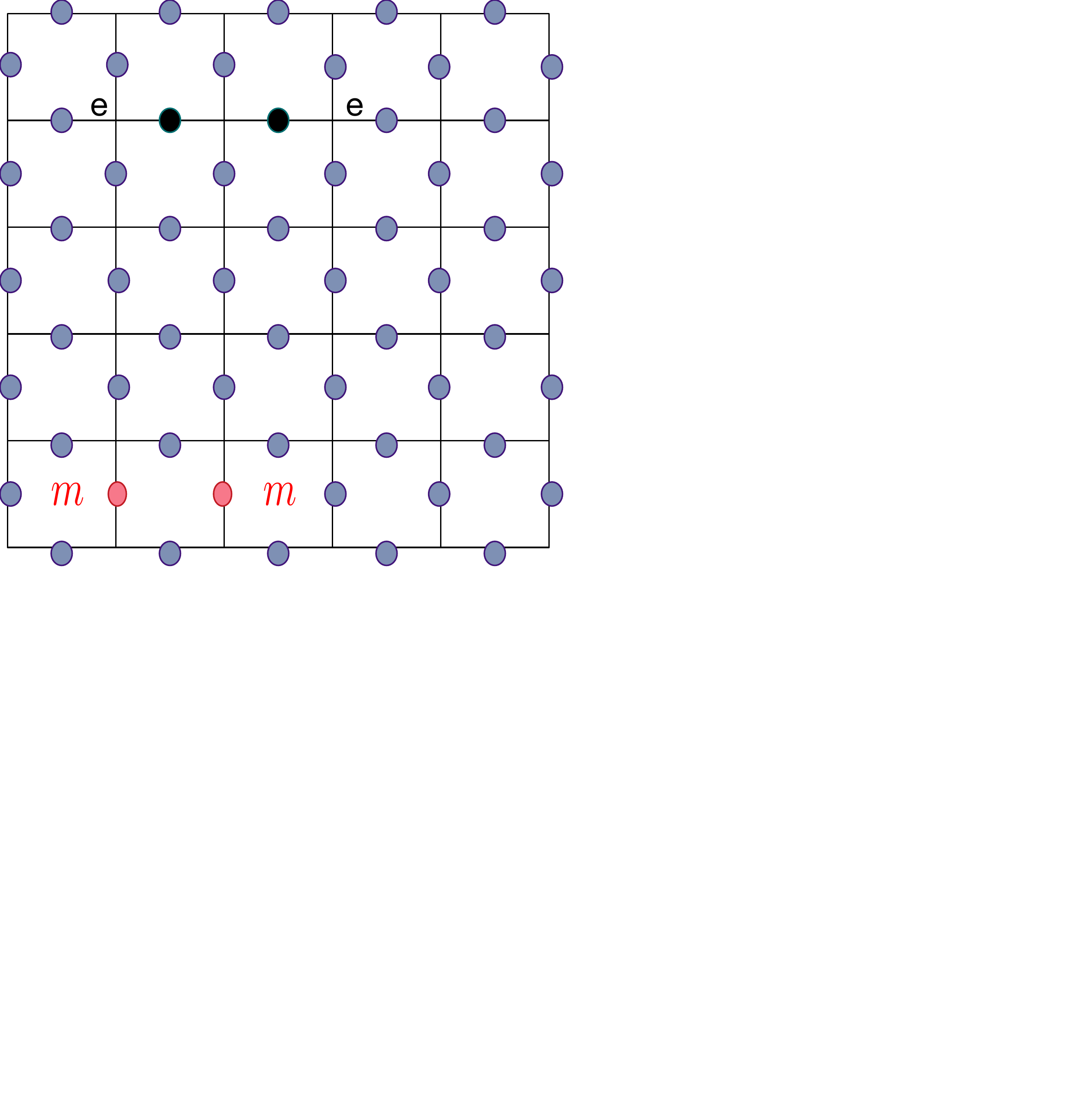} \quad \quad ~~~
  \includegraphics[width=0.18\textwidth]{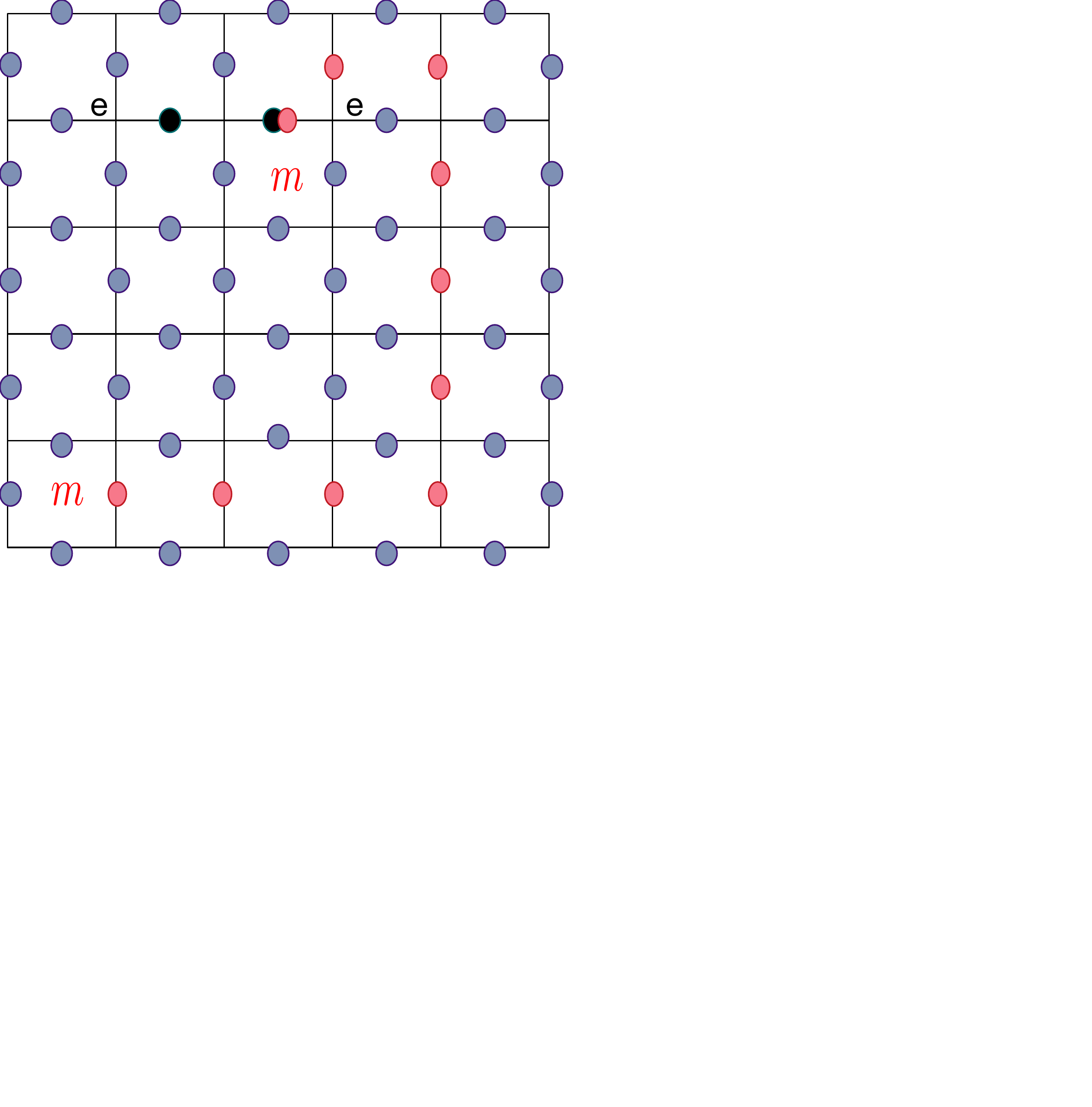} \quad \quad  ~~~
  \includegraphics[width=0.18\textwidth]{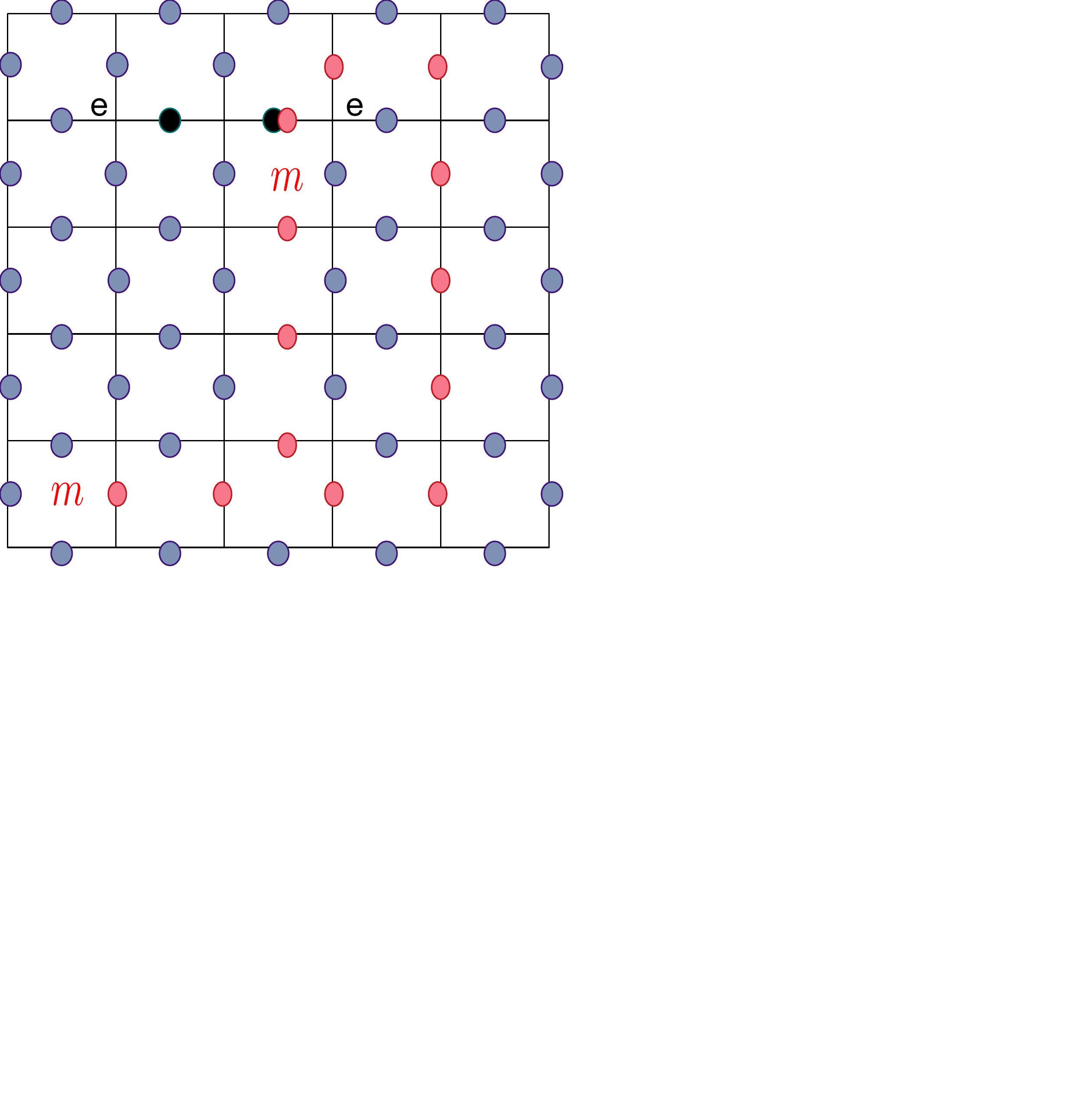} \quad \quad  ~~~
  \includegraphics[width=0.18\textwidth]{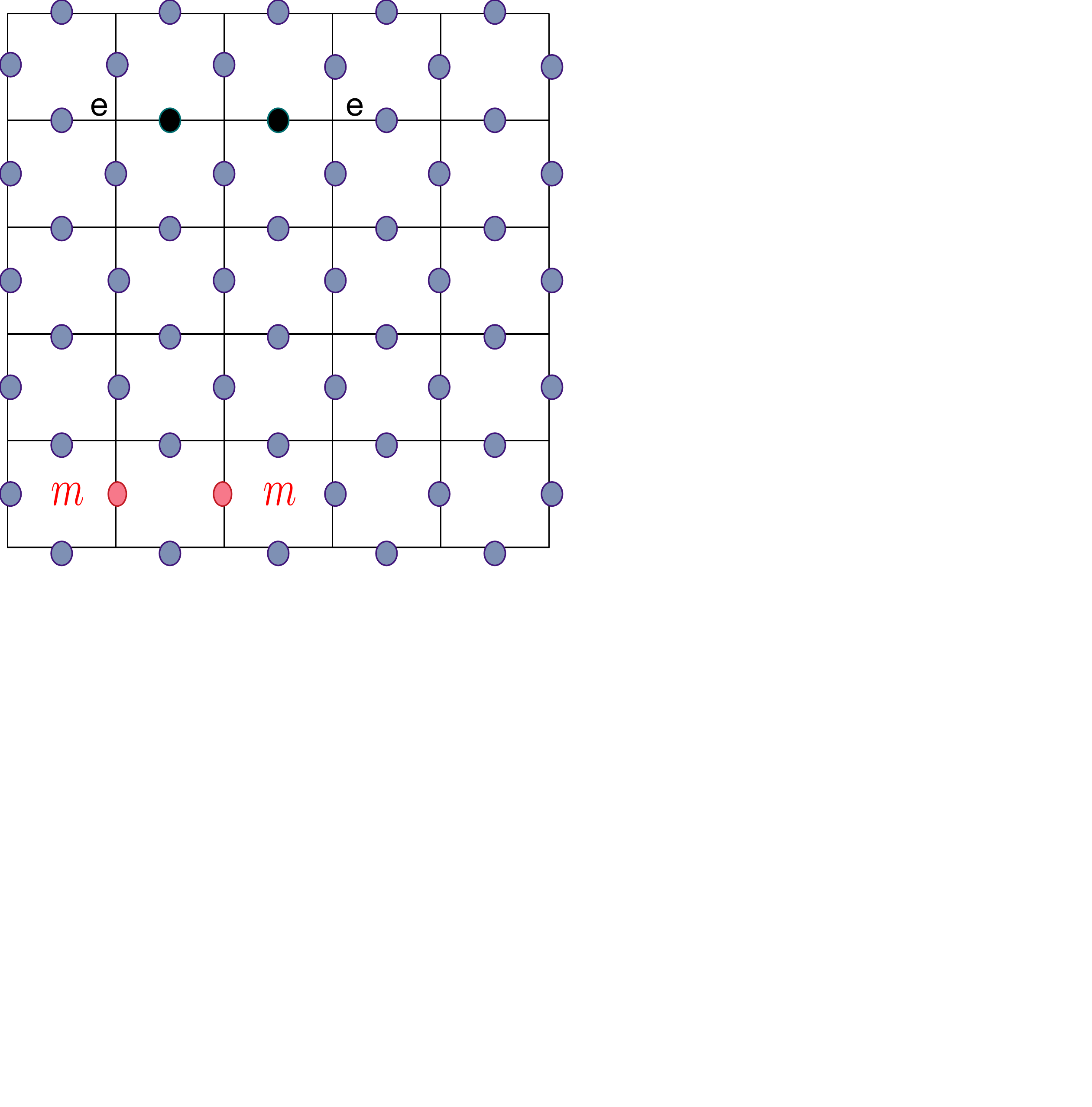}
\caption{The sequence of steps in taking an  $m$ charge around an $e$ charge by operating with the string
operator $\Pi_j\sigma_j^x$. In the second  step, we note that a $\sigma_x$ has to go past a $\sigma_z$,
leading to a negative sign. In going from the third to the last figure, a closed loop of $\sigma_x$ operators
has been removed, returning the configuration to the original configuration in the first figure.
 }\label{kitfig11}
	\end{figure}

 But now let us see consider the mutual statistics between $e$ and $m$ particles. 
   We first create pairs of $e$ and $m$ excitations at sites $i$ and $j$ by applying $\sigma_i^z\sigma_j^x|\zeta>$. We then separate the excitations by applying string operators as shown in Fig.\ref{kitfig11}(a)  and then we move an $e$-particle all around an $m$-particle, as shown in the series 
  of figures in Fig.\ref{kitfig11}(b),  Fig.\ref{kitfig11}(c) and Fig.\ref{kitfig11}(d).
   It is clear that there is one site at which the $\sigma^x$ has to be taken beyond a $\sigma^z$
  spin. This anti-commutation gives rise to a minus sign. The closed loop can be removed and the net effect of the
  process is that
  \beq
  \sigma_i^z\sigma_j^x|\zeta> \rightarrow  - \sigma_i^x\sigma_j^x|\zeta>~. 
 \eeq
 Now let $R_{em}$ be the operator that exchanges the $e$ and $m$ particles. We have found that the
 wavefunction for the creation of the two excitations, 
 \beq
 \psi ({\bf r}_e,{\bf r}_m)  \rightarrow R_{em}^2 \psi({\bf r}_e, {\bf r}_m) = - \psi({\bf r}_e, {\bf r}_m)
 \eeq
 since taking one particle completely around another is equivalent to two exchanges. Hence $R_{em} = e^{\pm i\pi/2} =\pm i$, which means that the $e$ and $m$ particles have mutual anyon statistics.
 
     %--------------- Fig 11 ----------
%--------------------------------------- 

 We have already seen that $R_{ee} = R_{mm} =1$.  The $e$ and $m$ particles are bosons under
 exchange. What about the $\epsilon$ particles? It is clear that if we take one $\epsilon$ particle
 completely around another, it is equivalent to taking an electron-monopole pair completely 
 around another electron-monopole pair ( since $\sigma_y|g.s> \sim  \sigma_x\sigma_z|g.s>$).
 In this case, the phase we expect to get is +1 since there will be two negative signs coming
 from anti commuting  a $\sigma_x$ through a $\sigma_z$ and vice-versa, so two anti-commutations
 altogether. But this is not enough to tell us whether a single exchange gives a +1 or a -1.
 But the fact that taking the $m$ particle around the $e$ particle gives rise to a -1 can be interpreted 
 as getting a negative sign when the $\epsilon$ particle is rotated through $2\pi$. This is the
 signature of a fermion. It is a `spinor' and requires a rotation through $4\pi$  to get back to itself.
 Hence, the $\epsilon$ particle is a fermion and we conclude that $R_{\epsilon\epsilon} = -1$.
 
 So now we have all the fusion and braiding rules for the excitations of the toric code.
 The particle content of the model is given by $I,e,m,\epsilon$.  The fusion rules are
 \bea
 &e\times e = I,~~~ m\times m = I,~~~ \epsilon\times \epsilon = I \nonumber \\
 & e\times m= \epsilon,~~~ e\times \epsilon  = m, ~~~m\times \epsilon = e
 \eea
 and the braiding rules are
 \beq
 R_{ee} = R_{mm} = 1, ~~~ R_{\epsilon\epsilon} = -1,~~~ R_{em} = i
 \eeq
 Hence, this model has mutual (abelian) anyonic statistics. We will leave this model here
 and now go on to study a model which has non-abelian anyon statistics.

\section{Non-abelian anyons}

In this section, we will study a model\cite{kitaev3} which has non-abelian anyon excitations.
Let me start with a brief explanation of why non-abelian anyons\cite{mooreread,mong} are of interest today,
other than being an exotic form of exchange statistics. 
The reason is that they are expected to be relevant to quantum computation\cite{alicea}. Quantum
computation requires the possibility of storing quantum information. This needs a  `protected'
portion of the Hilbert space which will not be disturbed by noise, temperature, etc,
as long as the length scales are below the gap which separates the `protected' states
from the rest of the states. This protection may be due to some symmetry or even due
to topology, which, in a sense acts like a robust symmetry, since it cannot be easily destroyed.
This leads us to the idea of topological order\cite{topologicalorder}, which exists if there 
is a degeneracy due to topology.
For instance, in the earlier section, we studied the toric code, which had only abelian (mutual) anyons
and no degeneracy on the plane. But the toric code on a torus has a four-fold degeneracy which
was topological and could be used for quantum computation.

In general, in models with non-abelian anyons, we will find that the ground state is degenerate,
and if the ground state is separated from all other states by a gap, then the ground state has topological
order and can be used for quantum computation.

With this motivation, let us now see\cite{preskill} what one means by non-abelian excitations.
Suppose we have $N$ degenerate states represented by $\psi_\alpha$ $,\alpha = 1\dots N$,
representing $N$ particles.  The idea is that if we now exchange particles 1 and 2, it can now
do more than just give a phase. It can rotate the state to another wave-function in the same
degenerate space. In other words, the column vector $\psi = (\psi_1\psi_2 \dots \psi_N)^T \rightarrow \psi' 
=U\psi$ where $U\equiv U_{\alpha\beta}$ is an $N\times N$ unitary matrix. If we now exchange two other particles, say 2 and 3, it could lead to $\psi \rightarrow \psi'= V\psi$  with $V$ another unitary
$N\times N$ matrix. If $U$ and $V$ do not commute, which, in general, they will not, the particles
are said to have non-abelian statistics. Clearly, to have non-abelian statistics, we need to have at
least 2 degenerate states, since otherwise, $U$ and $V$ are just phases and commute and
we have abelian anyons.

   One could think about generalising the physical model of any anyon - a charge orbiting around
   a flux - to the non-abelian case. In this case, the non-abelian charge would be a vector
  $ |q_i> = \begin{pmatrix} q_1  & q_2  & . & . &q_N \end{pmatrix}$ moving around a non-abelian flux
  and returning to its original position, but in the process, instead of just acquiring phase factors,
  \beq
  |q_i> \rightarrow |q_i^\prime> = \sum_j U_{ij} |q_j>~,
 \eeq
 where $U_{ij}$ is the non-abelian flux  matrix.
But unlike the abelian flux which was path independent, the non-abelian counterpart $U$ is
path dependent. It depends on where the path begins and ends as well as the contour.  If we think
of $U$ as a matrix belonging to the gauge group $U(N)$ or $SU(N)$,  it means that the
transport of charge around a flux  is gauge dependent because $U$ depends on the choice
of gauge. It is only the eigenvalues of $U$ (also called conjugacy class of the flux in group $G$)
which is gauge independent. So for the non-abelian anyons, the physical picture does not help
in simplifying or understanding the model and it is better to deal with the more abstract picture.

The main ingredients for a theory of non-abelian anyons are the following - \\
(1) We need a list of types of particles in the model. \\
(2) We need fusion rules - rules for fusing two constituents into one and also
for splitting a particle into two constituents, which is its inverse. \\
(3) We need rules for braiding two particles (equivalently exchanging two particles).

     %--------------- Fig 1 ----------
\begin{figure}
 \includegraphics[width=0.75\textwidth]{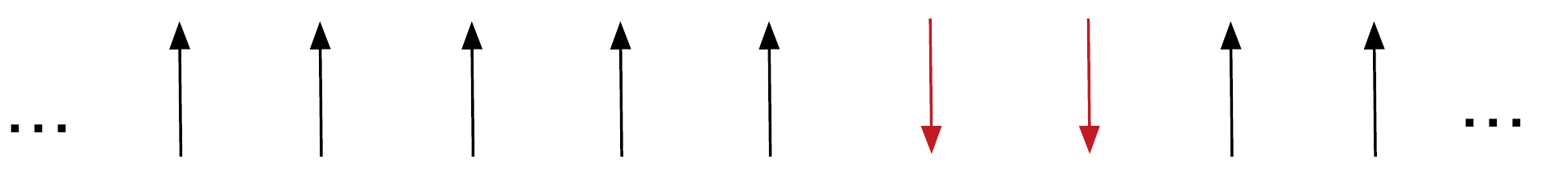} \\
  \includegraphics[width=0.75\textwidth]{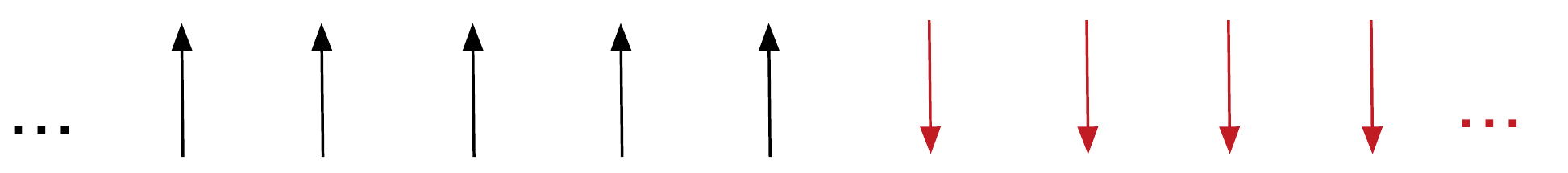} 
  \caption{The first denotes a local excitation, since it can be removed by flipping two spins locally,
  whereas the second one is a topological quasi-particle since it is protected by the boundary conditions
  which are fixed.
 }\label{nafig1}
	\end{figure}
%--------------------------------------- 

So let us now start with an abstract model. We first need a list of particles with their
charges.  Note that here by charges, we mean topological charges. In a condensed
matter system, one can have quasi-particle excitations which are local or which are
topological. For instance, in the toric code model that we studied in the last section,
a spin operator could be applied locally (at a given position) to create a pair
of excitations ($e$-type or $m$ type particles by applying $\sigma_x$ or $\sigma_z$).
So a pair of excitations is a local or trivial type of particle or equivalent to the identity.
But individually, the $e$ or $m$ particles carry a $Z_2$ charge which is topological.
Another example is that of spin flip excitations in an Ising  model as shown in Figs. \ref{nafig1}(a)
and \ref{nafig1}(b).
A single spin-flip
(or even a few spin flips) is a local excitation and is equivalent to the identity as
far as particle type is considered. But a domain wall is a topological excitation
that is protected by the boundary conditions and cannot be removed by local perturbations.
So let us list the topological particles in our system as $a,b,c, \dots$.

Next, we need to specify the fusion rules for these particles. The fusion algebra
is defined as
\beq
a \times b = \sum_c N_{ab}^c c
\eeq
where $N_{ab}^c \ge 1$. This equation simply means that if $N_{ab}^c=0$,then the particle $c$ is
not obtained and if $N_{ab}^c=1$, then the particle $c$ is obtained as a fusion product and if
$N_{ab}^c>1$, then $c$ can be obtained in $N_{ab}^c$ ways. Here, $a,b,c$ are just labels
of the different kinds of particles.
For abelian anyons,  an anyon with statistics parameter $\alpha_1$
will fuse with an anyon with statistics parameter $\alpha_2$ to yield a specific anyon with
statistics parameter $\alpha_3$ and $N_{\alpha_1\alpha_2}^{\alpha_3} = 1$ and is otherwise zero.
But this is not true for non-abelian anyons.  The fusing of two anyons could lead to different
types of particles with different probabilities. In that sense, fusion of particles is like a measurement.
Given two abelian anyons $a$ and $b$, their fusion is well-defined and leads to a unique
answer. But for non-abelian anyons, it does not lead to a single answer. The integers $N_{ab}^c$
define the probabilities of the outcome. The simplest non-abelian model will have $N_{ab}^c \ne 0$ for
at least two distinct values of $c$. The Hilbert space of two (or more) dimensions formed from
these distinct outcomes of fusing two particles is known as the topological Hilbert space of the pair
of non-abelian anyons.

Now, let us consider what happens when we have many non-abelian anyons. For abelian anyons,
each time you bring in a new particle, the fusion rules give you only one possible outcome.
But for non-abelian anyons, since even two anyons can have more than one outcome, a third anyon
can fuse with either of the outcomes and again give rise to more possible outcomes. So one can get
many fusion paths, since fusion is not unique.  Since sometimes, different paths can lead to the
same outcomes, there are consistency conditions that need to be satisfied called pentagon equations.
Once we also include braiding matrices in the game, there are also consistency conditions called hexagon
equations which need to be satisfied.

However, instead of going further ahead with the abstract analysis, we will now change gears and study
a concrete model with non-abelian excitations. The is the Kitaev one-dimensional toy model with unpaired
Majorana fermions. We will explicitly show that these Majorana fermions obey non-abelian statistics
under exchange.

\subsection{Kitaev model in one dimension}

The Hamiltonian for the Kitaev model in one dimension is given by\cite{kitaev3}
\beq
H = -\mu \sum_{x=1}^N n_x - \sum_{x=1}^{N-1} (t c_x^\dagger c_{x+1} +\Delta  c_xc_{x+1} + h.c.) \label{kitaev}
\eeq
where $c_x$ represents  spinless fermions on site $x$, $t$ is the amplitude of hopping to
 nearest neighbour sites, and $\Delta$ is the superconducting parameter and denotes $p$-wave pairing - $p$
 wave because the electrons are of the same kind (spinless or equivalently same projection of spin) - 
 on nearest neighbour sites are paired.  $\mu$ is the chemical potential and
 $n_x=c_x^\dagger c_x$ is the number operator, so that $N=\sum_x n_x$.
 
 Now, let us rewrite the Hamiltonian in terms of new operators called Majorana operators.
 \bea
 c_x &=&\frac{1}{2} (\gamma_{A,x} + i\gamma_{B,x}) \nonumber \\
 c_x^\dagger  &=&\frac{1}{2} (\gamma_{A,x} - i\gamma_{B,x})~. \label{split}
\eea
This implies that $\gamma_{A,x} = c_x +c_x^\dagger$ and  $\gamma_{B,x} = i(c_x - c_x^\dagger)$
are hermitean (self-conjugate) operators. This is the definition of Majorana operators.

Now, let us look at some properties of these  Majorana modes.  
We can check that they are fermions, in the sense that they anti-commute. More precisely, they
satisfy the algebra given by
\beq
\{\gamma_a,\gamma_b\} = \delta_{ab}, \quad \gamma_a^2 = \gamma_b^2 =1~,
\eeq
whereas genuine fermions satisfy
\beq
\{c_a,c_b^\dagger\} = \delta_{ab},  \{c_a,c_b\} = 0, \quad c_a^2=0~.
\eeq
Pairs of Majorana fermions ($\gamma_A$ and $\gamma_B$) can be combined to form genuine fermions which can form a single 2 level
system, depending on whether the fermion state is occupied or unoccupied. The next step is to  consider what
happens when we have $2N$ Majorana fermions. We can pair them up to make $N$ ordinary fermions -
\beq
q_x = \frac{1}{2} (\gamma_{A,x} +i\gamma_{B,x}),\quad q_x^\dagger = \frac{1}{2} (\gamma_{A,x} - i\gamma_{B,x})
\eeq
(the same equations used to split the fermions into Majorana modes given in Eq.\ref{split})
with the number operators at each site $x$ given as $N_x=q_x^\dagger q_x = 0,1$. This gives a $2^N$ dimensional
Fock space.

Why is it interesting to rewrite fermions in terms of pairs of Majorana fermions? Naively, this seems to
be something which can always be done, and does not lead to anything new. But if a pair of  Majoranas
can be spatially separated, then the fermion made from them is delocalised. It is hence, protected from 
local changes that affect only one of them and hence protected from decoherence. This is why
Majorana modes are expected to be relevant in quantum computation.

Now, let us get back to the Kitaev model. To understand the physics in a simple way, let us 
consider two simple limits, where the Hamiltonian
becomes particularly simple. First, consider the case when $\mu=0$ and $t=\Delta$. Here, we get
\beq
H = -it \sum_{x=1}^{N-1} \gamma_{B,x}\gamma_{A,x+1}~.
\eeq
In the other limit, we take $\mu<0$ and $t=\Delta=0$ and get
\beq
H=- \frac {\mu}{2}\sum_{x=1}^{N}(1+ i\gamma_{B,x}\gamma_{A,x})~.
\eeq
What do these two limits mean?

   %--------------- Fig 2 ----------
\begin{figure}
 \includegraphics[width=0.7\textwidth]{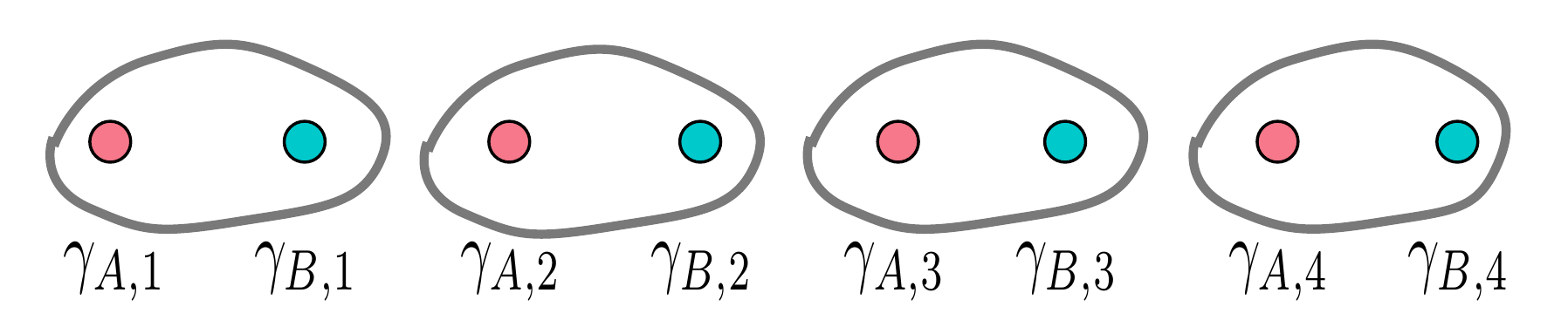} \\
  \caption{Here, the bonds are between Majorana modes at the same site. The ground state
  is unique and the end Majorana modes do not play any special role.
 }\label{nafig2}
	\end{figure}
%--------------------------------------- 

We first analyse the second case. Here, the fermion at each site is simply broken up  into two
Majorana fermions and the $\mu$ term simply couples them as shown in Fig.\ref{nafig2}.  In this case, there
is a unique ground state corresponding to the vacuum state with no fermions. Adding a fermion to the
system costs an energy $\mu$, so the system is gapped.

   %--------------- Fig 3 ----------
\begin{figure}
 \includegraphics[width=0.7\textwidth]{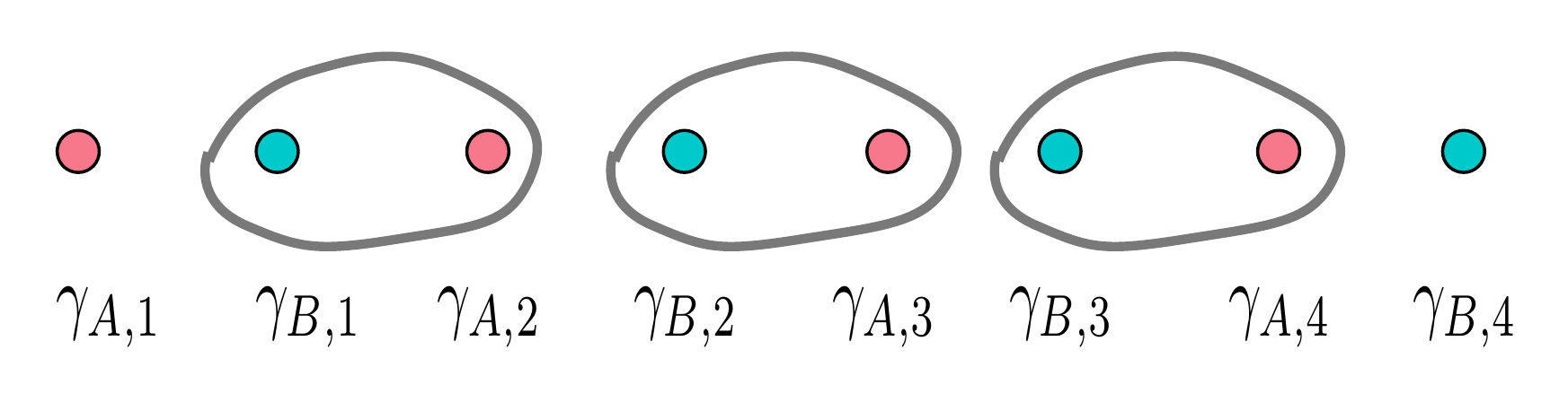} \\
  \caption{Here, the bonds are between Majorana modes on adjacent sites. There are unpaired
  Majorana modes at the two ends. The ground state is doubly degenerate depending on whether
  the fermion state formed from the unpaired Majorana modes is occupied or unoccupied.
 }\label{nafig3}
	\end{figure}
%--------------------------------------- 

The first limit, on the other hand, couples Majorana modes at adjacent sites. In terms of new fermions
$d_x = (\gamma_{B,x} + i\gamma_{A,x+1})/2$, the Hamiltonian can be rewritten as
\beq
H = 2t \sum_{x=1}^{N-1} d_x^\dagger d_x
\eeq
This clearly shows that the system has a gap and the cost of adding a fermion to the system is $2t$.
But the interesting point to note is that the Hamiltonian is completely independent of the two
Majorana modes $\gamma_{A,1}$ and $\gamma_{B,N}$ as shown in Fig.\ref{nafig3}, at the two ends
of the wire. These two Majorana modes can be combined to form a fermion as 
\beq
 c_M =\frac{1}{2} (\gamma_{A,1} + i\gamma_{B,N})~.
\eeq
But this is a highly non-local fermion, since $\gamma_{A,1}$ and $\gamma_{B,N}$ are localised
at opposite ends of the chain. Moreover, since this fermion is absent from the Hamiltonian, the energy
is the same whether or not this fermion state is occupied. So the ground state is degenerate. If $|0>$ is
the ground state, then $c_M^\dagger |0>$ is also a ground state. Note that $\gamma_a^2=1$  
implies that there is no Pauli principle for the Majorana modes - 
in fact, as we saw earlier, there is no notion of occupation number for a single Majorana mode. 
Number operators only exist for fermions formed from pairs of Majorana modes.
 Depending on the occupation or not of the zero energy  mode of the fermion  - i.e, of $c_M$ -
 there exists an odd or even number of fermions in the ground state referred to as 
`fermion parity'. To change the parity, electrons have to be added or removed from the superconductor.
This is unlike normal gapped superconductors, (e.g. the second limit), which have a unique ground state
with even fermion parity.

In the more general case\cite{leinjse}, when $\mu,t$ and $\Delta$ are non-zero, the general features of the topologically
trivial case with a unique ground state, and the topologically non-trivial case with the Majorana edge states
persist. Why do we call them topologically trivial and non trivial in the two cases? Well, in the trivial case
there are no edge states and in the non-trivial case, there are edge states.  In terms of the bulk properties of
the Kitaev chain, one can find the bulk quasiparticle spectrum by going to momentum space and rewriting
the Hamiltonian in Eq.\ref{kitaev} as
\beq
H = \sum_k  \zeta_k c_k^\dagger  c_k +\sum_k ( \Delta_k c_k  c_{-k} +h.c.)
\eeq
using $c_x =\frac{1}{\sqrt{N}} \sum_k e^{ikx}c_k$ and $c_x=c_{x+N}$, $\zeta_k = -2t\cos k-\mu$
and $\Delta_k =-2i \Delta \sin k$
We then find the quasi-particle spectrum by going to Nambu space and writing the Boguliobov-de Gennes
Hamiltonian as
\beq
H = (c_k^\dagger c_{-k} )
\begin{pmatrix}
\zeta_k &\Delta_k^* \\
\Delta_k & -\zeta_k \end{pmatrix}
\begin{pmatrix}
c_k^\dagger \\c_{-k} \end{pmatrix}
\eeq
and finally we get the spectrum $E_k = \sqrt{ (\zeta_k^2 +|\Delta_k|^2)}$.
Hence, the model is gapped, except when $\mu=-2t$ when $k= k_F=0$ or when $\mu=+2t$ when
$k=k_F=\pm \pi$. The lines $\mu = \pm 2t$ are where the system becomes gapless. For $\mu<2t$,
the system is topological and is adiabatically connected to the first limit with Majorana edge states.
For $\mu>2t$, the system is topologically trivial and is adiabatically connected to the second limit
with no edge states.

%Now, let us consider the statistics of the Majorana modes. For $2N$ Majorana modes, we saw that
%the Fock space is $2^N$ dimensional. So the wave-function has  $2^N$ components and under
%exchange of any two of the Majorana modes, it mixes all the $2^N$ components with each other.
%In other words, under exchange, the $2^N$ component wave function is multiplied by a unitary matrix.
%The exchange statistics is clearly non-abelian since arbitrary unitary matrices do not commute.
%

\subsection{Statistics of the Majorana modes}

   %--------------- Fig 4 ----------
\begin{figure}
 \includegraphics[width=0.4\textwidth]{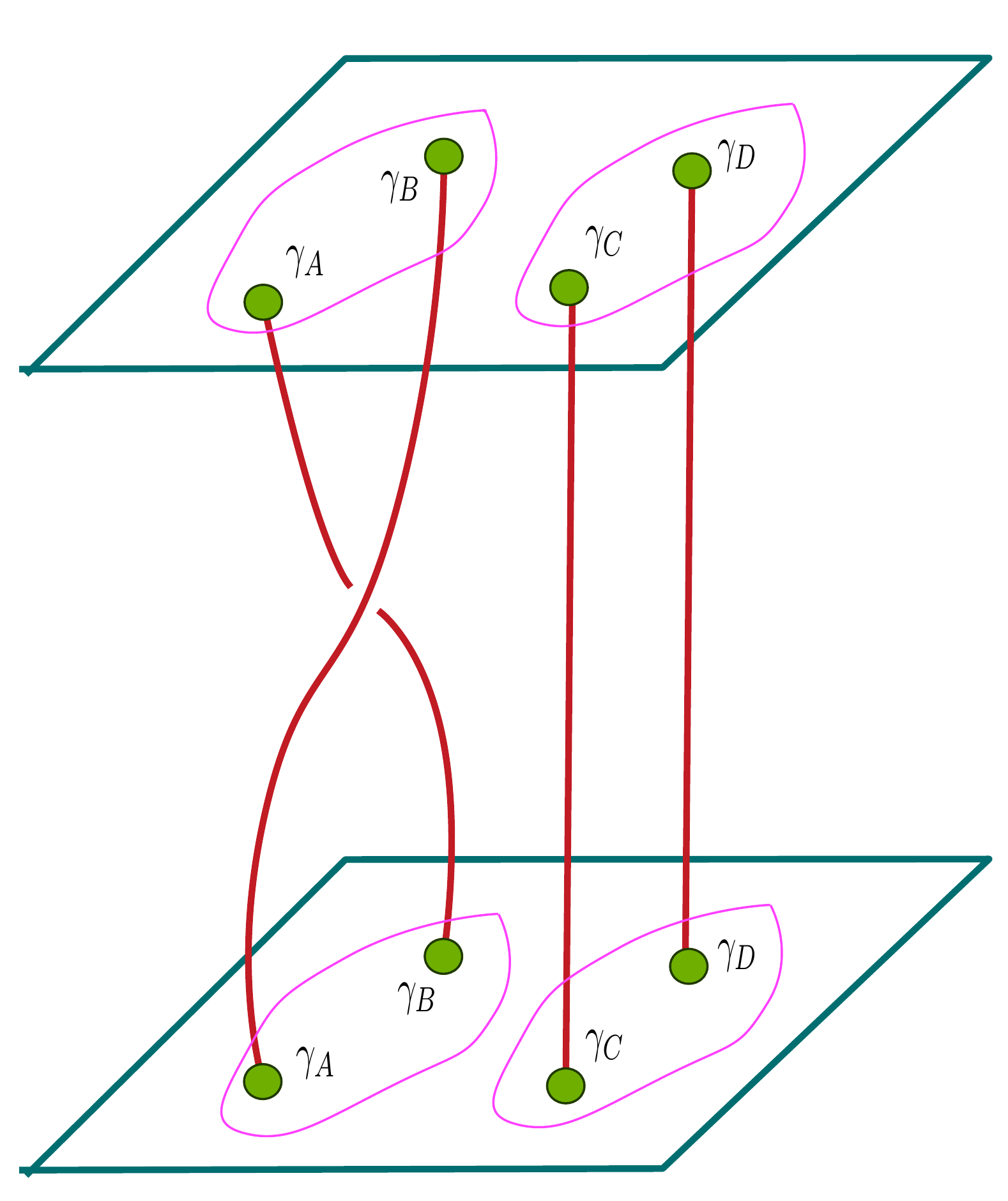} \\
  \caption{Here, we exchange the Majorana modes belonging to the same fermion.
  This only gives rise to a phase and hence abelian statistics.
 }\label{nafig4}
	\end{figure}
%--------------------------------------- 

Now, let us consider the statistics\cite{ivanov,berg} of the Majorana modes. 
Let us start with the simplest case where $N=1$. In this case, there are only two Majorana modes
and the braid group only has a single generator $\tau$. As we saw in the first section,  $\tau$ is 
the operator that  exchanges the Majorana
modes $A$ and $B$ (as shown in Fig.\ref{nafig4}) - 
\beq
\gamma_A \rightarrow \gamma_A^\prime = \tau^\dagger  \gamma_A \tau =e^{i\phi}\gamma_B~,
\eeq
but it does not fix the phase $\phi$ which is arbitrary.  We can choose it to be +1. But then the phase of
\beq
\gamma_B \rightarrow \gamma_B^\prime = \tau^\dagger  \gamma_B \tau=e^{i\phi}\gamma_A \label{exchange}
\eeq
is forced to be $-1$. This is because when there are only 2 Majorana modes, and  the system is isolated,
the fermion parity is forced to be conserved. The fermion state formed from the two Majorana modes
is either occupied or unoccupied and we can check that
\beq
i\gamma_A\gamma_B= (1-2c_M^\dagger c_M).
\eeq
So if the right hand side remains unchanged, then $i\gamma_A\gamma_B$ has to remain unchanged,
which is only possible, if we choose the phases as shown above, since $\gamma_A\gamma_B = -
\gamma_B\gamma_A$.  Here, 
we can choose the exchange operator to be of the form
\beq
\tau = \frac{1}{\sqrt {2}} (1+\gamma_A\gamma_B).
\eeq
It is easy to check that $\tau$ defined this way is unitary and that it actually carries out the
exchange by substituting for $\tau$ in Eq.\ref{exchange}. It is also easy to check that $\tau$
can be rewritten as ${\rm exp}(\pi\gamma_A\gamma_B/4)$. If we write it in terms of
the fermion number operator, 
\beq
\tau = e^{i\pi (1-2n)/4}, \quad {\rm where}  \quad n=c_M^\dagger c_M .
\eeq
Clearly, since $n$ does not change, the statistics parameter is abelian and it cannot rotate
states in the ground state manifold  ($|0>,c_M^\dagger|0>$).

Now, let us see what happens when $N=2$. Here, we have 4 Majorana modes $\gamma_i, i=A \dots D$
which can form 2  normal fermions -
\bea
&c_1 = \frac{1}{2}( \gamma_A+i\gamma_B), ~~c_1^\dagger  = \frac{1}{2}( \gamma_A-i\gamma_B) \nonumber \\
&c_2 = \frac{1}{2}( \gamma_C+i\gamma_D), ~~c_2^\dagger  = \frac{1}{2}( \gamma_C-i\gamma_D)
\eea
The degenerate states of the system are 
given by $|n_1,n_2> =  c_1^\dagger c_2^\dagger |0,0> = \{|0,0>,|1,0>,|0,1>,|1,1>\}$.
Operator $\tau_{AB}$ exchanges the Majoranas $A$ and $B$ keeping $C,D$ unchanged and 
operator $\tau_{CD}$ exchanges the Majoranas $C$ and $D$ keeping $A,B$ unchanged. Similarly,
we can define, $\tau_{AC}$, $\tau_{BD}$, etc.

   %--------------- Fig 5 ----------
\begin{figure}
 \includegraphics[width=0.4\textwidth]{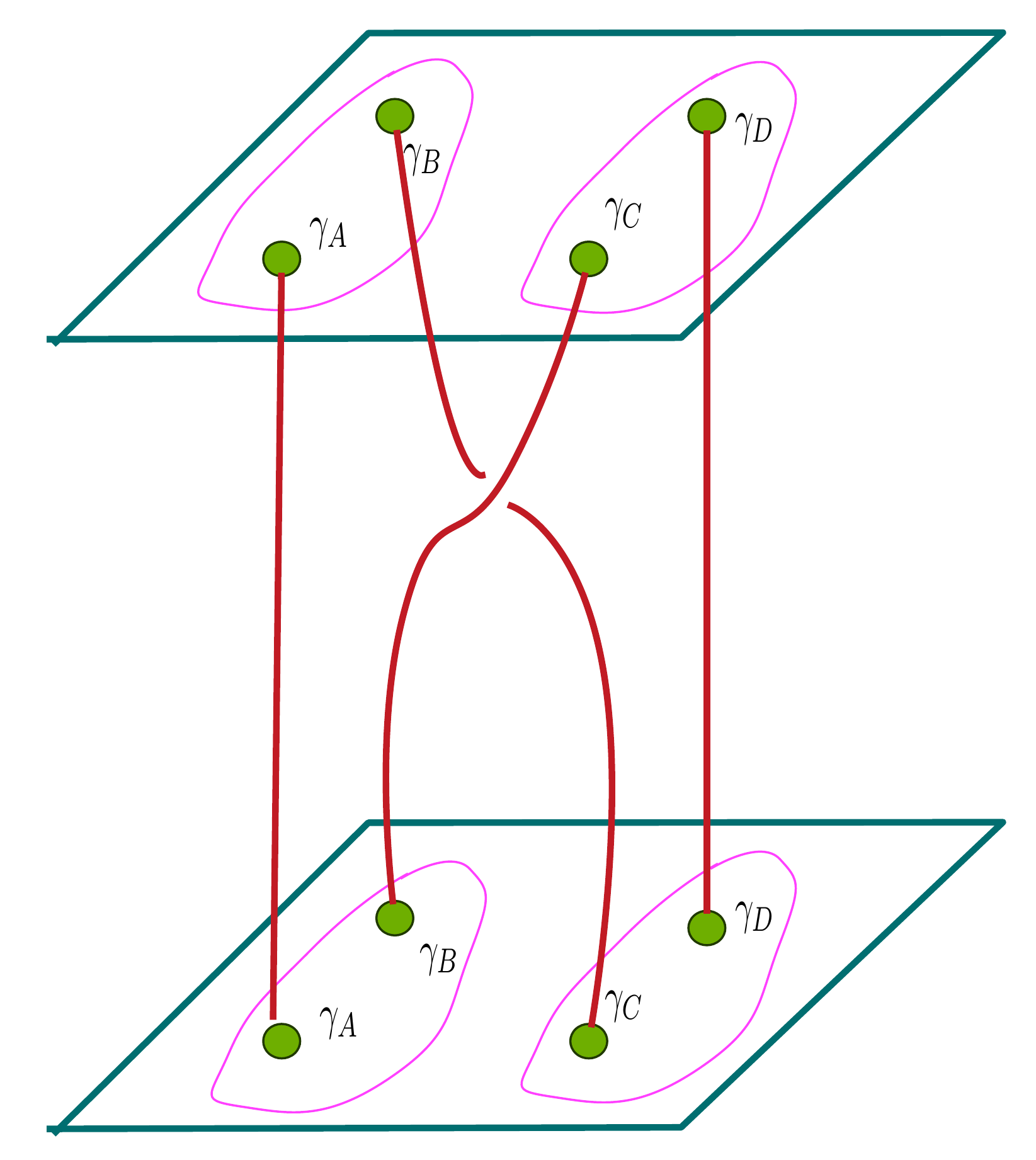} \\
  \caption{Here, we exchange the Majorana modes belonging to two different fermions. 
  This leads to a unitary rotation in the space of degenerate states and hence, to non-abelian statistics.
 }\label{nafig5}
	\end{figure}
%--------------------------------------- 

It is now clear that analogous to the $N=1$ case, if we exchange the two Majorana zero modes
from the same fermion, as shown in Fig.\ref{nafig4}, we will only get a phase - i.e., 
\bea
\tau_{AB}|n_1,n_2> &=& e^{{i\pi (1-2n_1)/4}}|n_1,n_2> \nonumber \\
\tau_{CD}|n_1,n_2> &=& e^{{i\pi (1-2n_2)/4}} |n_1,n_2>~.
\eea
In the first equation, $n_2$ comes along for a ride and in the second equation, $n_1$ comes along for a ride.
So both these operators are abelian operators.
But now, let us 
exchange one Majorana from one of the fermions with another  Majorana from the other fermion,  (as shown in Fig.\ref{nafig5}) -
\beq
\tau_{BC} = \frac{1}{\sqrt {2}} (1+\gamma_B\gamma_C).
\eeq
In terms of the fermions $c_1$ and $c_2$, this can be written as
\beq
\tau_{BC} = \frac{1}{\sqrt {2}} [1-i(c_1-c_1^\dagger) (c_2+c_2^\dagger)].
\eeq
 Now acting this on $|n_1,n_2>$ does
not lead to a phase. Instead, it leads to a rotation in the space of degenerate states given by
\beq
\tau_{BC}|n_1,n_2> = \frac{1}{\sqrt{2}} [ |n_1,n_2> +i (-1)^{n_1} |1-n_1,1-n_2> ]~.
\eeq
If we now consider sequential exchanges, it  is clear that different exchanges will not commute
with one another - the final state will depend on the order of the operations. This is what
is meant by saying that the Majorana particles have non-abelian statistics under exchange.

The derivation of the non-abelian statistics is not dependent on the details of how the exchange between
the particles is carried out, and hence it cannot be changed by disorder or local details. It is topologically stable.

To understand multiple non-abelian anyons, as we already mentioned, we need to understand
fusion paths, since the fusion rules do not need to unique results. These fusion paths represent
a basis of the degenerate ground state manifold, and are most conveniently studied in terms
of conformal blocks of the appropriate conformal field theory\cite{anyonsqc}.  But that is beyond the scope
of these lectures and we will stop here.

\section{Conclusion}

Let me conclude by repeating the main message of these lectures - understanding the notion
of anyons and non-abelian anyons is an exciting field today. The study of these excitations could lead to an understanding of concepts like decoherence and entanglement which are relevant in quantum computation. Work on  non-abelian states, in general, is still in its infancy. For young researchers, hence, this should be a useful and relevant topic of study at the crossroads of condensed matter physics and quantum information. For more information and references, there are many recent reviews\cite{stern,anyonsqc,alicea} available
 on the net.

%{\it Acknowledgments .-} 

\end{document}